\documentclass[12pt, a4paper]{article}
\usepackage{jheppub}
\usepackage{ifpdf}
\usepackage{graphicx}
\usepackage{microtype}
\usepackage{amsmath,amssymb}
\usepackage[utf8x]{inputenc}
\usepackage[T1]{fontenc}
\usepackage{ae}
\usepackage{aecompl}
\usepackage{titlesec}
\usepackage[export]{adjustbox}
\def\bfseries{\fontseries \bfdefault \selectfont \boldmath}
\usepackage{bm}
\usepackage{subfigure}
\usepackage[normalem]{ulem} 
\usepackage{xcolor,soul}
\usepackage{enumitem}
\definecolor{olive}{rgb}{0.3, 0.54, .1}

\DeclareMathOperator{\extdm}{d}
\newcommand{\extd}{\extdm \!} 
\titleformat*{\section}{\large\bfseries}
\titleformat*{\subsection}{\bfseries}
\titleformat*{\subsubsection}{\bfseries}
\newcommand{\eq}[2]{\begin{equation}
#1\label{#2}
\end{equation}}
\newcommand{\bes}{\begin{eqnarray*}}
\newcommand{\ees}{\end{eqnarray*}}
\newcommand{\bel}[1]{\begin{eqnarray}\label{#1}}
\newcommand{\be}{\begin{eqnarray}}
\newcommand{\ee}{\end{eqnarray}}
\newcommand{\bea}{\begin{eqnarray}}
\newcommand{\eea}{\end{eqnarray}}

\newcommand{\nn}{\nonumber}
\newcommand{\GN}{G_5}

\newcommand{\source}{j}

\newcommand{\mytitle}{Relativistic hydrodynamics with phase transition}

\title{\boldmath\mytitle}

 \author{F.~Taghinavaz}

 \affiliation{ School of Particles and Accelerator, Institute for Research in Fundamental Sciences (IPM), P.O. Box 19395-5531, Tehran, Iran.}
 \emailAdd{ftaghinavaz@ipm.ir}

\abstract{ Assessing the applicability of hydrodynamic expansions close to phase transition points is crucial from either theoretical or phenomenological points of view. We explore this within the gauge/gravity duality, using the Einstein-Klein-Gordon model, a bottom-up string theory construction. This model incorporates a parameter, $B_4$, that simulates different types of phase transitions in the strongly coupled field theory existing at the boundary. We thoroughly examine the thermodynamics and dynamics of time-dependent, linearized perturbations in the spin-2, spin-1, and spin-0 sectors. Our findings suggest that "hydrodynamic series breakdown near transition points" is valid exclusively for second-order phase transitions, not for crossovers or first-order phase transitions. Additionally, we observe that the high-temperature and low-temperature limits of the radius of convergence for the hydrodynamic series ($q^2_c$) are equal. We also discover that the relationship $(\text{Max}\vert q^2_c\vert)_{\text{spin-2}} < (\text{Max}\vert q^2_c\vert)_{\text{spin-0}} < (\text{Max}\vert q^2_c\vert)_{\text{spin-1}}$ is consistent for different spin sectors, regardless of the phase transition type. At the chaos point, we observe the emergence of pole-skipping behavior for both gravity and scalar perturbations at $\omega_n = - 2\pi T n i$. Lastly, comparing the chaos momentum with $q^2_c$, we find that $q^2_{ps} < q^2_c$, except for extremely high temperatures.
}

\begin{document}

\maketitle

\section{Introduction}\label{sec: Introduction}
Our knowledge of strongly interacting matter in extreme conditions has been growing in recent decades. Heavy-ion experiments at relativistic energies can create a hot and dense quark matter in which the composed particles can weakly interact with each other. This is the well-known Quark-Gluon Plasma (QGP) phase that is reachable in the early universe, inside compact stars, or at terrestrial experiments \cite{Rischke:2003mt, Annala:2019puf, Shuryak:2004tx}. In the limit of today’s energies, it has been confirmed that the QGP shows fluid-like or collective behaviors, and relativistic hydrodynamics (RH) as an effective theory is a powerful tool to describe such collective motions of heavy-ion particles \cite{Busza:2018rrf}.

Relativistic hydrodynamics has been remarkably successful in experimental applications. For instance, the flow patterns of emitted particles \cite{Heinz:2013th, Teaney:2000cw} or the particle yields seen at the detectors can be fully described by RH hybrid codes \cite{Song:2007ux, Jeon:2000wg}. Additionally, there is evidence to suggest that collective behavior can also be observed in the collision of lighter particles, such as in proton-proton or electron-electron interactions \cite{CMS:2010ifv, Jia:2017hbm, CMS:2016fnw}. This has led to the recognition of the "unreasonable effectiveness" of RH \cite{Noronha-Hostler:2015wft}.

However, experimental research has also shown that the QGP is a strongly interacting matter, rendering perturbative approaches invalid \cite{Shuryak:2003xe}. As a result, non-perturbative methods like the gauge/gravity duality have become essential for studying the QGP's properties. One famous example of this duality is the AdS/CFT correspondence where the type IIB supergravity on the AdS$_5 \times$S$^5$ is dual to the 4$-$dimensional supersymmetric Yang-Mills (SYM) theory living on the boundary of the  AdS$_5$ \cite{Maldacena:1997re, Witten:1998qj, Aharony:1999ti}. This duality maps a strongly coupled quantum gauge field theory to a weakly coupled classical gravity in one higher dimension. One of the remarkable predictions of the AdS/CFT is the universal ratio of shear viscosity to entropy density,  $\eta/s=1/(4\pi)$ \cite{Policastro:2001yc, Son:2002sd}, which agrees very well with the experimental data \cite{Heinz:2013th}.

Theoretically, the RH approach is an effective field theory that operates in the long wave-length regime of momenta where each conserved current can be expressed in terms of a gradient expansion \cite{Kovtun:2012rj, Romatschke:2017ejr}.  The coefficients of this series, known as transport coefficients, provide valuable insights into the underlying microscopic theory. Questions about the convergence of this series and the potential impact of divergences on physical outcomes have led to further research. For instance, the divergences of the RH series in the real space \cite{Heller:2013fn} can be handled by using the Borel-Pad\'e techniques \cite{Heller:2015dha, Heller:2016rtz,  Aniceto:2015mto, Aniceto:2018uik, Shokri:2020cxa}.
This can offer insights into the origin of these divergences and the potential region of convergence for the RH series. Moreover, the convergence of the gradient expansion in real space can be related to the radius of convergence of RH in momentum space \cite{Heller:2020uuy, Heller:2020jif}.

One of the key guidelines for today's strong interaction research is to explore the vicinity of critical points within the QCD phase diagram \cite{Lovato:2022vgq, Achenbach:2023pba}. This is a challenging area due to the multitude of degrees of freedom involved. Fluctuations play an important role near the transition points, necessitating modifications to the standard relativistic hydrodynamics (RH) framework \cite{Bluhm:2020mpc, Stephanov:2017ghc}. Thermodynamic properties in this region are not well understood, as lattice QCD simulations are hindered by the sign problem, leading to a reliance on effective models \cite{Bzdak:2019pkr, Parotto:2018pwx}. The BEST (Beam Energy Scan Theory) collaboration at the RHIC laboratory has been established with the primary goal of thoroughly examining the QCD phase diagram\cite{An:2021wof}.

The validity of the RH series near critical points is a crucial area of investigation. In this study, we examine the thermodynamics and dynamics of time-dependent, linearized perturbations within a holographic model known as the Einstein-Klein-Gordon model  \cite{Gubser:2008ny}. This phenomenological string theory model maps a strongly coupled scalar field theory on the boundary of spacetime to a weakly coupled gravity model by the AdS/CFT correspondence. The model includes a family of superpotentials parameterized by $B_4$, which can emulate various kinds of phase transitions of the boundary theory. For example, $B_4=0$ mimics the crossover phase transition, $B_4 = - 0.0098$ imitates the second-order phase transition, and any $B_4 < - 0.0098$ can provide us with the first-order phase transition. This is notable because introducing small perturbations in this context is akin to studying RH for systems undergoing different types of phase transitions. This holographic model lacks a known analytical solution, so numerical solutions are employed. The parametrization freedom is fixed by choosing the Gubser gauge $u:=\phi(u)$, resulting in a free parameter $\phi_H$, which represents the horizon value of the scalar field \cite{Gubser:2008ny}. This free parameter acts as an external source, with all quantities dependent on it, reflecting the many-body properties of the theory.

The paper is organized as follows. In section \ref{sec: Summary of main results}, the main results of the theoretical and numerical computations are reviewed. In section \ref{sec: model}, the ingredients of the Einstein-Klein-Gordon action, including the solutions for thermal states and thermodynamics derived from this background, are provided. To avoid mathematical complexities, details are provided in separate Appendices \ref{appendix-HRG}, \ref{appendix-nearhor}. In Appendices \ref{sec:low-T BH} and \ref{sec: short}, the low and high-temperature limits of the solutions and perturbations are investigated. In section \ref{sec: QNM-eqs}, the linearized equations of perturbations in the master formula framework are given. In section \ref{sec: remarks}, some important points are elaborated on. In section \ref{sec: crossover-results}, the QNM spectra and radius of convergence for the crossover phase transition are discussed. Distinct subsections are devoted to describing the results in the spin-2, spin-1, and spin-0 sectors. In section \ref{sec: secondorder-results}, the numerical outputs for the second-order phase transition are described, and different subsections are responsible for giving the details of each spin sector. Similar work is done for the first-order phase transition in section \ref{sec: firstorder-results}. In section \ref{sec: pole-skipping}, the pole-skipping feature of this model is studied in detail, and we compare the radius of convergence with the chaos momentum. Finally, the paper concludes with a summary and an outlook to further directions in section \ref{sec: conclusion}. 

\section{Summary of main results and method}\label{sec: Summary of main results}
The great advantage of the present paper is to explore the hydrodynamic series in various kinds of phase transition. Due to the diversity and intricacy of the outcomes, we review the upcoming results in this section. For more convenience, we itemize them below.
\footnote{We emphasize that the given results and observations belong only to the current paper. In this sense, they are not universal and it is better to investigate them in more holographic models with or without phase transition.}
\begin{itemize}[label={\textbf{(\alph*)}},leftmargin=1.5cm,align=left] 
\item[$\bullet$ \bf{Action:}]  
The Einstein-Klein-Gordon (EKG) model, a string theory framework, simulates phase transitions in a strongly coupled boundary scalar field theory. It breaks conformal symmetry through a non-zero scalar field expectation. Using AdS/CFT, we calculate the boundary's energy-momentum tensor and scalar field one-point function to determine phase transition types. The EKG action's parameter $B_4$ controls these transitions: $B_4=0$ for crossover, $B_4=-0.0098$ for second-order, and $B_4=-0.02$ for first-order transitions. The model's transition points are detailed in the table. \ref{table: Tcphic}. Employing the Gubser gauge to fix the residual symmetries, numerical solutions to the equations of motion yield thermodynamic variables like pressure ($p$), speed of sound ($c_s^2$), entropy ($s$), and transport coefficients ($\eta,\xi$). The ratio $\eta/s$ consistently equals $1/(4\pi)$, while $\xi/\eta$ peaks near transition points, varying by transition type. First-order transitions exhibit instability near the transition point, suggesting thermodynamic state tunneling, visible as negative $c_s^2$ values or $\xi/\eta$ plot twists. Near-horizon expansion and low-temperature solutions are discussed in separate appendices.
\item[$\bullet$ \bf{ Linearized Equations and numerical method:}]
The master formula approach efficiently derives linearized perturbation equations for various spin sectors, simplifying QNM numerical analysis \cite{Jansen:2019wag, Buchel:2021ttt}. Utilizing spectral methods with Chebyshev polynomials, this approach quickly converges and minimizes errors in QNM calculations for black holes and compact objects \cite{Grandclement:2007sb}. Numerical tasks are performed at two discretization points, with results within 0.1\% discrepancy selected. The outcomes include both hydro and non-hydro modes, the latter being crucial near collision points.
\item[$\bullet$ \bf{Radius of convergence:}]
To determine the radius of convergence for the hydrodynamic series in systems with phase transitions, we employ a method involving the analytical continuation of momenta into the complex plane ($q^2 = \vert q^2\vert  e^{i\theta}$) within the QNM spectra. This approach identifies the smallest points where modes begin to collide at specific $\theta$ and $\vert q^2\vert $, offering insights into the convergence behavior near phase transition points \cite{Withers:2018srf, Grozdanov:2019kge, Grozdanov:2019uhi}.
\item[$\bullet$ \bf{Summary of crossover results:}] 
\item[] In the \textbf{spin-2 sector}, modes don't mix for real $q^2$, meaning identical high-temperature and low-temperature positions of QNMs due to the non-transition nature of the crossover. Collisions occur between the two lowest non-hydro modes and $q^2_c$ increases near the crossover phase transition. Furthermore, there is a high-$T$/low-$T$ duality in $q^2_c$ plots which refers to a nearly identical radius of convergence at high and low temperatures.

In the \textbf{spin-1 sector}, high-$T$ and low-$T$ modes don't mix, but collisions can happen at real and positive $q^2$, involving hydro modes and the lowest gravity non-hydro modes. $q^2_c$ also increases near the transition point.

In the \textbf{spin-0 sector}, modes appear in pairs due to coupled gravity and scalar field perturbations. Collisions for $q^2_c$ involve the hydro mode and the closest scalar non-hydro mode at low-$T$ and high-$T$, and the nearest gravity non-hydro mode at middle points. $q^2_c$ increases near the transition point, similar to the spin-1 sector.
We show in Fig. \ref{fig:1} the $q^2_c$ versus temperature in the crossover EoS for different spin sectors.

\begin{figure}
    \centering
    \includegraphics[width=0.333\textwidth, valign=t]{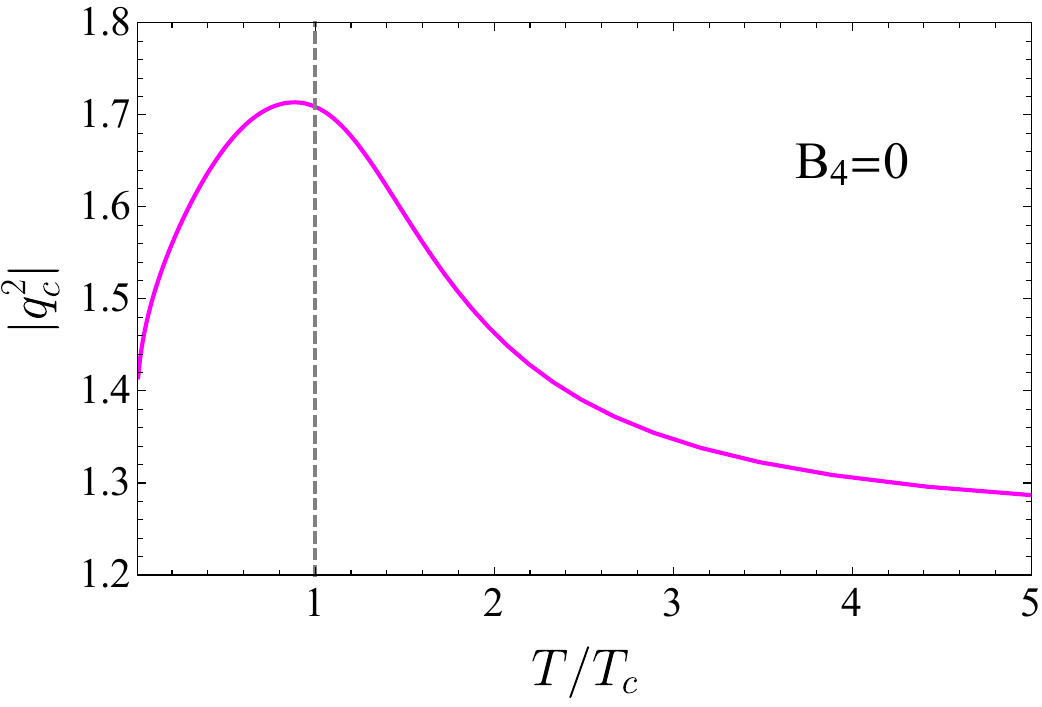}
    \includegraphics[width=0.323\textwidth, valign=t]{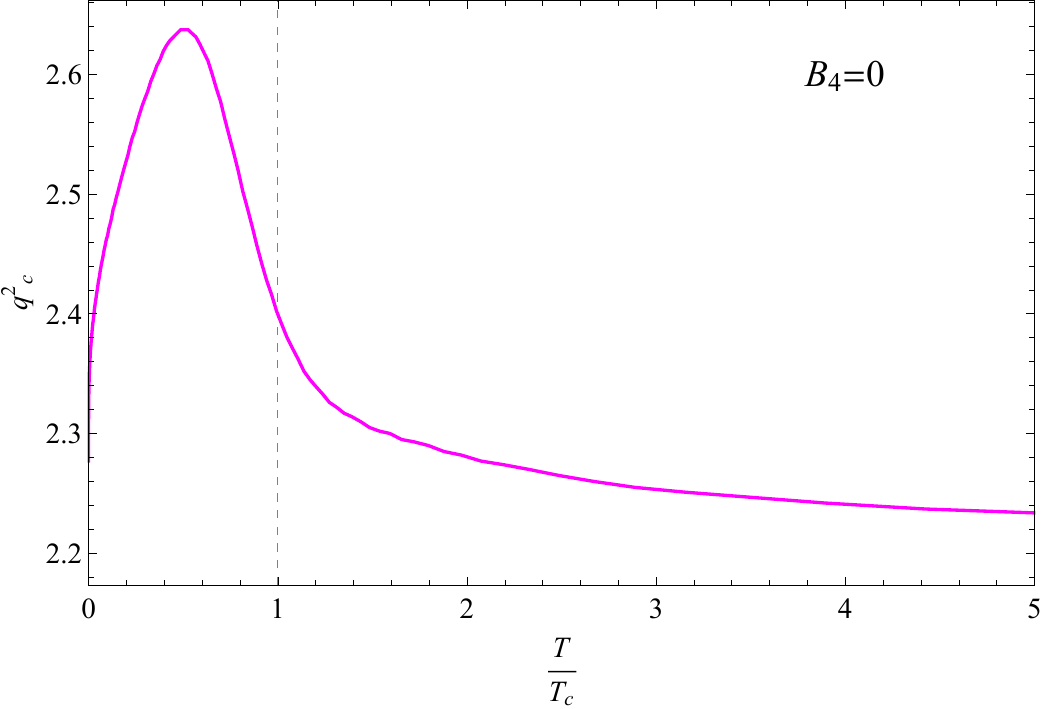}
     \includegraphics[width=0.323\textwidth, valign=t]{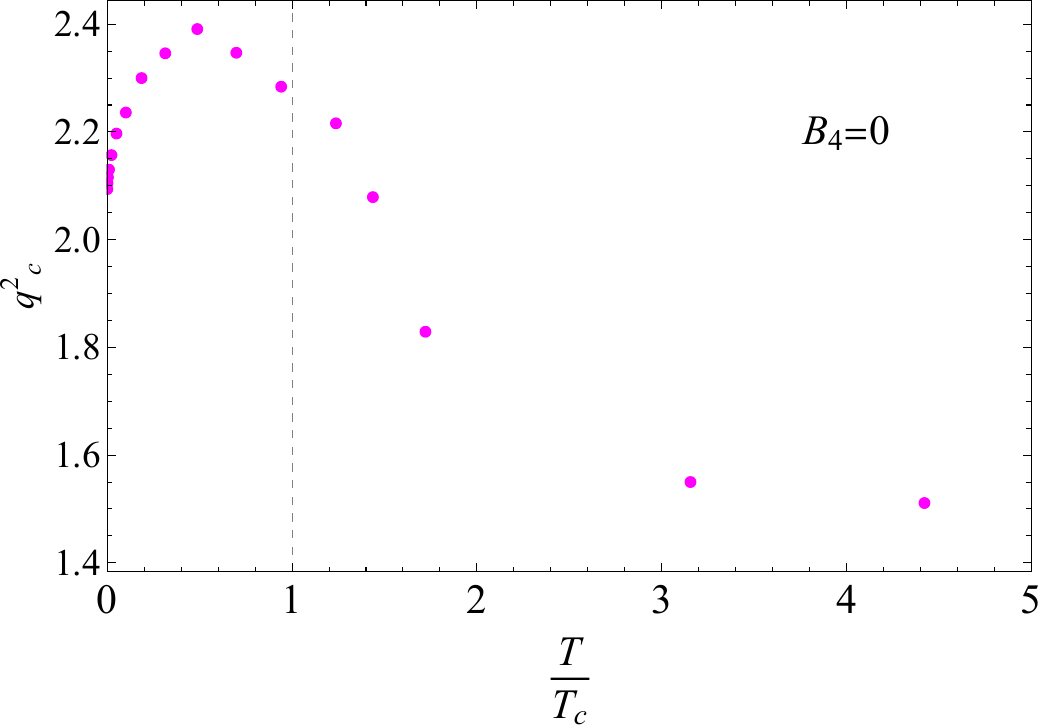}
    \caption{Plots of the radius of convergence in terms of $T/T_c$ for different spin sectors in the crossover EoS. The results for spin-2, spin-1, and spin-0 are sketched from left to right.}
    \label{fig:1}
\end{figure}
\item[$\bullet$ \bf{Summary of second-order results:}] 
\item[] In the \textbf{spin-2 sector}, for real $q^2$, the second lowest mode diverges at low-$T$, while others converge to 5D-AdS black hole results. This follows $\Omega_{i+1}(T \gg T_c) \to \Omega_i(T\ll T_c)$ for $i > 2$. No collisions occur at positive $q^2$, and the first collision is between non-hydro modes at $\theta = \pi$. Near the transition point, $q^2_c$ decreases (Figs. \ref{fig: SO-Real-Momenta-EoS-Im-Re-Omega}-\ref{fig: convergence-radius-phiH-2nd}).

In the \textbf{spin-1 sector}, collisions happen for positive $q^2$ between hydro modes and the nearest gravity non-hydro modes. Near the second-order phase transition, $q^2_c$ decreases, with identical high-$T$/low-$T$ values (Figs. \ref{fig: SO-Real-Momenta-EoS-Im-Re-Omega-spin1}-\ref{fig: collision-SO}).

In the \textbf{spin-0 sector}, the number of escaping modes doubles due to coupled equations. At real $q^2 \neq 0$, high-$T$/low-$T$ limits of hydro modes are the same. Collisions for the radius of convergence are similar to the crossover spin-0 case. At very low temperatures, gravity and scalar non-hydro modes pair with equal frequencies, except for the escaping modes.

We could say that the paradigm "breakdown of the hydro series near the transition points" seems to be true in the second-order phase transition. This fact is seen in Fig. \ref{fig:2} where we present the $q^2_c$ for different spin sectors in the second-order EoS.

\begin{figure}
    \centering
    \includegraphics[width=0.333\textwidth, valign=t]{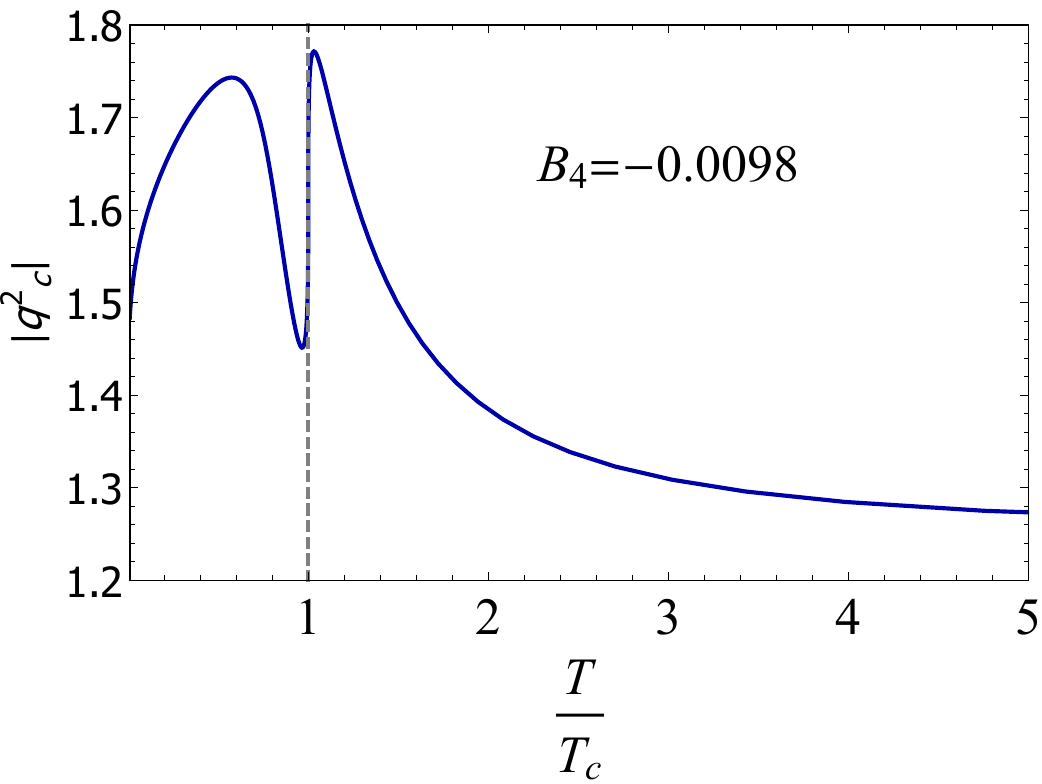}
    \includegraphics[width=0.323\textwidth, valign=t]{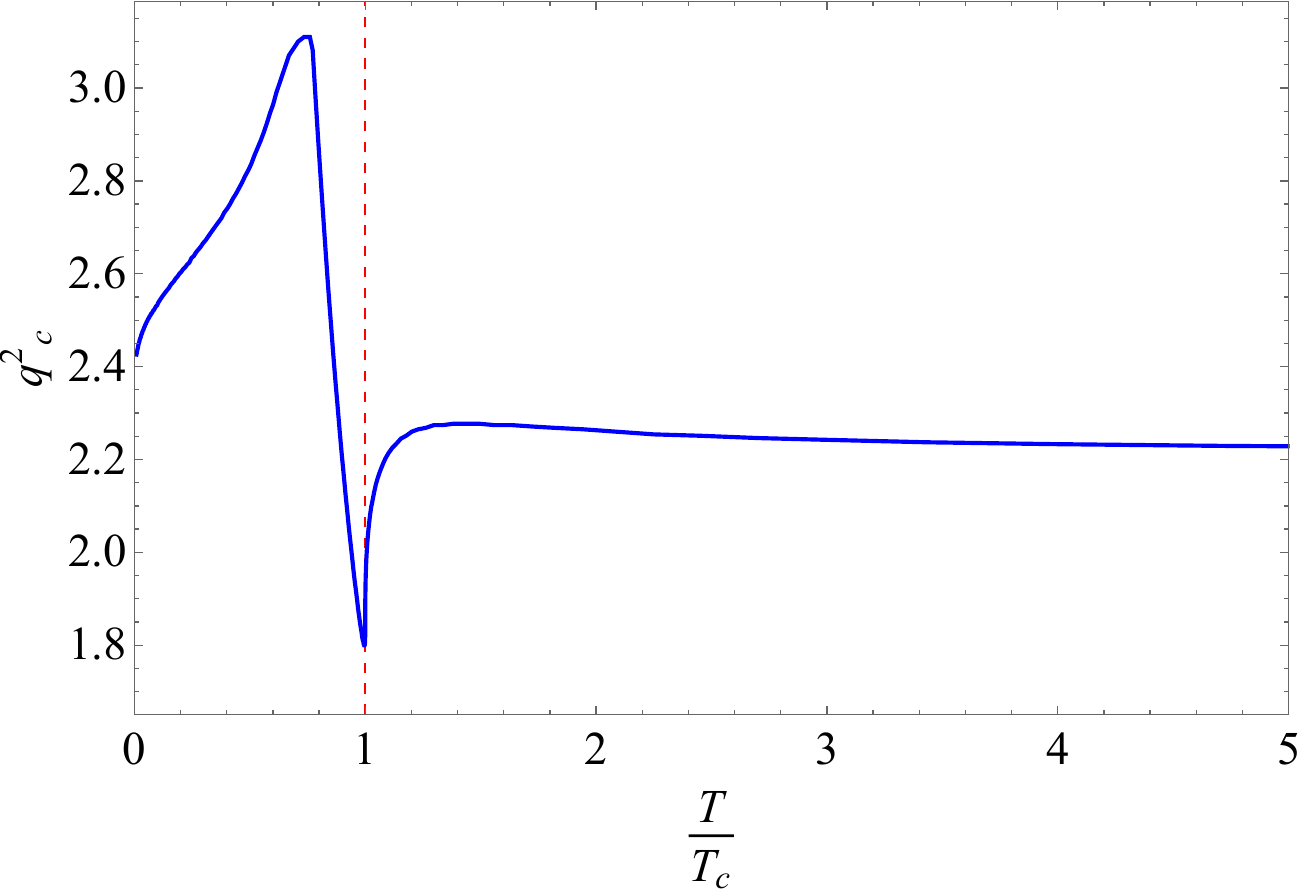}
     \includegraphics[width=0.323\textwidth, valign=t]{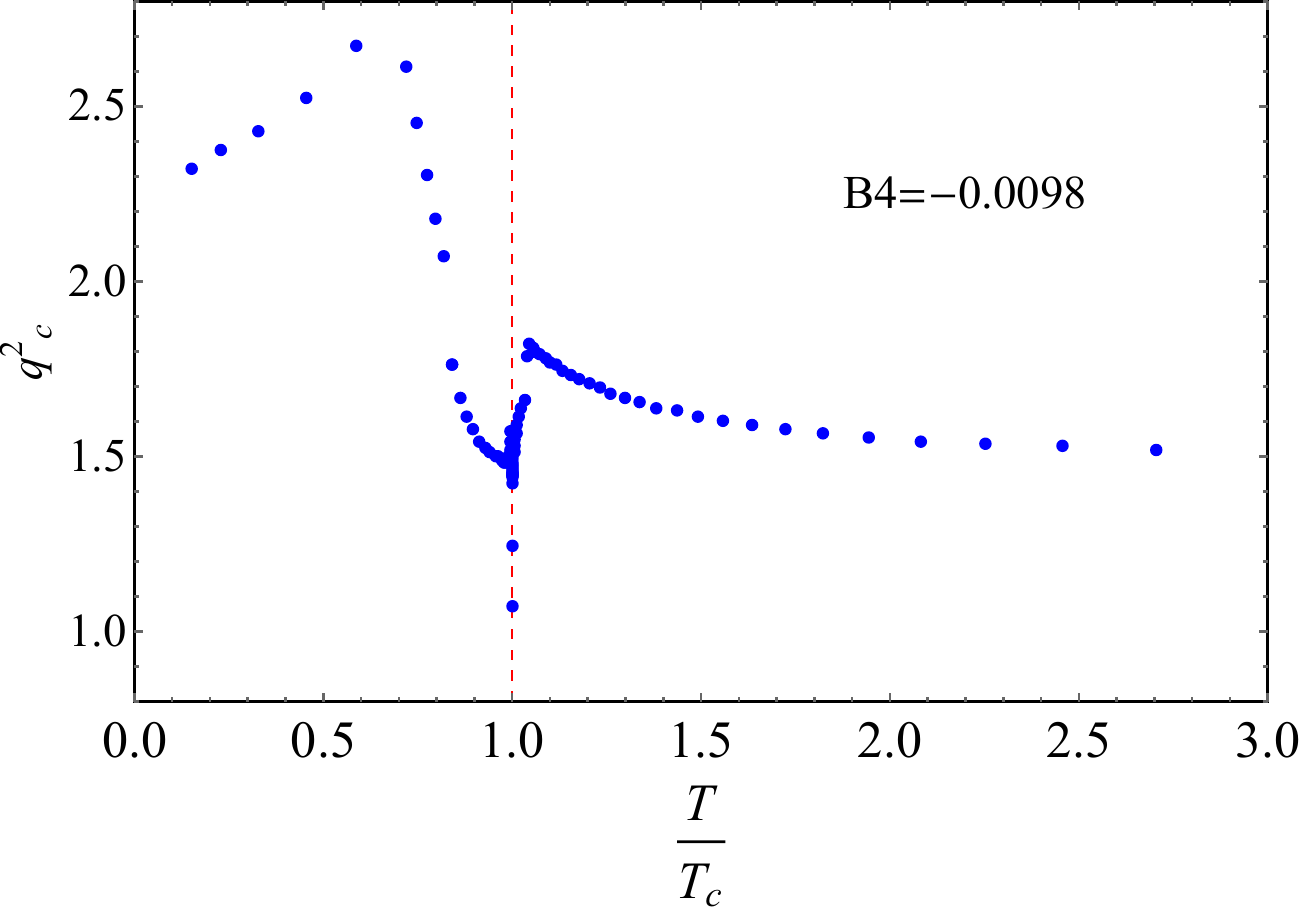}
    \caption{Plots of the radius of convergence in terms of $T/T_c$ for different spin sectors in the second-order EoS. The results for spin-2, spin-1, and spin-0 are sketched from left to right.}
    \label{fig:2}
\end{figure}
\item[$\bullet$ \bf{Summary of first-order results:}] 
\item[] In the \textbf{spin-2 sector}, the number of escaping modes exceeds those in the second-order case. Collisions occur for negative $q^2$ between two non-hydro modes. For the first-order transition, marked by three $\phi_H$ values, $q^2_c$ increases near the largest $\phi_H$, with no clear pattern elsewhere. High-$T$/low-$T$ equality in $q^2_c$ curves is observed (Figs. \ref{fig: FO-Real-Momenta-EoS-Im-Re-Omega}-\ref{fig: convergence-radius-phiH-1st}).

In the \textbf{spin-1 sector}, hydro modes at $q^2 \neq 0$ are the same at low and high temperatures. Collisions happen for positive $q^2$ between the gravity hydro and the smallest gravity non-hydro mode. Near the largest $\phi^c_H$, $q^2_c$ increases, with no clear behavior elsewhere (Figs. \ref{fig: FO-Real-Momenta-EoS-Im-Re-Omega-spin1}-\ref{fig: rc-shear-fo}).

In the \textbf{spin-0 sector}, the escaping mode behavior is complex for real $q^2$. Around $3 \lesssim \phi_H \lesssim 8$, the real part of the hydro modes vanishes, and the imaginary part splits—a known property of mode collision in hydrodynamics. This allows $q^2_c$ to occur for positive values. Similar to previous first-order results, an increase in $q^2_c$ near the largest $\phi^c_H$ is observed (Figs. \ref{fig: FO1-Real-Momenta-EoS-Im-Re-Omega-spin0}-\ref{fig: rc-FO-spin-0}).

We could say that The paradigm "breakdown of the hydro series near the transition points" seems to be not true in the first-order phase transition. This reality is visible in Fig. \ref{fig:3} where we depict the $q^2_c$ in terms of $T/T_c$ in the first-order EoS for different spin sectors.

\begin{figure}
    \centering
    \includegraphics[width=0.333\textwidth, valign=t]{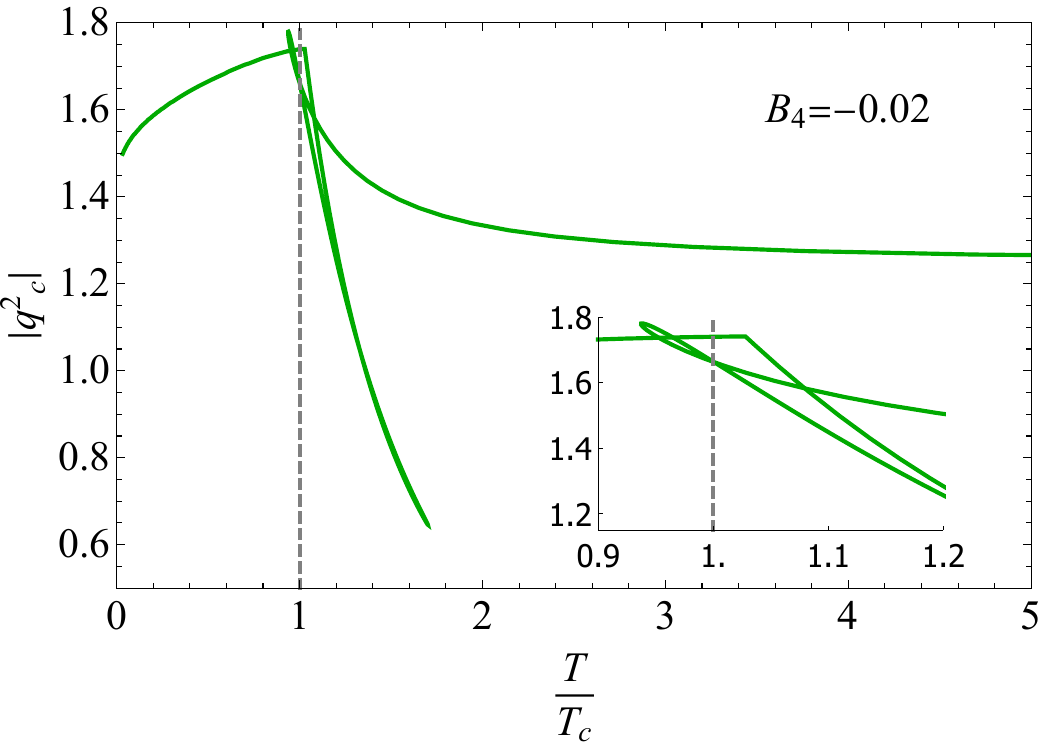}
    \includegraphics[width=0.323\textwidth, valign=t]{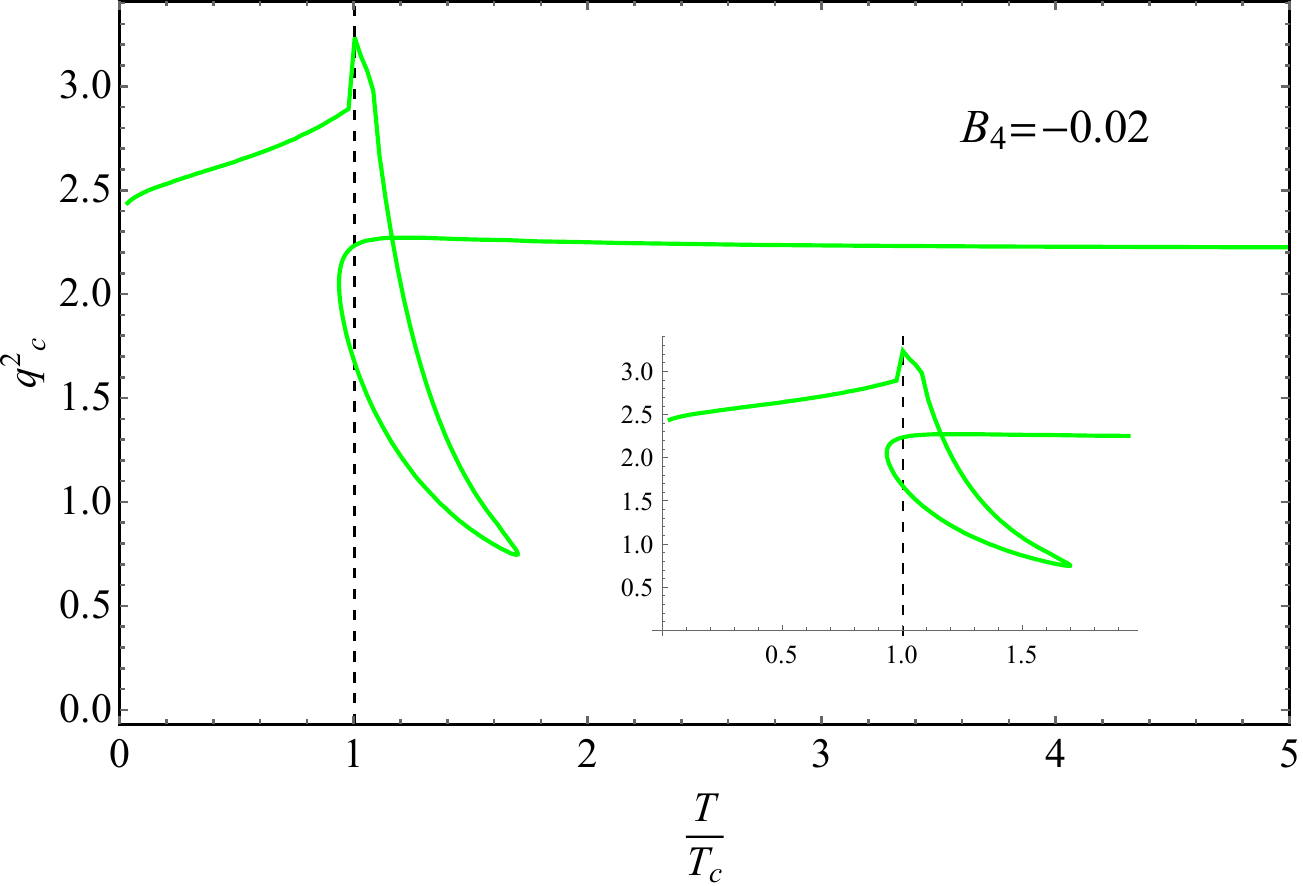}
     \includegraphics[width=0.323\textwidth, valign=t]{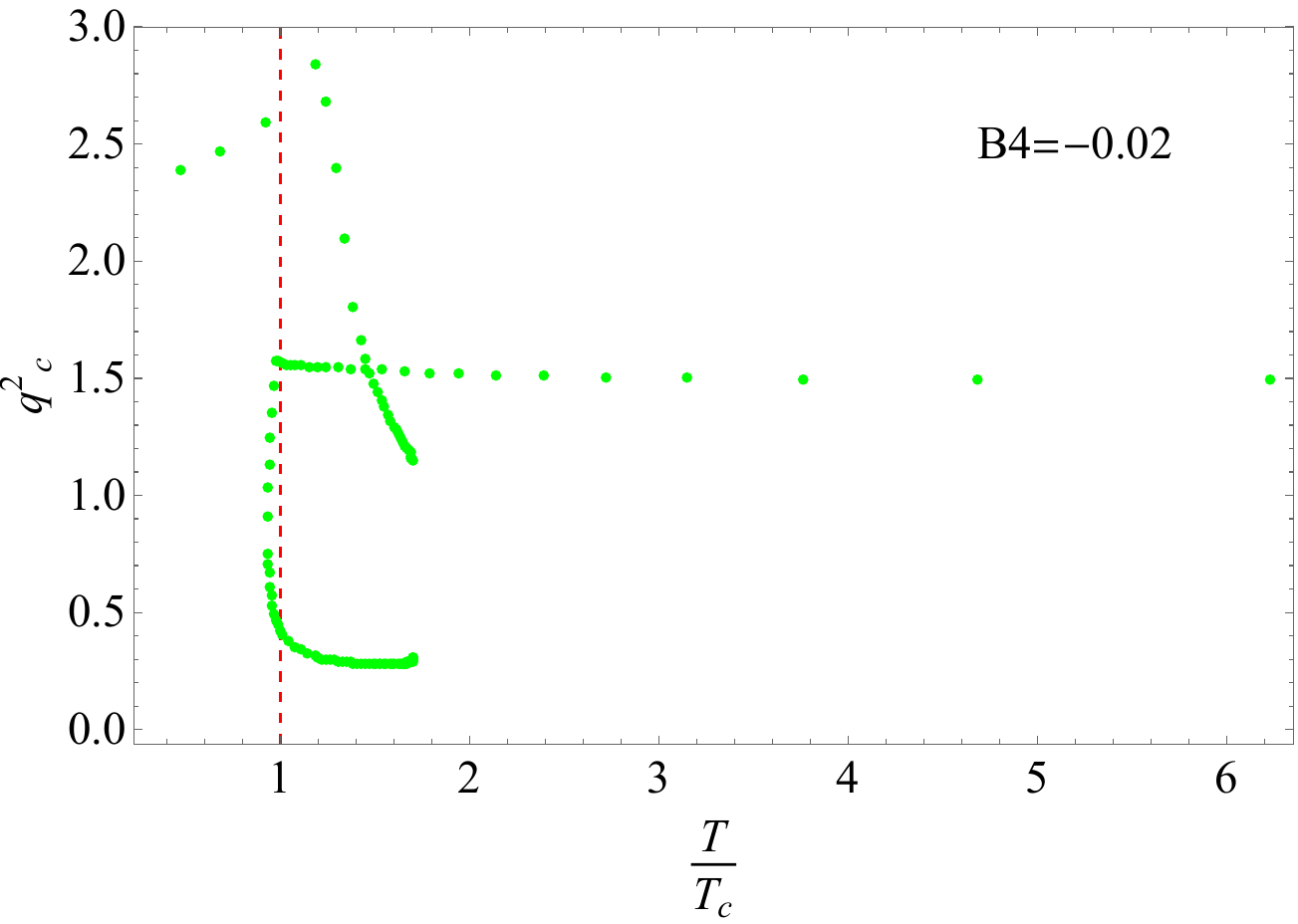}
    \caption{Plots of the radius of convergence in terms of $T/T_c$ for different spin sectors in the first-order EoS. The results for spin-2, spin-1, and spin-0 are sketched from left to right.}
    \label{fig:3}
\end{figure}
\item[$\bullet$ \bf{Results for pole-skipping:}]
At the chaos point, namely  $\omega = i \lambda_L = 2 \pi T i$ and $k_\star = i k_0 = i \lambda_L/v_B$, it has been shown that the $"v v"$ component of Einstein's equation in the spin-0 sector becomes identically zero which is a sign of multivaluedness of $G^R_{T_{0 0} T_{0 0}}(\omega, k)$ \cite{Grozdanov:2017ajz, Blake:2017ris, Blake:2018leo}. This feature is the so-called pole-skipping and it has great importance because chaos which is a far-from-equilibrium behavior can be examined by the near-equilibrium properties of solutions. Additionally, having multiple results around $\omega_n = - 2 \pi T n i$ are seen for scalar field perturbations \cite{Amrahi:2023xso, Grozdanov:2023txs}. We compare the $q^2_{ps} = k_0^2/(2 \pi T)^2$ with $q^2_c$ and find that at high temperatures there is room for $q^2_{ps} = q^2_c$. Besides that, always $q^2_{ps} < q^2_c$ which marks hydro validity even on the chaos point. We find that in the  region $ 1 \lesssim \phi_H \lesssim 6 $, the inequality  $(q^2_{ps})_{\text{FO}} < (q^2_{ps})_{\text{SO}}< (q^2_{ps})_{\text{CO}}$ holds. Furthermore, $q^2_{ps}$ can find the location of the transition point.
\item[$\bullet$ \bf{Further results:}] 
We find analytic expressions for temperature and entropy at very low temperatures and show how $\phi_H$ can violate the conformal symmetry. Moreover, we investigate the $q^2_c$ in the spin-2 sectors at very high temperatures and find $q^2_c = 1.486\, e^{i \theta_c}$ with $\theta_c = 0.98 \pi$ regardless of the kinds of phase transition. We observe that for very large $\vert q^2\vert$ the lowest non-hydro scalar modes approach the sound mode of the 5D-AdS-Schwarzschild black hole. 
\end{itemize}

\section{Holographic model}\label{sec: model}

In this work, we consider deformed holographic Conformal Field Theories(CFTs) that are dual to five-dimensional Einstein gravity with a minimally coupled massive self-interacting scalar field, which we refer to as the Einstein-Klein-Gordon (EKG) model. In this section, we summarize the gravity side of the holographic model that we study in the rest of this work. 

In section \ref{sec: Action} we present the EKG action and the explicit form of the scalar potential.  In section \ref{sec: BH-numerics} we discuss solutions corresponding to thermal states in the dual field theory.
For completeness, in Appendix \ref{appendix-HRG} we derive the explicit form of the counter-term that renders the n-point functions of local boundary operators well-defined and explicitly derive the one-point functions of the boundary theory. Also, in Appendix \ref{appendix-nearhor} we provide the near-horizon solutions of bulk fields. Moreover, in Appendix \ref{sec:low-T BH} we derive thermal gas solutions, corresponding to vacuum solutions of the boundary theory.  Then we present a method to compute the low-temperature solutions perturbatively. 

\subsection{Action}\label{sec: Action}
The total action of the EKG model is given as follows \cite{Gubser:2008ny, Ecker:2020gnw}
\begin{equation}\label{eq: action}
S_{\textrm{tot}}=\frac{1}{16\pi \GN}\int_{\mathcal{M}}\!\!\extd^{5}x\sqrt{-g}\, \Big(R-\frac{1}{2}g^{M N} \partial_M \phi \partial_N \phi - V(\phi)\Big) + \frac{1}{8\pi \GN}\int_{\partial\mathcal{M}}\!\!\!\!\extd^4x\sqrt{-\gamma}\,K+S_{\textrm{ct}},
\end{equation}
where $G_5$ is the five-dimensional Newton's constant. The first term represents the bulk contribution which includes the Ricci scalar $R$ of the bulk geometry with metric $g_{\mu \nu}$ lying on a manifold $\cal M$ as well as the kinetic part for the scalar field $\phi$ and the self-interacting potential $V(\phi)$. The second term is the  Gibbons-Hawking-York boundary term, which lives on the boundary of $\mathcal{M}$, denoted by $\partial \mathcal{M}$. The trace of the extrinsic curvature of the induced metric,   $\gamma_{\mu \nu}$, is represented by  $K = \frac{1}{2} \gamma^{\mu \nu} \partial_r \gamma_{\mu \nu}$. The last term is the counter-term that ensures the variational principle is well-defined and the on-shell action is finite.

For an asymptotically AdS$_5$ spacetime with the AdS radius $L=1$, the potential’s asymptotic expansion at small  $\phi$ should be as follows
\begin{equation}\label{eq:potenital-expansion}
V(\phi)=-12+\frac{1}{2}m^2\phi^2+V_4\phi^4+\ldots.
\end{equation}
For simplicity, we assume the $\mathbb{Z}_2$ symmetry,  $V(\phi)=V(-\phi)$. According to the usual AdS/CFT dictionary, the conformal weight $\Delta$ of the dual operator on the boundary is related to the mass of the scalar field as $m^2 = \Delta (\Delta - 4)$.
We restrict ourselves to potentials that can be written globally in terms of a superpotential $W(\phi)$
\be
V(\phi)=-\frac{1}{3}W(\phi)^2+\frac{1}{2}W'(\phi)^2\label{V2W}.
\ee 
Having this form is very useful. First, it simplifies the process of finding solutions, such as the ground state (domain wall) solution or even numerical thermal state solutions. Second, the  $S_{\textrm{ct}}$ in Eq. \eqref{eq: action} are explicitly known in terms of the superpotential (see appendix \ref{appendix-HRG}). Third, near-boundary solutions in Fefferman–Graham or Gubser-gauge do not contain logarithmic terms due to the absence of the trace anomaly. This makes the numerical analysis of QNMs more stable on such backgrounds.

We are working with a family of superpotentials that are  characterized by a single parameter $B_4$
\be\label{eq:superpotential}
W(\phi)=-6-\frac{1}{2}\,\phi^2+B_4\,\phi^4.
\ee
The associated potential \eqref{V2W} reads
\eq{
V(\phi)=-12-\frac{3 }{2}\,\phi ^2-\frac{1}{12}\, \phi ^4+\frac{B_4 \left(24 B_4+1\right)}{3}\,  \phi ^6-\frac{B_4^2}{3} \, \phi ^8\,.
}{eq:angelinajolie} 
In what follows, we will explore how different kinds of phase transitions, including first-order, second-order, and crossover, can be reached by varying $B_4$. Linearized perturbations are studied on top of these backgrounds, and hence the hydrodynamics for various thermodynamics can be analyzed.

\subsection{Thermal states}\label{sec: BH-numerics}
The equations of motions for the metric and scalar fields in  the action \eqref{eq: action} can be written as
\begin{align}\label{eq: EoMs}
R_{M N}-\frac{1}{2}\partial_M\phi\,\partial_N\phi-\frac{1}{3} V(\phi) g_{M N} &= 0,\nn\\
\frac{1}{\sqrt{-g}}\partial_M\left(\sqrt{-g} g^{M N} \partial_N \phi \right)-\frac{\partial V(\phi)}{\partial \phi} &= 0.
\end{align}
To describe a more general (thermal) state, we make the following ansatz for the metric in the Eddington--Finkelstein coordinates \cite{Gubser:2008ny}
\bea\label{eq: metric}
\extd s^2=e^{2A(u)} \left(-H(u) \extd t^2 + \extd x^2+\extd y^2+ \extd z^2\right)+ 2 e^{A(u)+B(u)}\extd u \extd t.
\eea
 The ansatz $A, B$, and $H$ are functions of the radial coordinate $u$ only. This ansatz possesses solutions that are $SO(3)$ invariant (invariant in the $x,y,z$ directions). The black-brane geometry corresponds to a simple zero of $H(u)$ at some $u = u_h$, with a regular event- and Killing horizon at $u = u_h$. This leads to finite temperature and entropy density of the dual field theory states. The boundary is located at $u=0$.

It is noteworthy that the ansatz \eqref{eq: metric} has residual gauge freedom, namely reparametrizations of the radial coordinate. We  fix this freedom by using the Gubser gauge \cite{Gubser:2008ny}, where the radial coordinate is identified with the corresponding value of the scalar field
\begin{equation}\label{eq: GubserGauge}
u := \phi(u).
\end{equation}
From here on, we provide the solutions in the Gubser gauge. According to the ansatz \eqref{eq: metric}, the equations of motion for $A,B$ and $H$ are
\begin{align}
 H\left(B'-4 A'\right)-H'+e^{2B}V' &= 0,\label{eq:EOM1}\\
 6\left(A'B'-A''\right)-1 &= 0,\label{eq:EOM2}\\
 H''+(4A'-B')H' &=0, \label{eq:EOM3}\\
 6 A' H'+H \left(24 A'^2-1\right)+2 e^{2 B} V &=0, \label{eq:EOM4}
\end{align}
where prime denotes derivation for the $u$. These equations can be rephrased as a single master equation for $G(\phi)\equiv A'(\phi)$ 
\begin{align}\label{eq: master}
18 G^2 G' V''+9 G V' \left(G' \left(6 G'+8 G^2+1\right)-2 G G''\right)+V \left(G' \left(6 G'+24 G^2+1\right)-6 G G''\right)=0.
\end{align}
For a given potential $V(\phi)$, a solution of \eqref{eq: master} allows us to express $A,B$, and $H$ in terms of the function $G$. According to Eqs. \eqref{eq:EOM1}, \eqref{eq:EOM2}, and \eqref{eq:EOM4}, the solution for these functions can be obtained as:
\begin{align}
&A(\phi)=A(\phi_h)+\int\limits_{\phi_h}^\phi \mathrm{d}\phi' G(\phi')\,,\nn\\
&B(\phi)=B(\phi_h)+\log\frac{G(\phi)}{G(\phi_h)}+\int\limits_{\phi_h}^\phi \frac{\mathrm{d}\phi'}{6G(\phi')}\,,\nn\\
&H(\phi)=-\frac{e^{2B(\phi)}(V(\phi)+3 G(\phi) V'(\phi))}{3 G'(\phi)}\,,
\end{align}
where $\phi_h$ represents the horizon value of the scalar field.

For certain simple choices of $V$ it is possible to solve the second order ordinary differential equation \eqref{eq: master} in the closed form \cite{Gubser:2008ny}. However, in general, the master equation \eqref{eq: master} needs to be solved numerically. In that case, it is useful to extract the  divergent asymptotic behavior of the master field $G$ inherited from the asymptotic behavior of $A$
\begin{equation}\label{eq:div-nearhor}
A(\phi)=\frac{\log(\phi)}{\Delta-4}+\tilde{A}(\phi), \quad B(\phi) = \log\left(\frac{1}{\phi (4 - \Delta)}\right) + \tilde{B}(\phi), \quad  G(\phi)=\frac{1}{(\Delta -4)\phi}+\tilde{G}(\phi),
\end{equation}
where $\tilde{A}(\phi), \tilde{B}(\phi)$ and $\tilde{G}(\phi)$ remain finite at the boundary $\phi\to 0$. As discussed in  \cite{Gubser:2008ny} such a near boundary behavior of the fields corresponds to a relevant deformation of the CFT$_4$, namely
\be
\mathcal{L}=\mathcal{L}_{\textrm{CFT}_4}+\source^{4-\Delta}\mathcal{O}_\phi\ .
\ee
To find the equation of state (EOS), we set the source $\source$ of the dual operator to one in units of AdS radius, $\source=1$. This leaves the horizon value of the scalar field $\phi_h$ (which is equal to the horizon radius in the Gubser gauge) as the only free parameter. 
\subsection{Thermodynamics}\label{sec:thermo}
In this section, we examine the thermodynamics of the system for different choices of the $B_4$  resulting in different types of phase transitions. We do this for the deformation of an operator with fixed conformal weight $\Delta=3$ and modify the quartic term of the potential \eqref{eq:superpotential} by choosing different values for $B_4$. We provide the near-horizon solution for $A, B$, and $H$ in terms of $V$ in Appendix \ref{appendix-nearhor}.

The entropy density and the temperature of the boundary system can be expressed in terms of horizon data as follows \footnote{Hereafter, we work in units $8 \pi G_5 = 1$.}
\begin{equation} \label{eq: thermodynamics}
s=2 \pi \,e^{3A(\phi_h)}, \qquad T=\frac{1}{4\pi}\,e^{A(\phi_h)-B(\phi_h)} H'(\phi_H) = \frac{1}{4\pi}\,e^{A(\phi_h) + B(\phi_h)}\left\vert V'(\phi_h)\right\vert\,.
\end{equation}
Also, the speed of the sound of the boundary theory can be computed from the horizon information 
\be\label{eq:cs}
c_s^2=\frac{\extd\  \ln{T}}{\extd\ \ln{s}} = \frac{\frac{\extd\ \ln T}{\extd\ \phi_H}}{\frac{\extd\ \ln s}{\extd\ \phi_H}}.
\ee
The energy density, \footnote{By `density' we mean that we divide by the trivial but infinite volume along the black brane.  
}
pressure and the expectation value of the deformation operator are expressed in Appendix \ref{appendix-HRG}. The transport coefficients are extensively discussed in \ref{sec: remarks}.

\begin{figure}
    \centering
    \includegraphics[width=0.323\textwidth, valign=t]{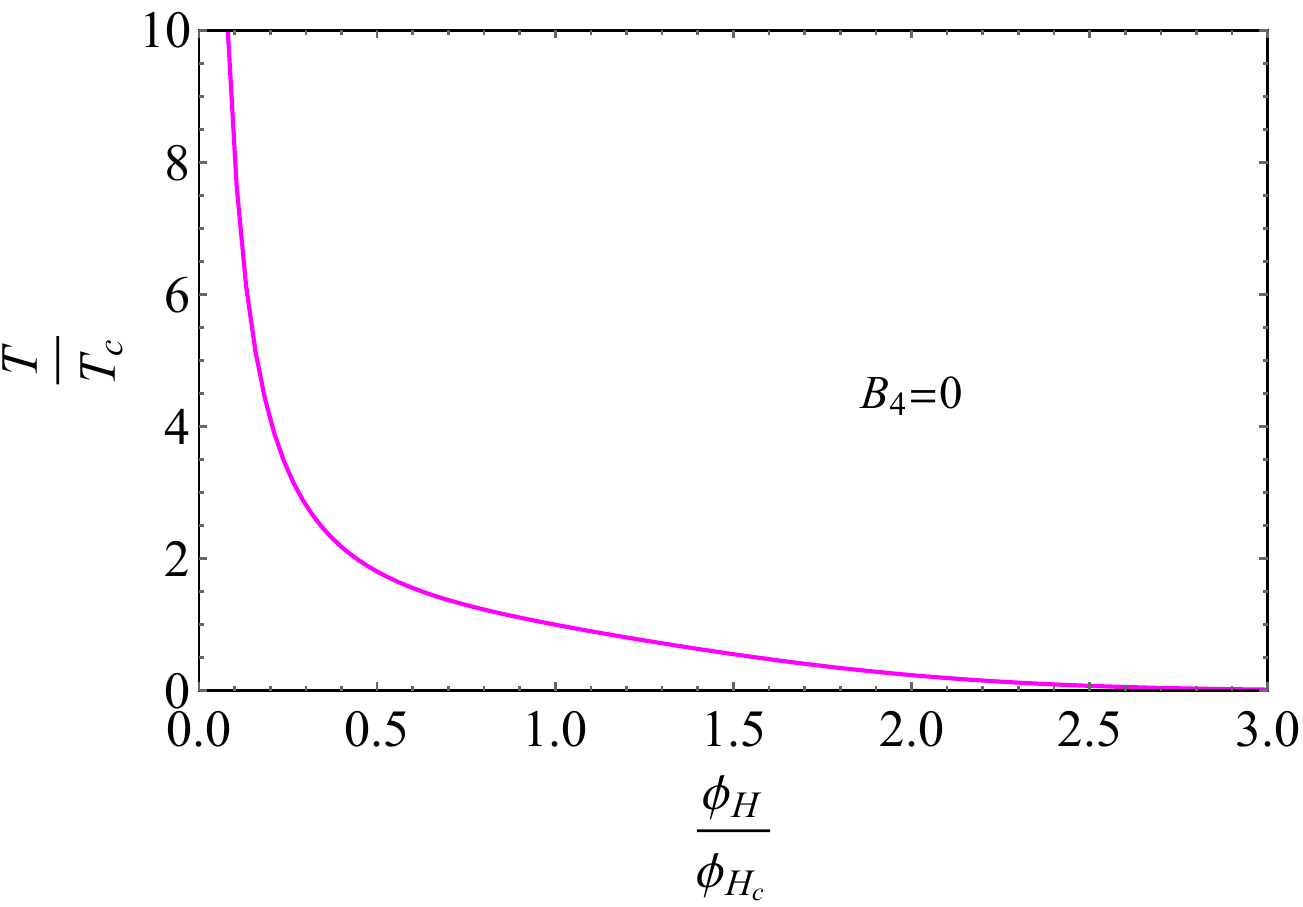}
    \includegraphics[width=0.323\textwidth, valign=t]{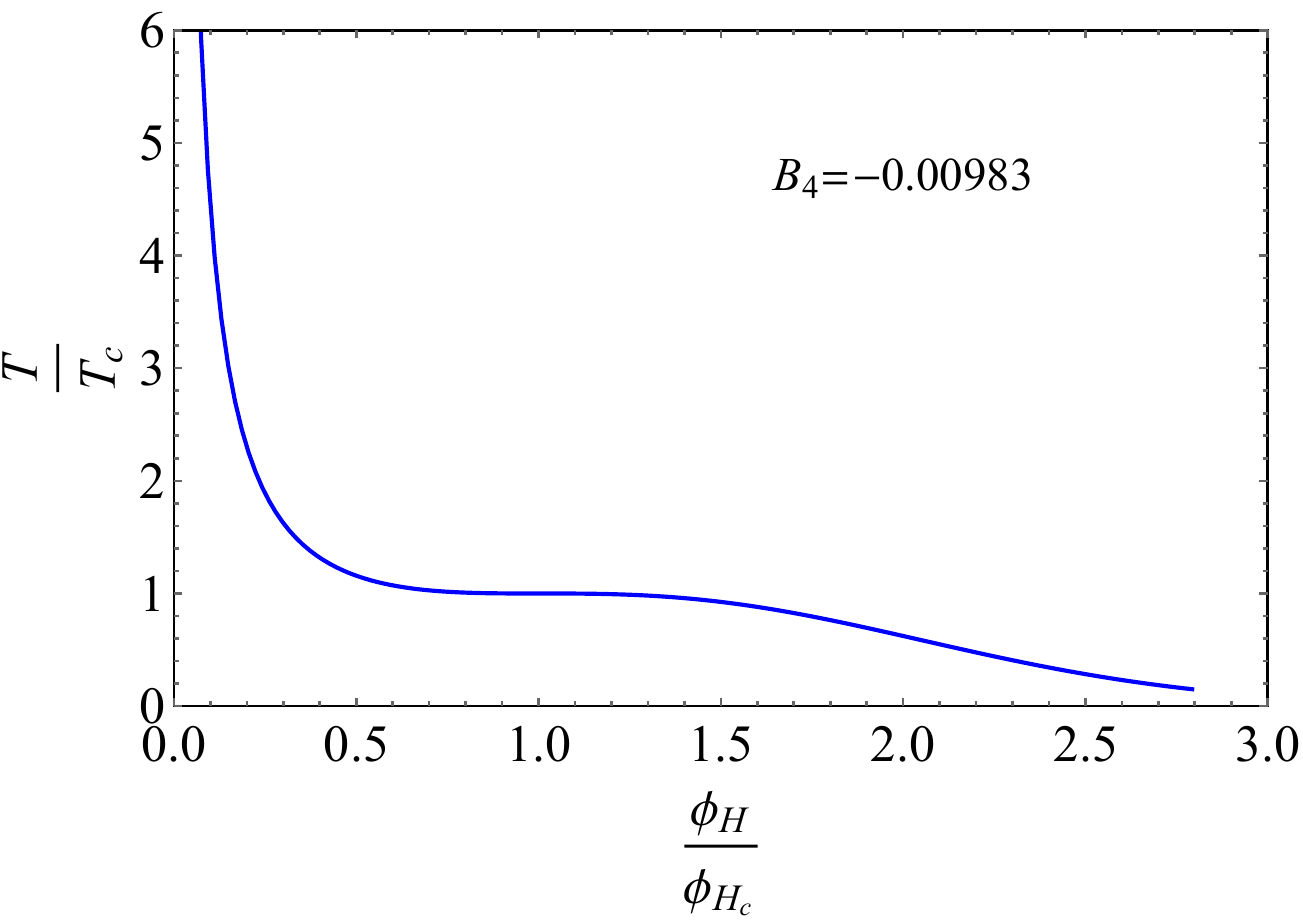}
     \includegraphics[width=0.323\textwidth, valign=t]{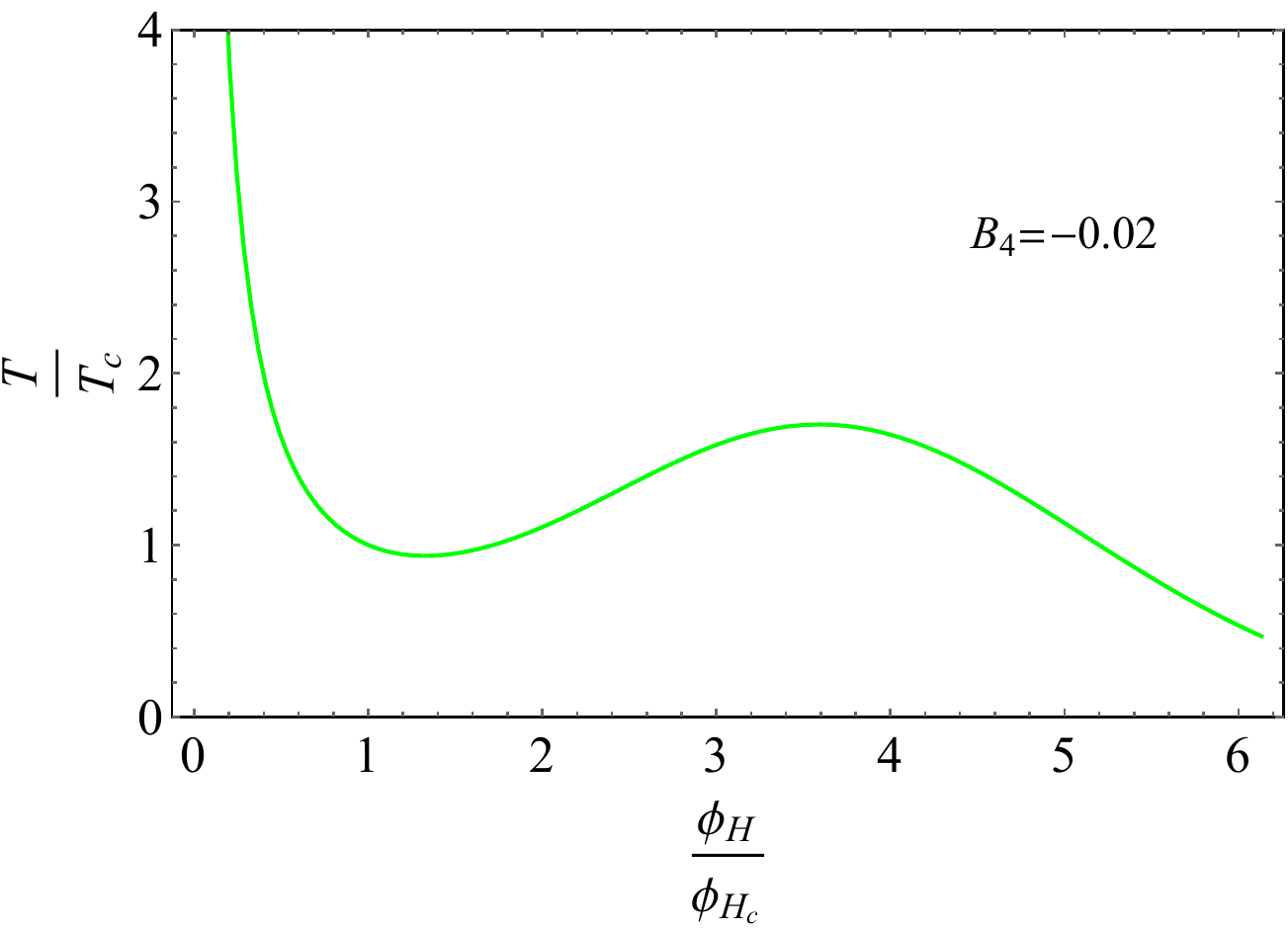}
    \caption{Dimensionless plots of $T(\phi_H)$ for the cross-over, second-order, and first-order phase transition from left to right, respectively.}
    \label{fig:Tphi}
\end{figure}

In Fig. \ref{fig:Tphi}, we show we show the temperature $T$ as a function of the horizon value of the scalar field $\phi_H$ for different choices of $B_4= (0, -0.0098, -0.02)$ corresponding to cross-over, second-order and first-order phase transition, respectively. In the cross-over EoS, the temperature has a smooth fall-off with $\phi_H$, while in second-order and first-order EoSs, it has a complex pattern near the phase transition point. The temperature has a flat profile in the second-order phase transition, while in the first-order phase transition, there is a dip and high corresponding to the stable and unstable branches, which is a unique property of every first-order phase transition.

\begin{figure}
    \centering
    \includegraphics[width=0.323\textwidth, valign=t]{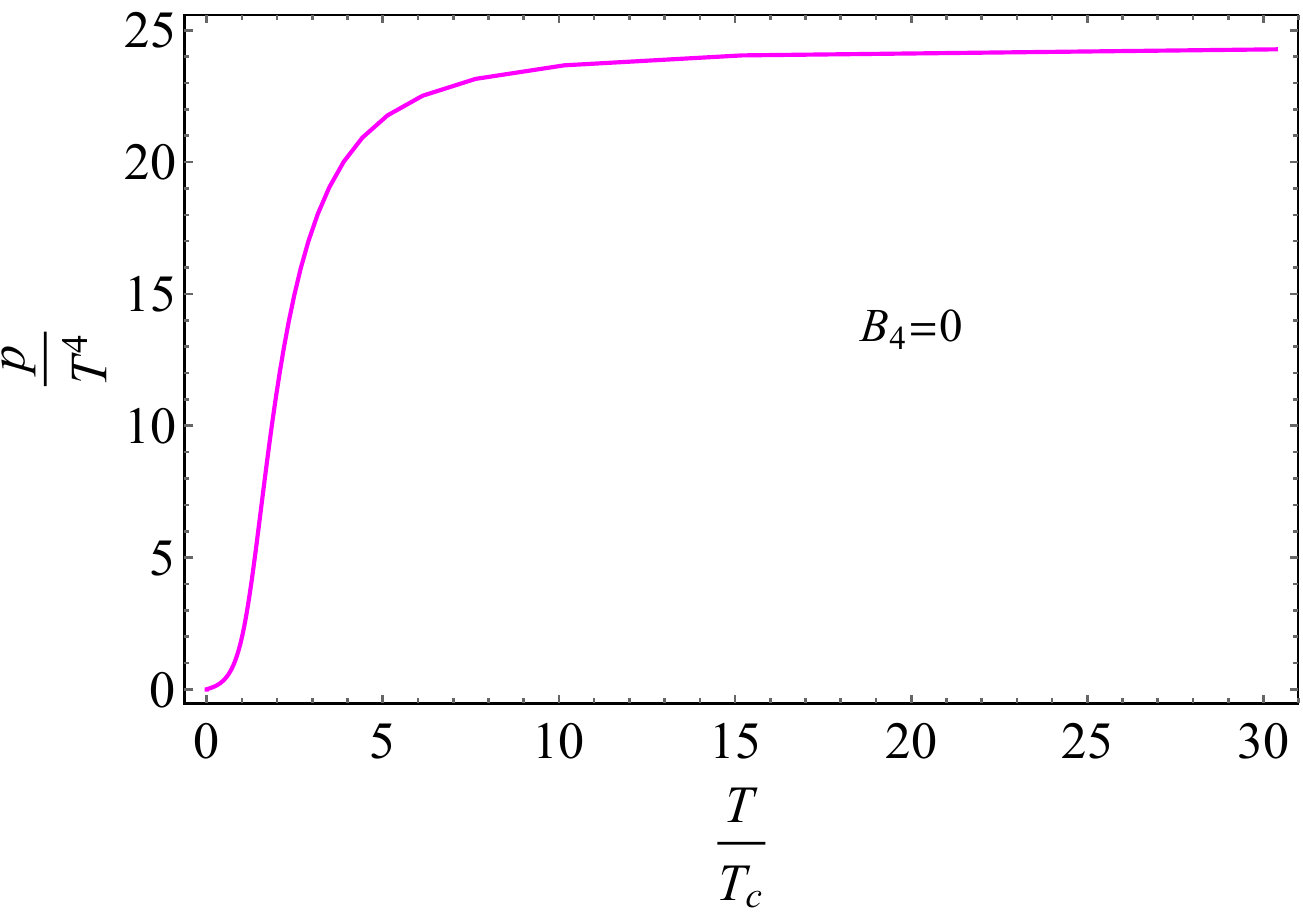}
    \includegraphics[width=0.323\textwidth, valign=t]{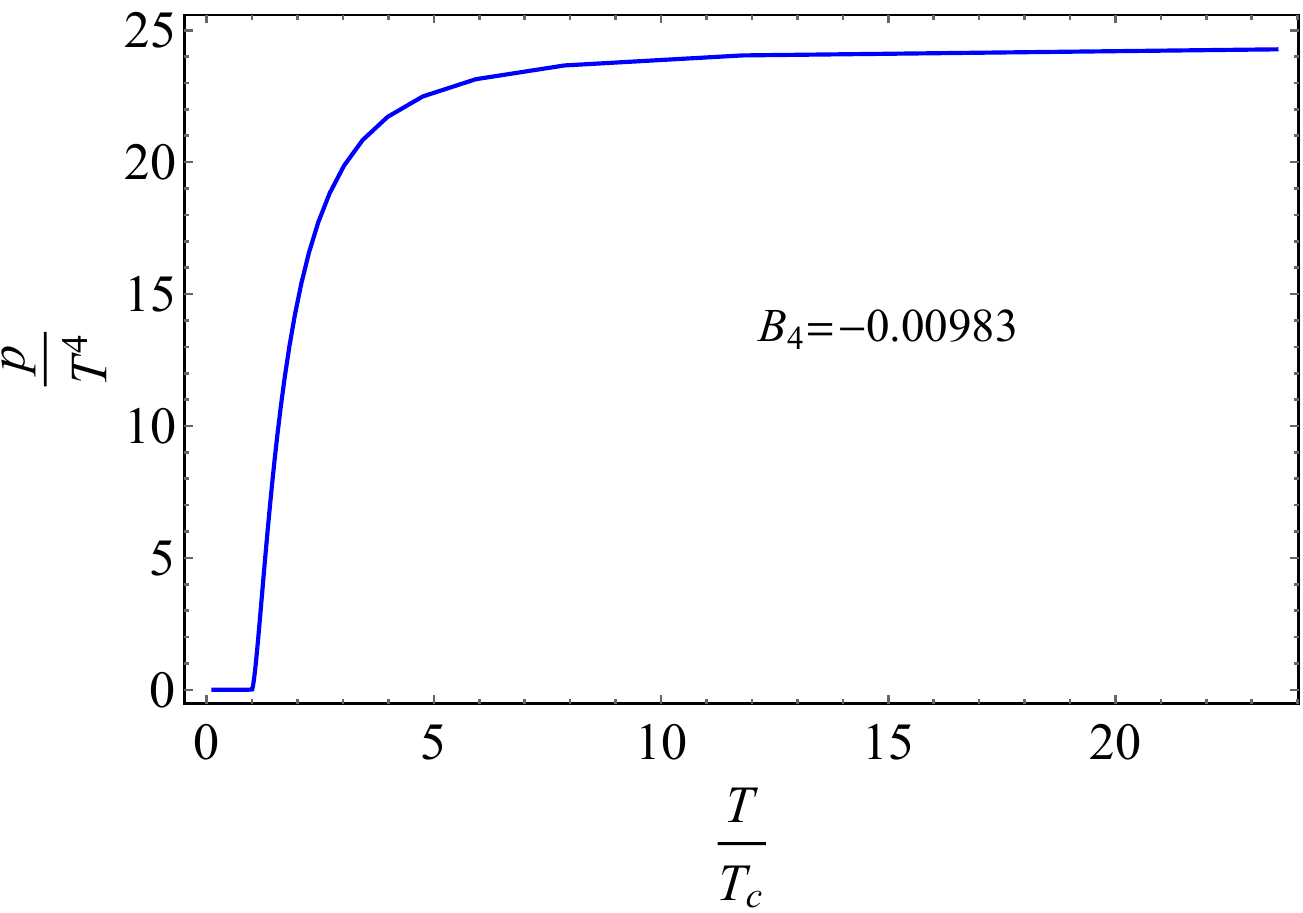}
     \includegraphics[width=0.323\textwidth, valign=t]{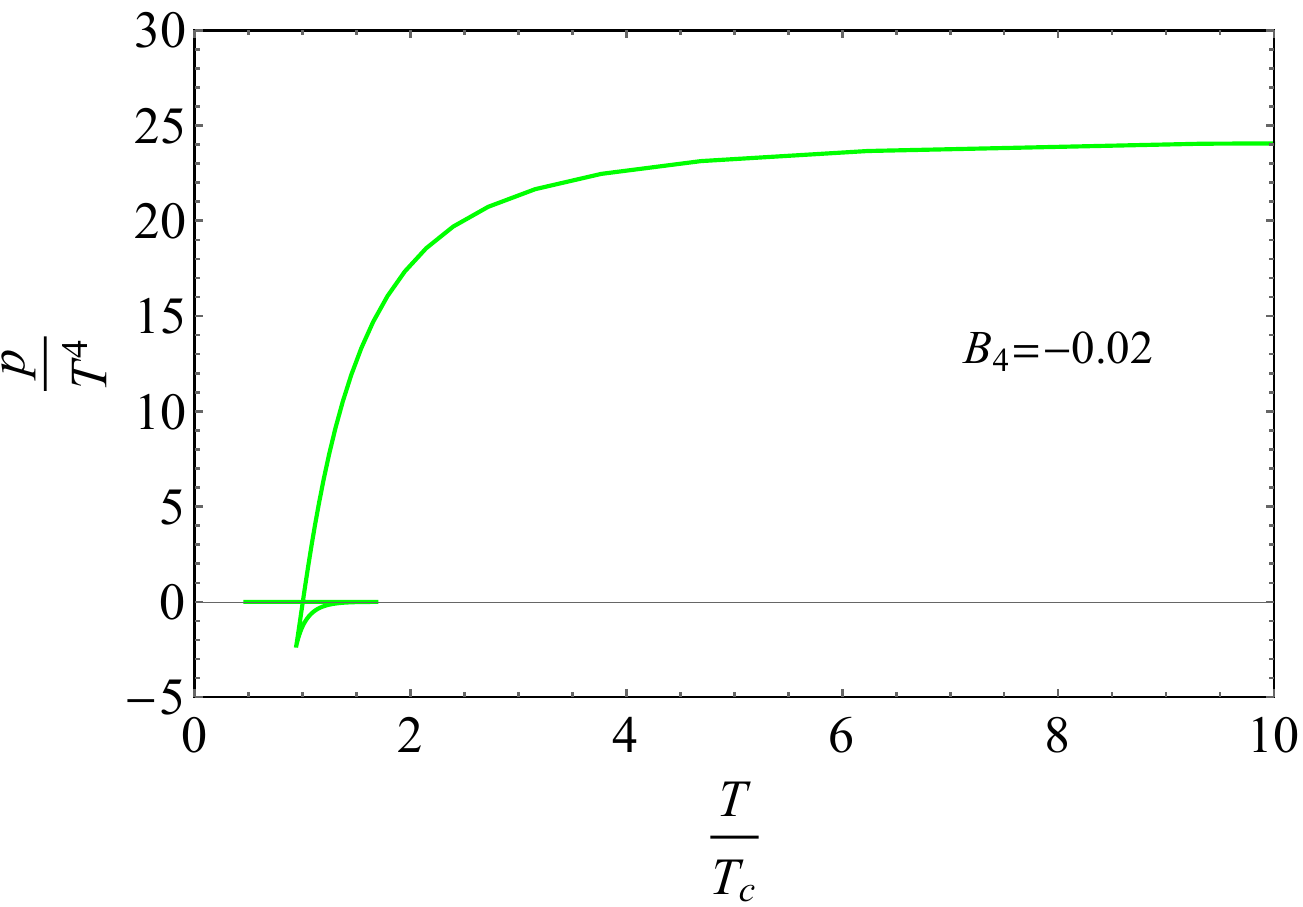}
    \caption{Plots of $\frac{p}{T^4}$ for cross-over, second-order, and first-order phase transition from left to right, respectively. Negative values of pressure near the phase transition points in the first-order plot reflect the possibility of damping the thermodynamic states.}
    \label{fig: pT}
\end{figure}

In Fig. \ref{fig: pT}, we sketch $p/T^4$ v.s. $T/T_c$ for each phase transition. As expected, in the cross-over and second-order phase transitions, pressure rises near the phase transition point due to the liberation of many degrees of freedom. In contrast, in the first-order phase transition,  pressure drops to negative values due to the thermodynamic instability of states. This instability arises from the multiple available states that can tunnel between them. At high temperatures, the pressure approaches the Steffan-Boltzmann value, indicating that in this limit, the corresponding states in all cases are close to the CFT$_4$. 

\begin{figure}
    \centering
  \includegraphics[width=0.323\textwidth, valign=t]{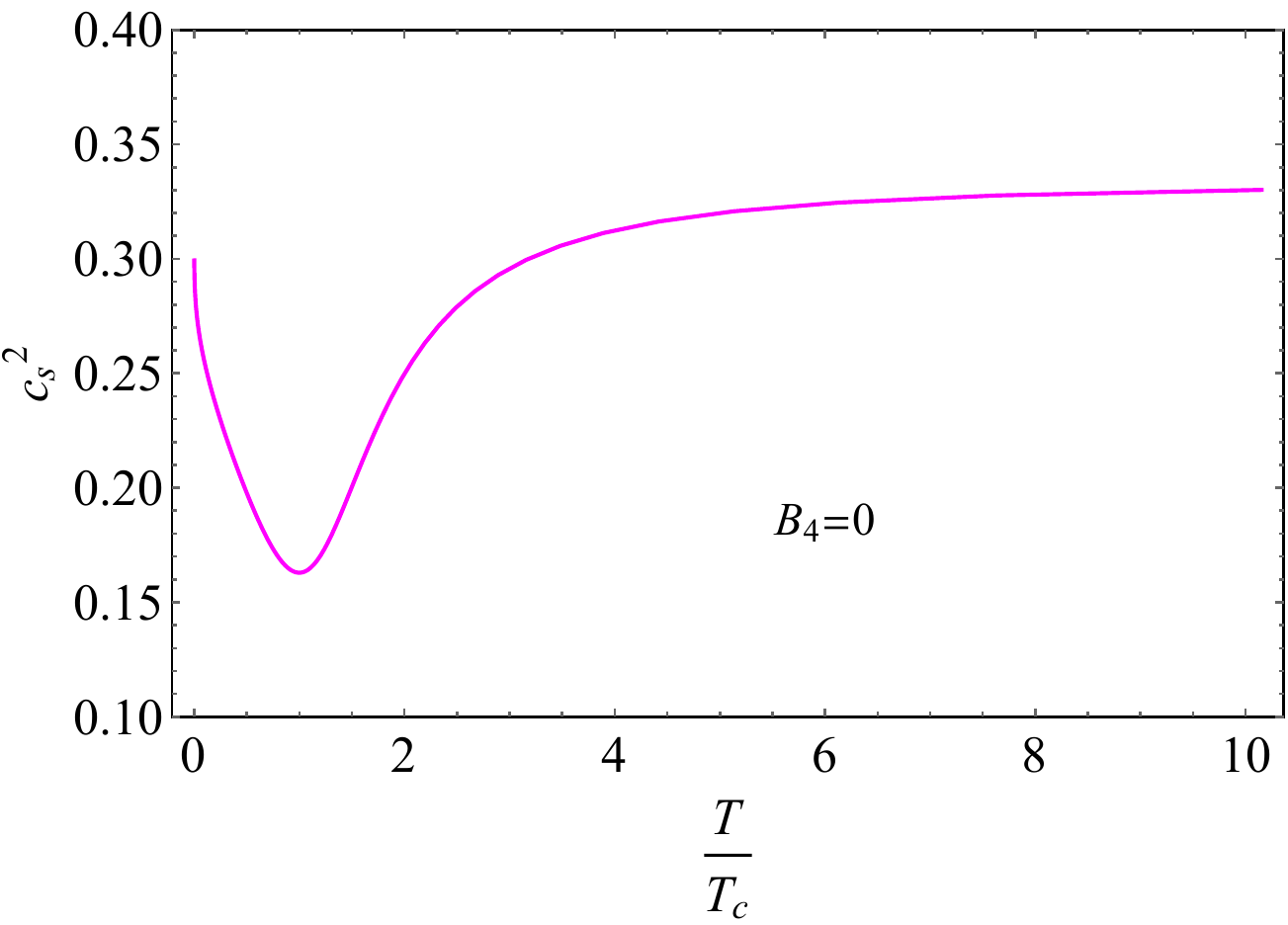}
  \includegraphics[width=0.323\textwidth, valign=t]{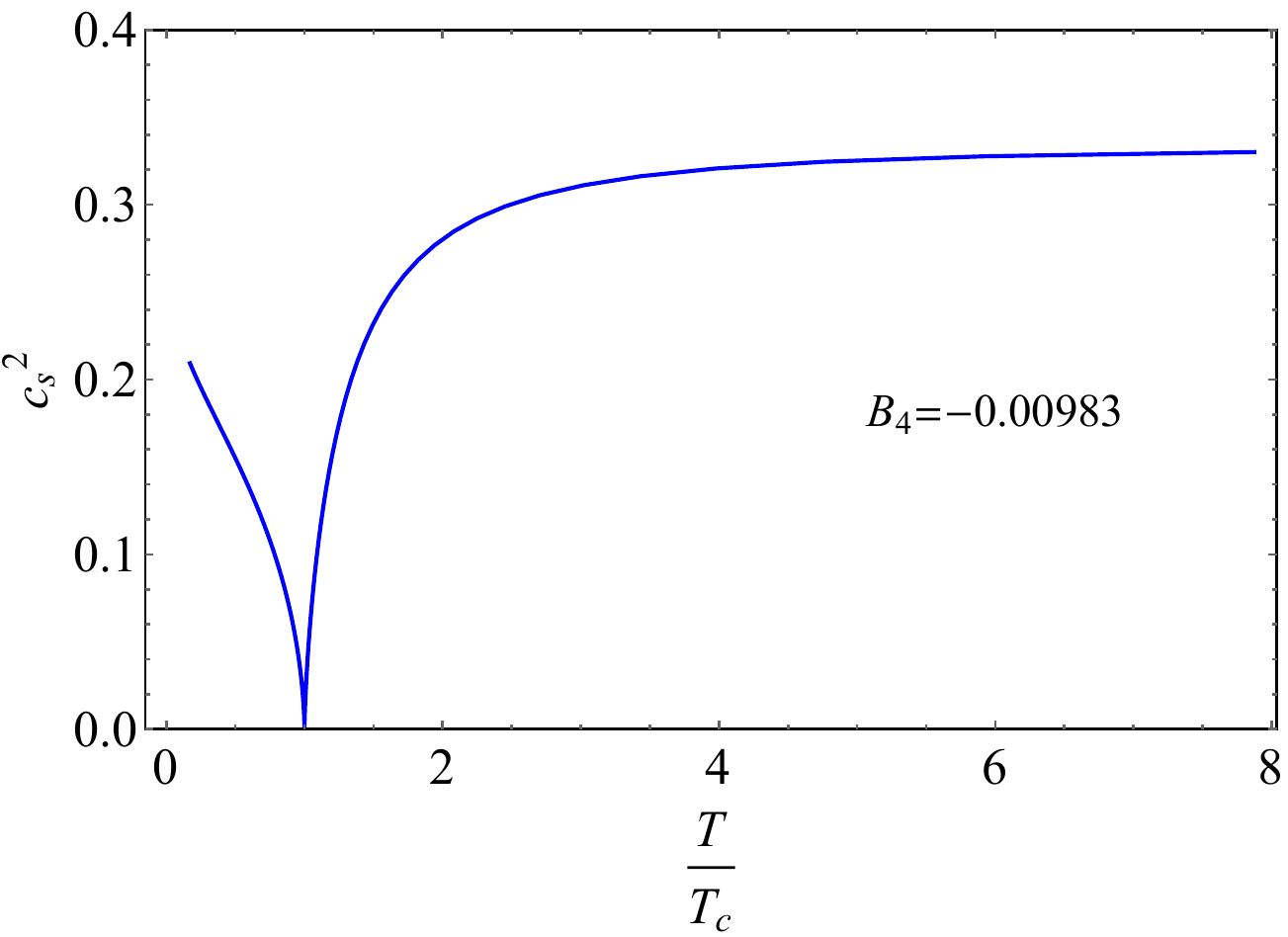}
  \includegraphics[width=0.323\textwidth, valign=t]{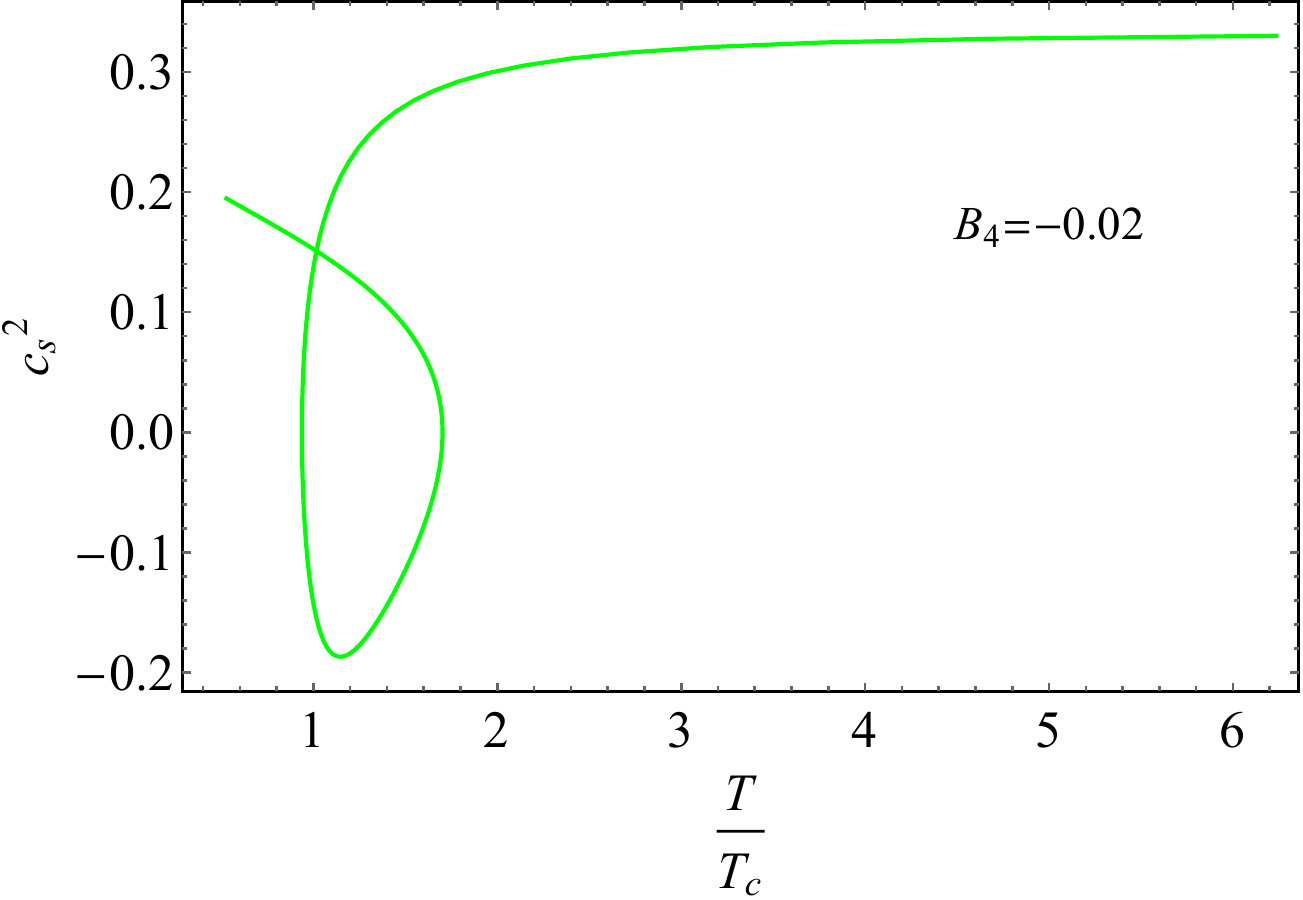}\\
  \includegraphics[width=0.323\textwidth, valign=t]{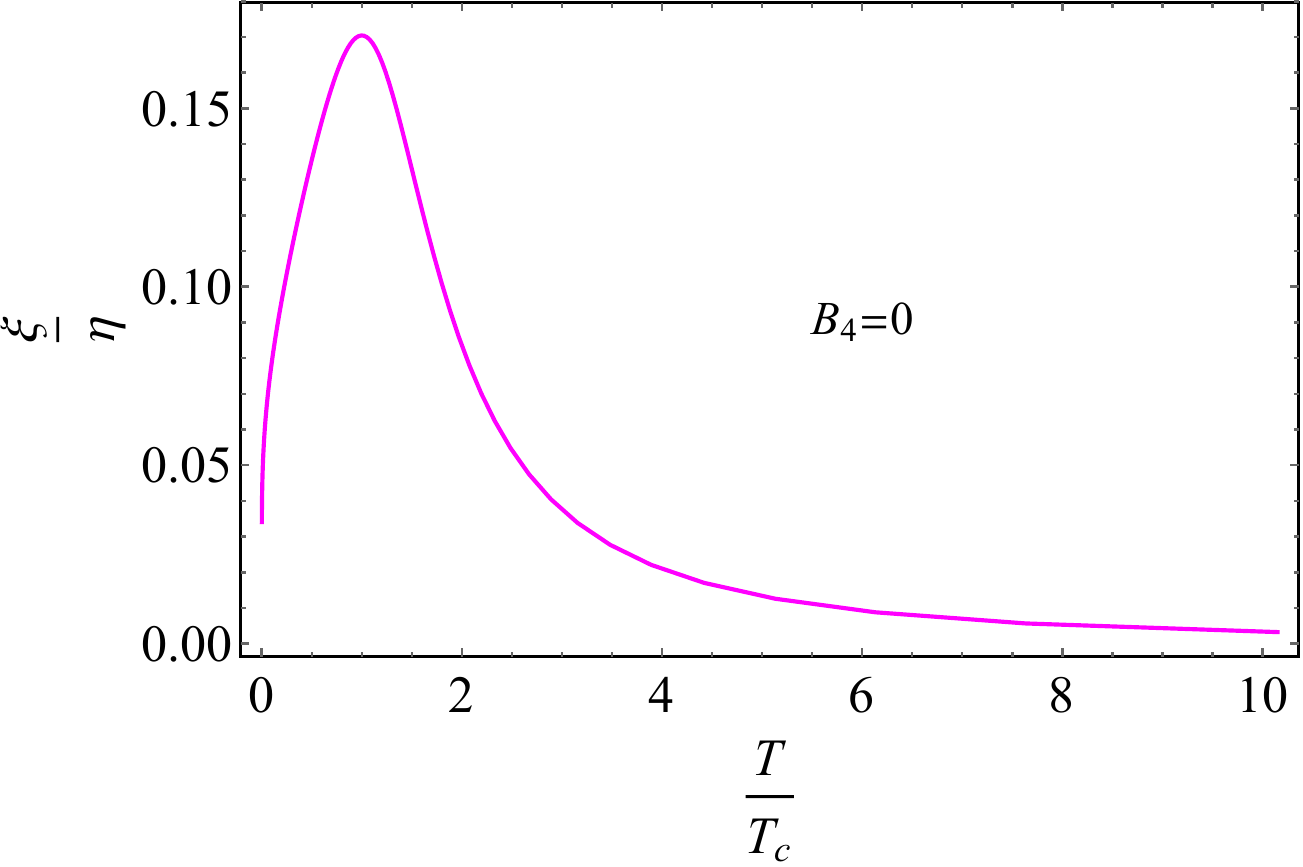}
  \includegraphics[width=0.323\textwidth, valign=t]{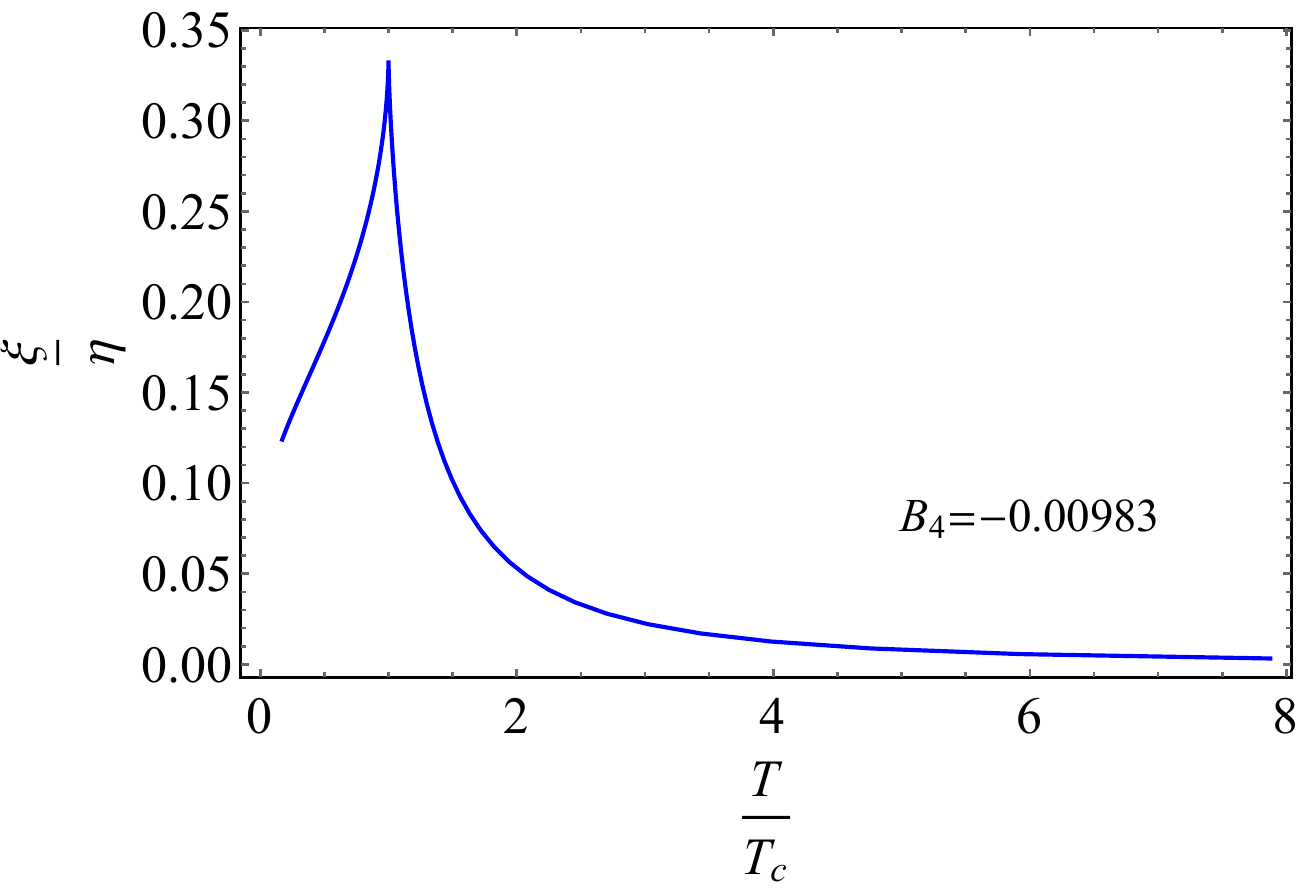}
  \includegraphics[width=0.323\textwidth, valign=t]{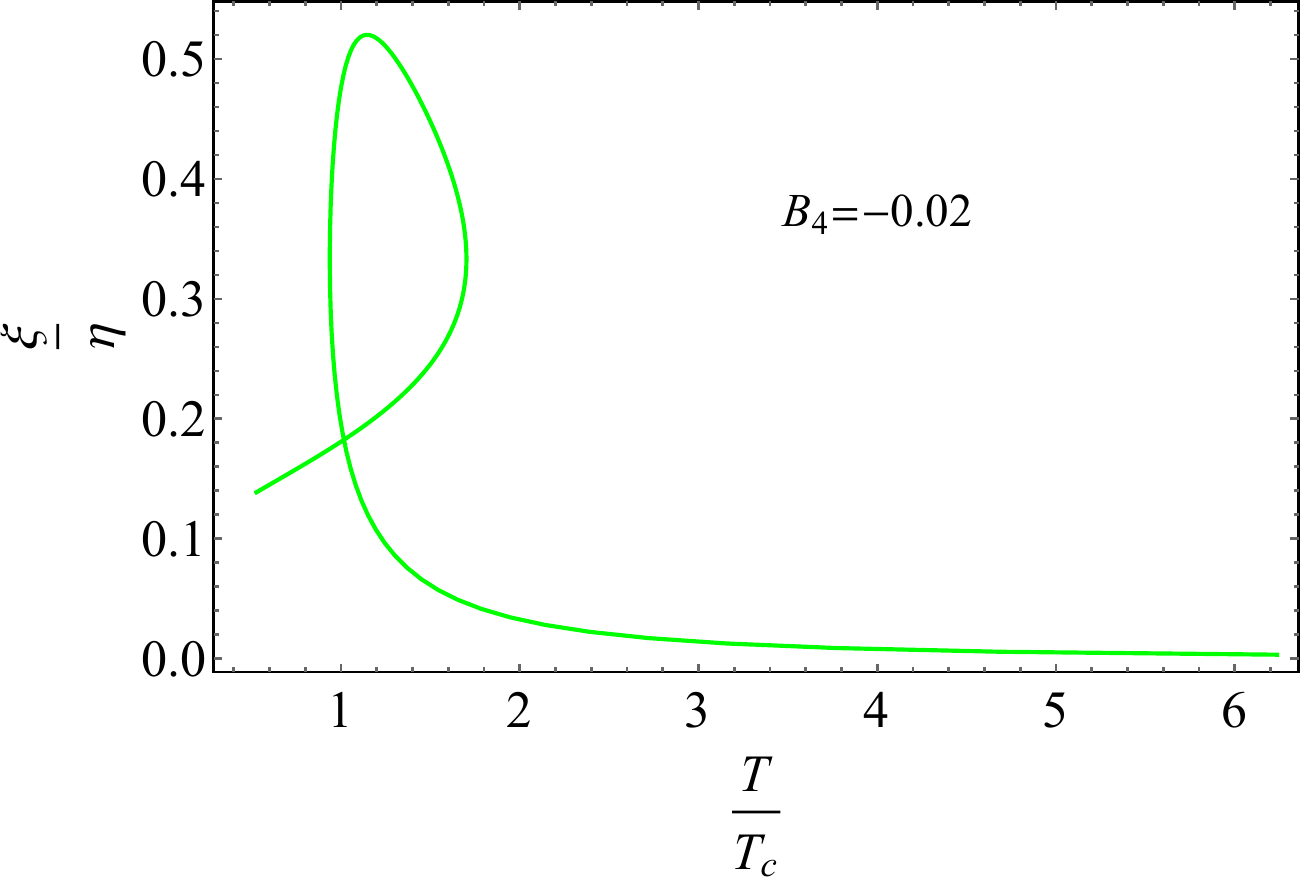}
    \caption{Plots of  $c_s^2(\frac{T}{T_c})$ and $\frac{\xi}{\eta}(\frac{T}{T_c})$ for different phase transitions $B_4 = (0, -0.0098, -0.02)$ from left to right. The top (bottom) panel corresponds to  $c_s^2(\frac{T}{T_c})$ ($\xi/s$) for the cross-over, second-order, and first-order phase transitions from left to right, respectively. The minimum (maximum) in figures $c_s^2(\frac{T}{T_c})$ ($\xi/s$) reflects the transition points.}
    \label{fig: cs2-zeta}
\end{figure}

 To derive the values of critical temperature and scalar field, namely the $T_c$ and $\phi_c$, we examine the minimum points of $c_s^2$ or equivalently the maximums of $\xi/s$. In Fig. \ref{fig: cs2-zeta}, we sketch $c_s^2(\frac{T}{T_c})$ and $\frac{\xi}{s}(\frac{T}{T_c})$ for different phase transitions. The transition points for different phase states are shown in Table \ref{table: Tcphic}.  The values of $T_c$ and $\phi_c$ are given in a unit where $L_{ADS} = 1$ and $8 \pi \GN = 1$. In the crossover, the transition point is softer than the first-order, and in the first-order, there is a region where  $c_s^2 < 0$ or $c_s$ is imaginary, signaling the possibility of transition between the stable thermodynamic states. Quasinormal modes in this part are unstable, which is known as spinoidal instability \cite{Chomaz:2003dz}. In all kinds of transitions in the low-temperature limit, we have $c_s^2(T\rightarrow0)\rightarrow1/3$ and $\xi/\eta(T\rightarrow0)\rightarrow0$.
\begin{table}
\centering
\begin{tabular}{|l|c|c|}
\hline
& $T_c \approx$ & $\phi_c \approx$ \\
\hline
$B_4 = 0$ (Cross-over) & $0.08888$ & $3.779$ \\
\hline
$B_4 = -0.00983$ (Second-order)& $0.11463$ & $5.371$ \\
\hline
$B_4 = -0.02$ (First-order)& $0.1447$ & $(2.446, 4.202, 12.715)$ \\ 
\hline
\end{tabular}
\caption{Table of transition points for different kinds of phase transitions. The values of $T_c$ and $\phi_c$ are given in units of $L_{ADS} = 1$ and $8 \pi \GN = 1$.}\label{table: Tcphic}
\end{table}

\begin{figure}
    \centering
    \includegraphics[width=0.323\textwidth, valign=t]{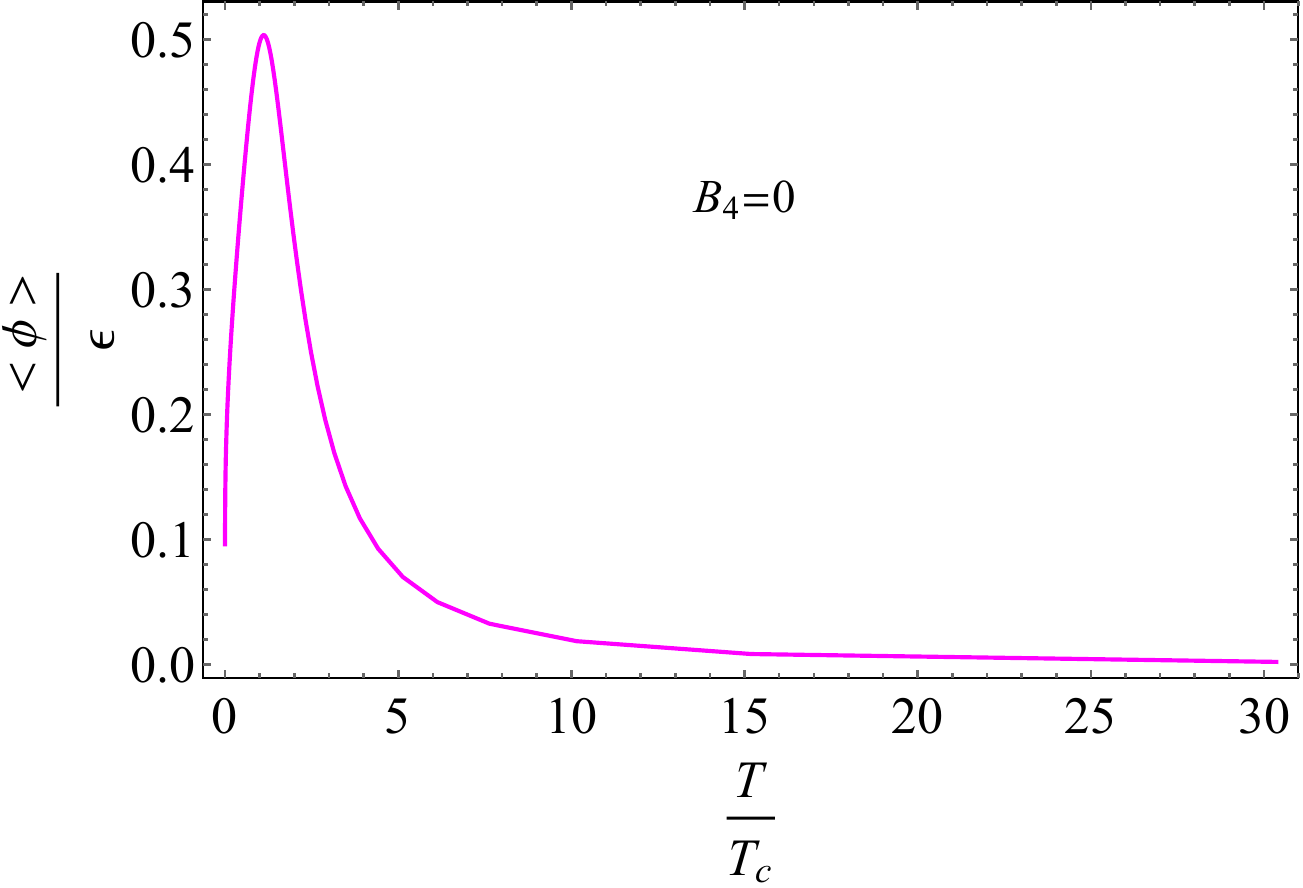}
    \includegraphics[width=0.323\textwidth, valign=t]{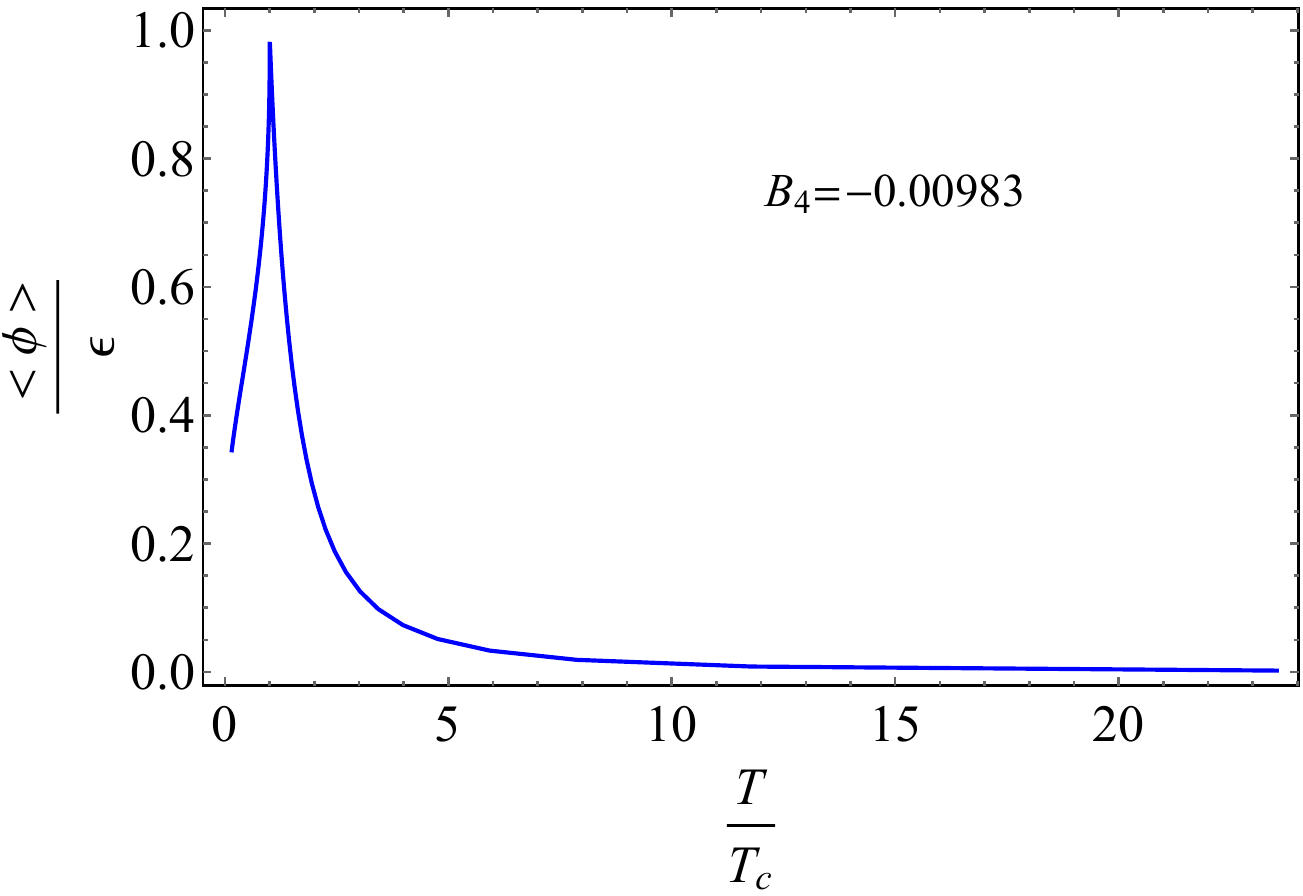}
     \includegraphics[width=0.323\textwidth, valign=t]{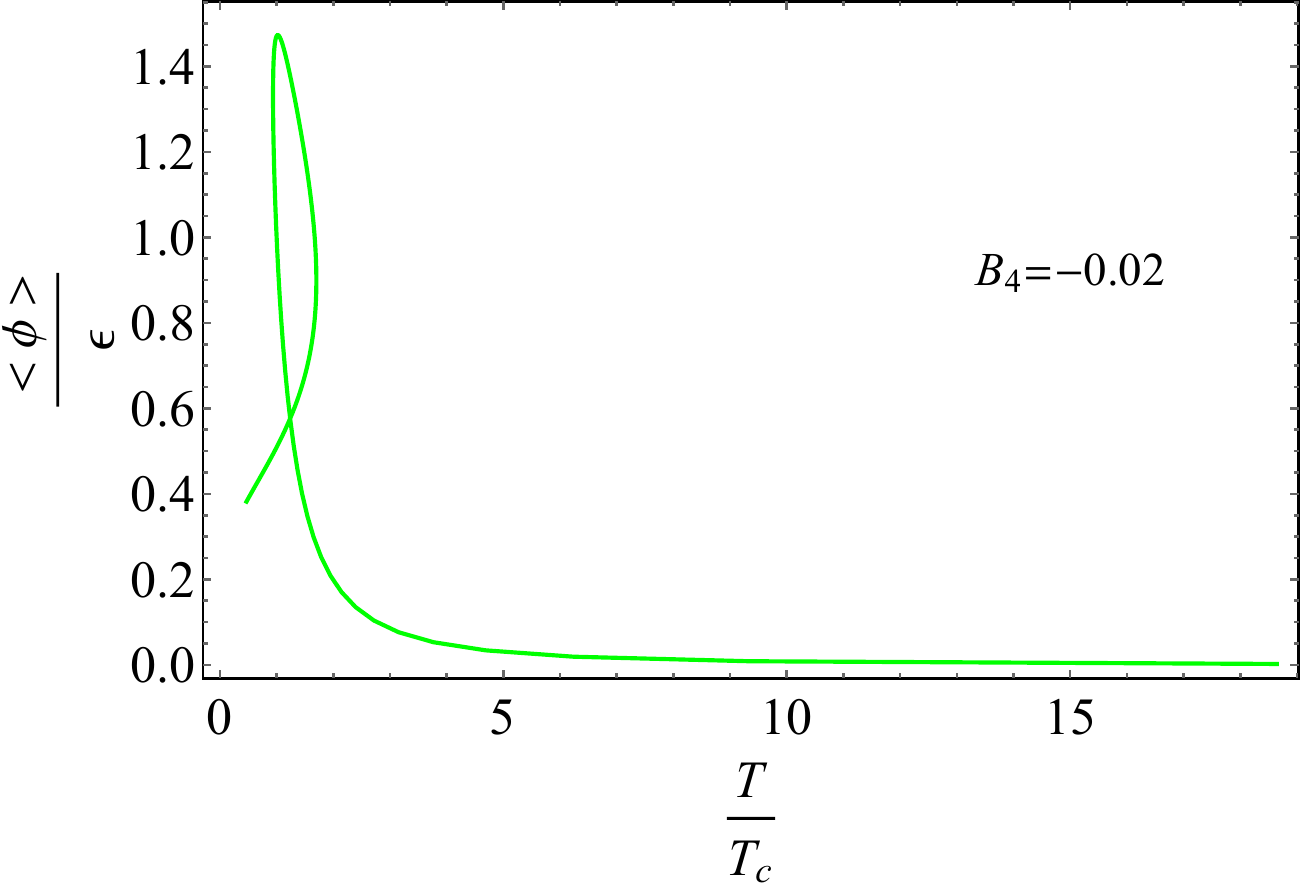}
    \caption{The sketch of dimensionless ration $\frac{\langle \mathcal{O}_\phi(T) \rangle}{\epsilon}$ for $B_4= (0, -0.0098, -0.02)$ from left to right.}
    \label{fig: ophiT}
\end{figure}

Another important figure to examine is Fig. \ref{fig: ophiT}, which shows $\langle \mathcal{O}_\phi(T) \rangle$ for different transitions. It peaks around the transition point, where conformal symmetry breaks strongly according to Eq. \eqref{ward}. However, at high and low temperatures, it vanishes, and conformal symmetry is restored.

For $B_4>-0.00983491$, the system undergoes a smooth crossover as the temperature increases. For $B_4=-0.00983491$, we find a second order phase transition, and $c_s^2$ vanishes at the critical temperature $T_c\approx 0.11463$. The entropy density shows critical behavior close to $T_c$ as
\begin{align}\label{eq:fit}
s(T)=s_0+s_1\left(\frac{T-T_c}{T_c}\right)^{1-\tilde{\gamma}}.
\end{align}
We estimate the critical exponent $\tilde{\gamma}=2/3$ for the system \cite{Gubser:2008ny, Ecker:2020gnw}. For $B_4<-0.00983491$ the system undergoes a first-order phase transition. On the gravity side, there exist three different black brane solutions around the critical point. These solutions have the same Hawking temperature but different free energies. In this family, we choose $B_4 = - 0.02$ as an example, which leads to a first-order phase transition at $T_c \approx 0.1447$ between a large and small black brane geometry.  
\section{Linearized equations}\label{sec: QNM-eqs}
The equations of motion for perturbations in specified channels proportional to the underlying symmetries are analyzed to study the linear response of the system to external sources \cite{Kovtun:2005ev}. In this section, we formulate the equations of motion for invariant perturbations and represent corresponding boundary conditions. We then discuss the results in the following parts of the paper.

In general, the perturbations on top of the background solutions can be written in the following form \footnote{Due to the $SO(3)$ symmetry, we choose the momentum direction to be aligned in the $z$ direction.}
\bea
&&g_{M N}(u, t, z) = g^{(0)}_{M N}(u) + h_{M N}(u)e^{-i\omega t + i k z},\nn\\
&&\phi(u, t, z) = u + \psi(u)e^{-i\omega t + i k z},
\eea
where $g^{(0)}_{M N}(u)$ is the metric components in the Eq. \ref{eq: metric}. The perturbations $h_{M N}(u)$ and $\psi(u)$ can be classified according to the infinitesimal
diffeomorphism transformations, i.e. $x^A\mapsto x^A +\xi^A$, where $\xi_A=\xi_A(u)e^{-i\omega t + i k z}$ is the infinitesimal displacement vector. The metric and scalar fields vary in a familiar way
\be\label{eq:diff}
h_{M N}\mapsto h_{M N} - \nabla_M \xi_N - \nabla_N\xi_M~, \hspace{35pt} \phi\mapsto \phi -\xi^A\nabla_A\phi~.
\ee
To obtain the invariant perturbations, we have to combine them to remain intact under the diffeomorphism transformations.   This enables us to decompose the perturbations into spin-2, spin-1, and spin-0 sectors due to the $SO(3)$ symmetry rules which are inherent in the Eq. \eqref{eq: metric}. Moreover, it is convenient to write the equations of linearized perturbations in a master form \cite{Buchel:2021ttt, Jansen:2019wag}
\begin{align}\label{eq: mastereq}
    \square \Phi^{(s)}_{h} - W^{(s)}_{h, h'} \Phi^{(s)}_{h'} = 0.
\end{align}
Here, $\square = \frac{1}{\sqrt{-g}}\partial_M \left(\sqrt{-g} g^{M N} \partial_N\right)$ and  $"s"$ and $"h"$ refer to the spin  and helicity number, respectively. $W^{(s)}_{h, h'}$ is the master and symmetric potential that couples different helicity states in a given spin sector. This technique facilitates numerical computations and we use the Eq. \eqref{eq: mastereq} for each spin sector in our numerical computations. 

In the spin-2 sector, the only independent perturbation is $h_{x y}(u)$ which is invariant under the diffeomorphism transformations of the Eq. \eqref{eq:diff}. The resulting equation is \cite{Buchel:2021ttt}
\begin{align}\label{eq:spin-2}
   &\square \Phi^{(2)}(u) = 0,\\
   & \square = H(u) e^{-2 B(u)} \partial^2_u +  \left(V'(u)+2 i \omega  e^{-A(u)-B(u)}\right) \partial_u + e^{-2 A(u)} \left(-k^2+3 i \omega  e^{A(u)-B(u)} A'(u)\right).\nn
\end{align}
We can define $\Phi^{(2)}(u) \equiv h^x_{y}(u) = e^{-2 A(u)} h_{x y}(u)$. Indeed, the dynamics of the spin-2 mode is identical to a massless scalar field equation.

In the spin-1 sector, the existing perturbations are $(h_{t x}(u), h_{u x}(u), h_{x z}(u))$ and they can be combined in the following way to make invariant fields 
\begin{align}\label{eq: spin-1-p}
    \mathfrak{h}_{t x}(u) = k h_{t x}(u) + \omega h_{x z}(u), \qquad \mathfrak{h}_{u x}(u) = i k h_{u x}(u) - h'_{x z}(u) + 2 A'(u) h_{x z}(u).
\end{align}
In relation to the spin-1 master scalar field $\Phi^{(1)}(u)$, we can write the above expressions as follows \cite{Buchel:2021ttt}
\begin{align}
\mathfrak{h}_{t x}(u) = H(u) e^{3 A(u)-B(u)} \left(3 \Phi^{(1)}(u) A'(u)+\Phi'^{(1)}(u)\right), \quad \mathfrak{h}_{u x}(u) = - \frac{ e^{A(u) + B(u)}}{H(u)}\partial_t \Phi^{(1)}(u).
\end{align}
We set $\mathfrak{h}_{u z}(u) = 0$ since time dependence is factored out and perturbations are completely $"u"$ dependent. 
 Following the steps of \cite{Buchel:2021ttt} in the spin-1 sector will lead us to this relation
\begin{align}\label{eq:spin-1}
\square \Phi^{(1)}(u) - W^{(1)}(u) \Phi^{(1)}(u) = 0,
\end{align} 
where 
\begin{align}
& W^{(1)}(u) =  \frac{1}{2} e^{-2 B(u)} \left(-6 A'(u) H'(u)-6 H(u) A'(u)^2+H(u)\right).
\end{align}

In the spin-0 sector, we work with the perturbations

$\left(h_{t t}(u), h_{u u}(u), h_{t u}(u), h_{u z}(u), h_{t z}(u), h_{x x}(u) = h_{y y}(u), h_{z z}(u)\right)$,  and there are five invariant combinations out of these perturbations\cite{Buchel:2021ttt, Jansen:2019wag} 
\footnote{In the Eqs. \eqref{fluc:soundmaster} the $g_{i j}$ components refer to the background metric in the Eq. \eqref{eq: metric}. Also, we express the invariant fields \eqref{fluc:soundmaster} in the Eddington-Finkelstein coordinates that are derived by coordinate transformations to the Schwarzschild forms \cite{Jansen:2019wag}.}

\begin{align}\label{fluc:soundmaster}
    &\mathfrak{h}_{t t}(u) = k^2 h_{t t}(u) + 2 \omega k h_{t z}(u) + \omega^2 h_{zz}(u) + \left( k^2 \frac{g'_{t t}(u)}{g_{t t}(u)}- \omega^2\right) h_{x x}(u),\nn\\
    & \mathfrak{h}_{t u}(u) =  h_{t u}(u) - \frac{g_{u t}(u)}{g_{t t}(u)} h_{t t}(u) + \frac{i}{k} \left(\partial_u + i \omega \frac{g_{u t}(u)}{g_{t t}(u)} - i \frac{g'_{t t}(u)}{k g_{t t}(u)}\right) h_{t z}(u) \nn\\
    & \hspace{1cm} + \frac{i \omega g_{u t}(u)^2}{g_{t t}(u) g'_{x x}(u)} h_{x x}(u) + \frac{i \omega}{2 k^2} \left(- \partial_u - i \omega \frac{g_{u t}(u)}{g_{t t}(u)} + \frac{g'_{t t}(u)}{g_{t t}(u)}\right) (h_{x x}(u) - h_{z z}(u)),\nn\\
    &\mathfrak{h}_{u u}(u) = h_{u u}(u) - 2 \frac{g_{u t}(u)}{g_{t t}(u)} h_{t u}(u) + \frac{g_{u t}(u)^2}{g_{t t}(u)^2} h_{t t}(u) - \frac{2 g_{u t}(u)^2}{g_{t t}(u) g'_{x x}(u)} \left(\partial_u + i \omega \frac{g_{u t}(u)}{g_{t t}(u)} \right)h_{x x}(u),\nn\\
    & \hspace{1cm} + \left(\frac{g_{u t}(u)^2}{g_{t t}(u)^2 } \left(\frac{g'_{t t}(u)}{g'_{x x}(u)} + \frac{g_{t t}(u)}{g_{x x}(u)}\right)- \frac{2 g_{u t}(u)^2 g_{x x}(u)}{3 g_{t t}(u) g'_{x x}(u)^2}\right) h_{x x}(u), \nn\\
    &\mathfrak{h}_{u z}(u) = h_{u z}(u) - \frac{g_{u t}(u)}{g_{t t}(u)} h_{t z} - \frac{i k g_{u t}(u)^2}{g_{t t}(u) g'_{x x}(u)} h_{x x}(u) \nn\\
    &\hspace{1cm} - \frac{i}{2 k} \left(\partial_u + i \omega \frac{g_{u t}(u)}{g_{t t}(u)} - \frac{g'_{x x}(u)}{g_{x x}(u)}\right)(h_{x x}(u) - h_{z z}(u)),\nn\\
    &\boldsymbol{\phi}(u) = \psi(u) - \frac{h_{x x}(u)}{g'_{x x}(u)}.
\end{align}
We can write these invariant perturbations in terms of the master fields $\Phi^{(0)}_2$ and $\Phi^{(0)}_0$ \cite{Buchel:2021ttt}. If we choose a gauge in which master fields do not depend on time, then the above combinations reduce to the following relations 
\begin{align}
       &\mathfrak{h}_{t u} = 0,\nn\\
    &\mathfrak{h}_{u u}(u) = \Phi^{(0)}_0(u) \frac{e^{2 B(u)}  \left(\frac{12 k^2 e^{2 B(u)} H(u) A'(u)}{3 e^{2 A(u)} A'(u) H'(u)+2 k^2 e^{2 B(u)}}-6 H(u) A'(u)\right)}{18 H(u)^2 A'(u)^2}+ \frac{e^{2 B(u)} \Phi'^{(0)}_2(u)}{\sqrt{3} H(u) A'(u)}\nn\\
    & + \Phi^{(0)}_2(u) \frac{e^{2 B(u)}  \left(\frac{24 \sqrt{3} k^2 e^{2 B(u)} H(u) A'(u)^2}{3 e^{2 A(u)} A'(u) H'(u)+2 k^2 e^{2 B(u)}}-3 \sqrt{3} A'(u) H'(u)+\sqrt{3} H(u)\right)}{18 H(u)^2 A'(u)^2},\nn\\
    &\mathfrak{h}_{u z}(u) = \Phi^{(0)}_0(u)\frac{ e^{2 (A(u)+B(u))}}{3 e^{2 A(u)} A'(u) H'(u)+2 k^2 e^{2 B(u)}} + \frac{\sqrt{3} e^{2 A(u)}\Phi'^{(0)}_2(u)}{2 k^2} \nn\\
    & + \Phi^{(0)}_2(u) \frac{e^{2 B(u)} \left(3 e^{2 A(u)} A'(u) H'(u)+12 e^{2 A(u)} H(u) A'(u)^2+2 k^2 e^{2 B(u)}\right)}{2 \sqrt{3} H(u) A'(u) \left(3 e^{2 A(u)} A'(u) H'(u)+2 k^2 e^{2 B(u)}\right)},\nn\\
     &\mathfrak{h}_{t t} = e^{2 A(u)} H(u) \bigg(e^{-2 B(u)} H(u) \left(\mathfrak{h}_{u u}(u) -2 \mathfrak{h}'_{u z}(u)\right)+2 \mathfrak{h}_{u z}(u) \left(2 A'(u) e^{- 2 B(u)} H(u) - V'(u)\right)\bigg),\nn\\
     &\boldsymbol{\phi}(u) = - \Phi^{(0)}_0(u) + \frac{\Phi^{(0)}_2(u)}{2 \sqrt{3} A'(u)}.
\end{align}
We call the $\Phi^{(0)}_2(u)$ mode as the sound mode, while the $\Phi^{(0)}_0(u)$ is called a non-conformal mode since it is intimately related to the scalar field. The equations of motion for the master field in the spin-0 sector can be written as follows \cite{Buchel:2021ttt}
\begin{align}\label{eq:soundmaster}
    \square \Phi^{(0)}_2(u) - W^{(0)}_{2 2}(u) \Phi^{(0)}_2(u) - W^{(0)}_{0 2}(u) \Phi^{(0)}_0(u) =0,\nn\\
    \square \Phi^{(0)}_0(u) - W^{(0)}_{0 0}(u) \Phi^{(0)}_0(u) - W^{(0)}_{0 2}(u) \Phi^{(0)}_2(u) =0,
\end{align}
where
\begin{align}\label{eq:Wsound}
  & W^{(0)}_{2 2}(u) = \frac{ k^4 C^{(0)}_{2 2, k^4} + k^2 C^{(0)}_{2 2, k^2} }{3 \left(3 e^{2 A(u)} A'(u) H'(u)+2 k^2 e^{2 B(u)}\right)^2},\nn\\
  & W^{(0)}_{0 2}(u) = \frac{k^4 C^{(0)}_{0 2, k^4} + k^2 C^{(0)}_{0 2, k^2} }{\sqrt{3} \left(3 e^{2 A(u)} A'(u) H'(u)+2 k^2 e^{2 B(u)}\right)^2},\nn\\
  & W^{(0)}_{0 0}(u) = \frac{  k^4 C^{(0)}_{0 0, k^4} + k^2 C^{(0)}_{0 0, k^2} + C^{(0)}_{0 0, k^0} }{6 \left(3 e^{2 A(u)} A'(u) H'(u)+2 k^2 e^{2 B(u)}\right)^2}.
\end{align}
The coefficients are given below
\begin{align}
& C^{(0)}_{2 2, k^4} = - 8 e^{2 B(u)} \left(6 A'(u) H'(u)+H(u) \left(6 A'(u)^2-1\right)\right),\nn\\
 &   C^{(0)}_{2 2, k^2} = - 72  e^{2 A(u)} A'(u)^2 H'(u) \left(3 H(u) A'(u)+H'(u)\right),\nn\\
 & C^{(0)}_{0 2, k^4} = -8  e^{2 B(u)} \left(H(u) \left(A'(u)-B'(u)\right)+H'(u)\right),\\
 & C^{(0)}_{0 2, k^2} = -2 e^{2 A(u)} H'(u) \left(H(u) \left(-6 A'(u) B'(u)+18 A'(u)^2+1\right)+6 A'(u) H'(u)\right),\nn\\
 &C^{(0)}_{0 0, k^4} = - 8 e^{2 B(u)} \left(H(u) \left(12 A'(u) B'(u)+3 B''(u)-6 B'(u)^2-1\right)+6 B'(u) H'(u)\right),\nn\\
 &C^{(0)}_{0 0, k^2} = -12 e^{2 A(u)} H'(u) \bigg(H'(u) \left(12 A'(u) B'(u)-1\right) \nn\\
 &\hspace{1cm} +H(u) \left(A'(u) \left(24 A'(u) B'(u)+6 B''(u)-12 B'(u)^2-1\right)+2 B'(u)\right)\bigg),\nn\\
 &C^{(0)}_{0 0, k^0} = -3 e^{4 A(u) - 2 B(u)} H'(u)^2 \bigg(18 H(u) A'(u)^2 B''(u)\nn\\
 &\hspace{1cm} + \left(6 A'(u) B'(u)-1\right) \left(H(u) \left(-6 A'(u) B'(u)+12 A'(u)^2+1\right)+6 A'(u) H'(u)\right)\bigg),\nn
\end{align}

An analysis of the equations \eqref{eq:soundmaster} near the conformal boundary leads to the asymptotic behavior
as $r\sim0$
\be\label{form}
\Phi^{(0)}_2(u)\sim A_1 + B_1\, u^{\frac{4}{4-\Delta}}~,
\hspace{40pt}
\Phi^{(0)}_0(u)\sim A_2\, u + B_2\, u^{\frac{\Delta}{4-\Delta}}~.
\ee
Transformation to the usual Fefferman-Graham coordinates close to the boundary, 
$u\mapsto\rho^{4-\Delta}$, reveals 
that $\Phi^{(0)}_2(\rho)$ has the asymptotic of 
metric components like the perturbations considered in \cite{Kovtun:2005ev}. This perturbation corresponds to the sound mode of the theory. On the other hand,  $\Phi^{(0)}_0(\rho)$ has the asymptotic of the 
background scalar field $\phi$ and is similar to the 
case studied in \cite{Benincasa:2005iv}.
The right boundary conditions for the QNM spectrum
are $\left(A_1=0, A_2=0 \right)$.
Examining the Eqs. \eqref{eq:spin-2} and \eqref{eq:spin-1} has shown that the spin-2 mode $\Phi^{(2)}(u)$ and spin-1 mode perturbations $\Phi^{(1)}(u)$ have similar
asymptotic behavior as $\Phi^{(0)}_2(u)$ near the boundary. Thus, it demands a standard Dirichlet boundary condition at $u=0$.

We will use dimensionless frequencies $\Omega\equiv\frac{\omega}{2\pi T}$ and dimensionless momenta $q\equiv \frac{k}{2\pi T}$ to present our results. This helps us to make a reasonable comparison for different regimes of temperatures and various phase structures. In what follows, to illustrate the results better, we separate the sections according to the kinds of phase transitions, and in each section, we clarify the corresponding results of each spin sector.

\section{Some remarks}\label{sec: remarks}
Let us explain some points before we dive into the numerical results. In practice, finding the Quasinormal Modes (QNMs) is nothing but solving a generalized eigenvalue equation. We use the spectral method to solve this task. This method tries to write the solution of the differential equation as a sum of certain “basis functions” and then chooses the coefficients in the sum to satisfy the differential equation. Rapid decay of errors and fast convergence of solutions are among the benefits of this approach. We use spectral discretization with Chebyshev polynomials to solve the complex equations \eqref{eq:spin-2}, \eqref{eq:spin-1}, and \eqref{eq:soundmaster} \cite{Grandclement:2007sb}. The polynomial character of the resulting matrix equation enables us to determine the frequency points for a given $q^2$, i.e., $\Omega(q^2)$ by evaluating the determinant of the matrix and setting it to zero. To find physical solutions with the right boundary conditions given in Eq. \eqref{form}, we fit the tail of the function to a form obtained from the small $u$ analysis. To avoid getting unphysical modes and to have larger modes, we perform the above procedure by two different series of Chebyshev functions. Then we select the outputs that differ only by $0.1\%$ discrepancy. The choice of summation numbers depends on $\phi_H$. For smaller values, they are selected to be ($N_1, N_2$)=($15,20$), while for larger values of $\phi_H$, they have to be chosen at least in ($N_1, N_2$)=($30,40$). It is noteworthy that all modes for which $\text{Re}\,\Omega \neq 0$ come in pairs due to the parity symmetry 
\begin{align}
    \Omega(q^2) = \pm \text{Re} \,\Omega(q^2) + i \, \text{Im}\,  \Omega(q^2).
\end{align}
Apart from this doubling, there is another degeneracy in the spin-0 sector because of the coupling between $\Phi^{(0)}_2$ and $\Phi^{(0)}_0$. According to the hydrodynamic description, the spin-2 mode equation \eqref{eq:spin-2} has no hydro mode, namely $\lim_{q \to 0} \Omega \neq 0$ and all modes are non-hydro, per se. However, in the spin-1 and spin-0 equations, we get either the hydro or non-hydro modes.

Non-hydro modes are universal properties once we investigate Green's functions \cite{Heller:2015dha}. Indeed, in analyzing (high order) hydrodynamics one finds poles/cuts in the Borel plane which exactly correspond to the lowest nonhydrodynamic QNM. This reveals that non-hydro excitations have to be included for the self-consistency of the theory. The importance of these modes will become obvious when we approach the critical points since 
the lowest QNMs become comparable to the
hydrodynamic ones near the critical points. So, the applicability of the effective hydrodynamic
description is to be questioned. These phenomena will be the focus of the present paper.  We find that they become very important in the vicinity of a transition point.

 To derive the transport coefficient from the usual ADS/CFT dictionary, we have to do some tasks on the bulk solutions \cite{Son:2002sd} and match the result with the hydro two-point functions \cite{Kovtun:2012rj}. For our model, this procedure ends with the following for shear and bulk transports
\begin{align}
    &\eta = \lim_{\omega \to 0} \lim_{u \to 0}  \frac{1}{2 \omega}\, e^{4 A(u) - B(u)} H(u)\, \text{Im} \left(h^x_{y}(u)\, h'^x_y(u)\right),\nn\\
    &\xi = - \lim_{\omega \to 0} \lim_{u \to 0}  \frac{1}{2 \omega}\, e^{4 A(u) - B(u)} H(u)\, \text{Im} \left(h^x_{x}(u)\, h'^x_x(u)\right).
\end{align}
The fluctuations $(h^x_y(u), h^x_x(u))$ should have incoming solutions near the horizon and regularity conditions on the boundary surface.  For numerical backgrounds such as we did in the present paper, it is a tedious job to get these kinds of solutions. However, from the hydro mode or QNMs in low momenta regime, we could find them.  We know that in the spin-1 and spin-0 sectors, the hydro modes have the following dispersion relation
\begin{align}\label{eq:hydromode}
    \Omega^{(1)}(q^2) \approx - i 2 \pi \frac{\eta}{s} q^2, \qquad \Omega^{(0)}(q^2) \approx \pm c_s \vert q\vert - i 2 \pi T \Gamma_s q^2.
\end{align}
Superscript indices refer to the spin numbers and  $\Gamma_s$ is the sound attenuation constant
\begin{align}
    \Gamma_s = \frac{1}{2 T} \left( \frac{4 \eta}{3 s} + \frac{\xi}{s}\right).
\end{align}
Therefore, finding the QNMs in low momenta and matching them with the polynomial behavior \eqref{eq:hydromode} will give us the $\eta/s, \xi/s$ as well as the $c_s$. In all cases, we found $\eta/s = 1/(4 \pi)$ and $c_s$ coincides with the Eq. \eqref{eq:cs}. The results are shown in Fig. \ref{fig: cs2-zeta}.
\section{Crossover phase transition}\label{sec: crossover-results}
In this section, we illustrate the outcomes for the crossover phase transition with $B_4 = 0$ according to the numerical setups we stated earlier. Specific results for each spin sector are given in separate subsections, from spin-2 to spin-0 consecutively. To find the collision between the hydro and non-hydro modes or detect the radius of convergence in the hydrodynamic series in each part, we apply the complex momentum paradigm in the QNM spectra \cite{Withers:2018srf, Grozdanov:2019uhi}. We obtain the QNMs for a complex momentum $q^2 = \vert q^2\vert e^{i \theta}$ and by varying $\vert q^2\vert$ and $\theta$, we can find the position of mode touching. This shows the location where the modes possess a singularity in the complex momenta plane.  

\subsection{Spin-2 sector}

\begin{figure}
    \centering
    \includegraphics[width=0.40\textwidth, valign=t]{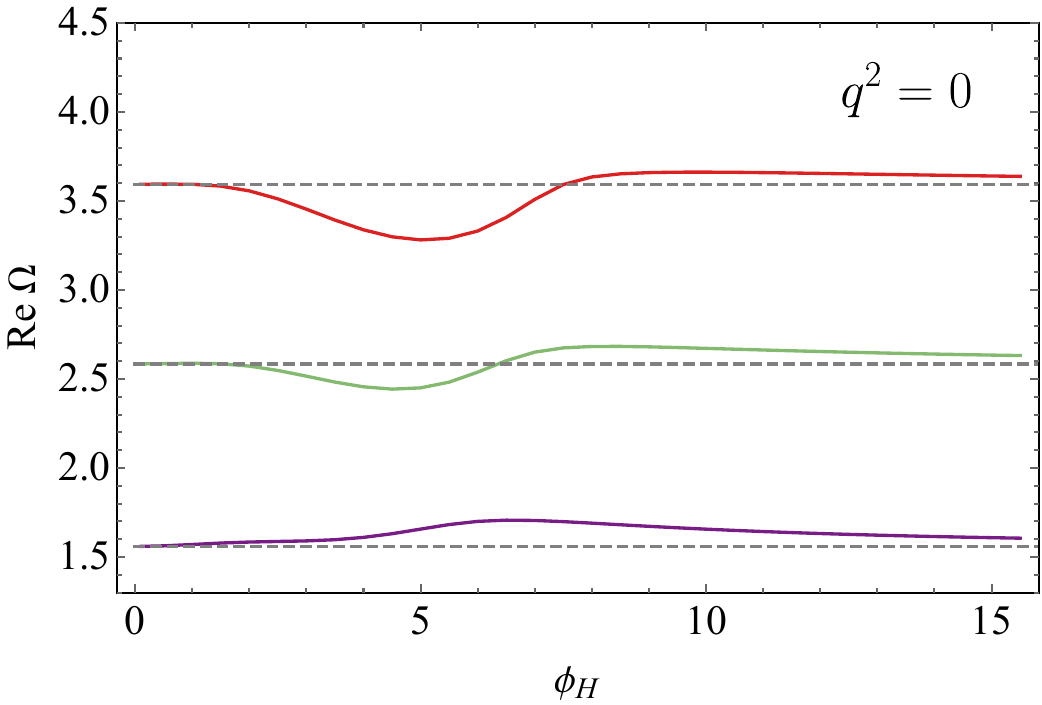}
    \includegraphics[width=0.40\textwidth, valign=t]{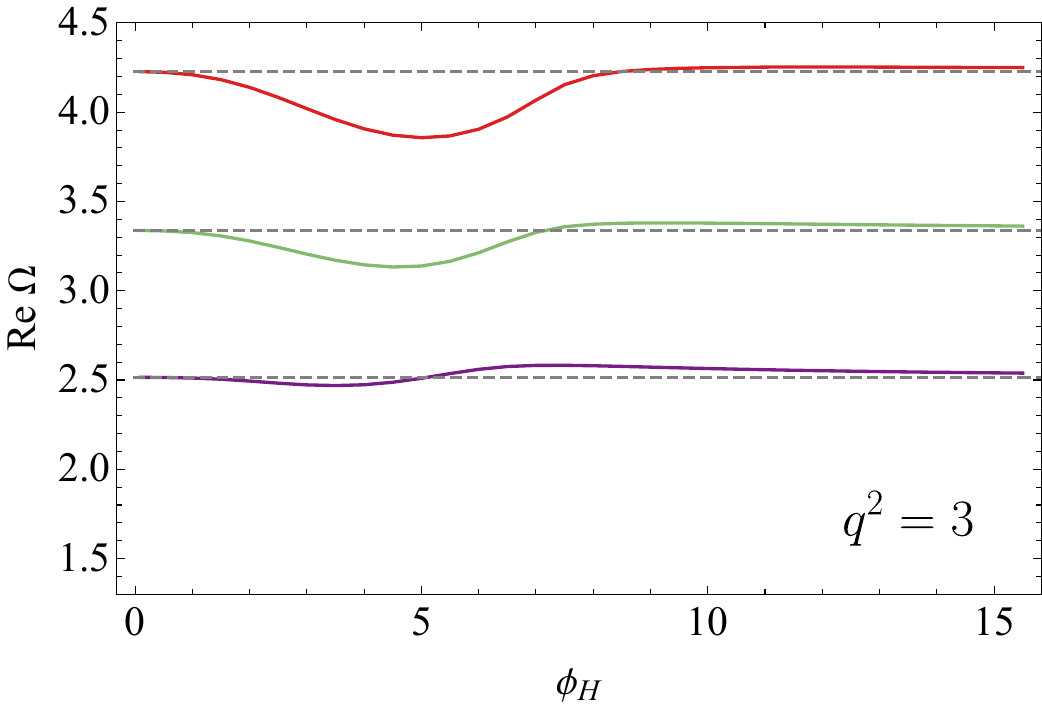}\\
    \includegraphics[width=0.40\textwidth, valign=t]{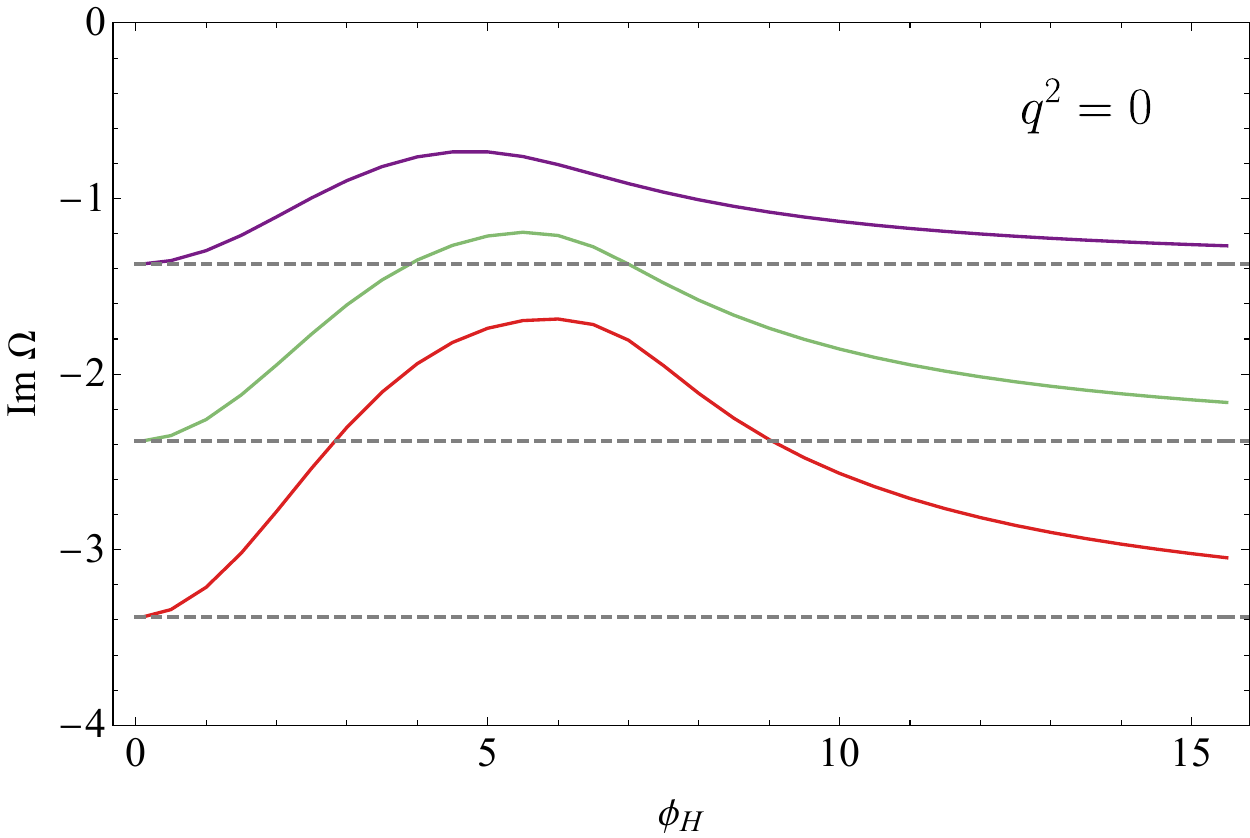}
    \includegraphics[width=0.40\textwidth, valign=t]{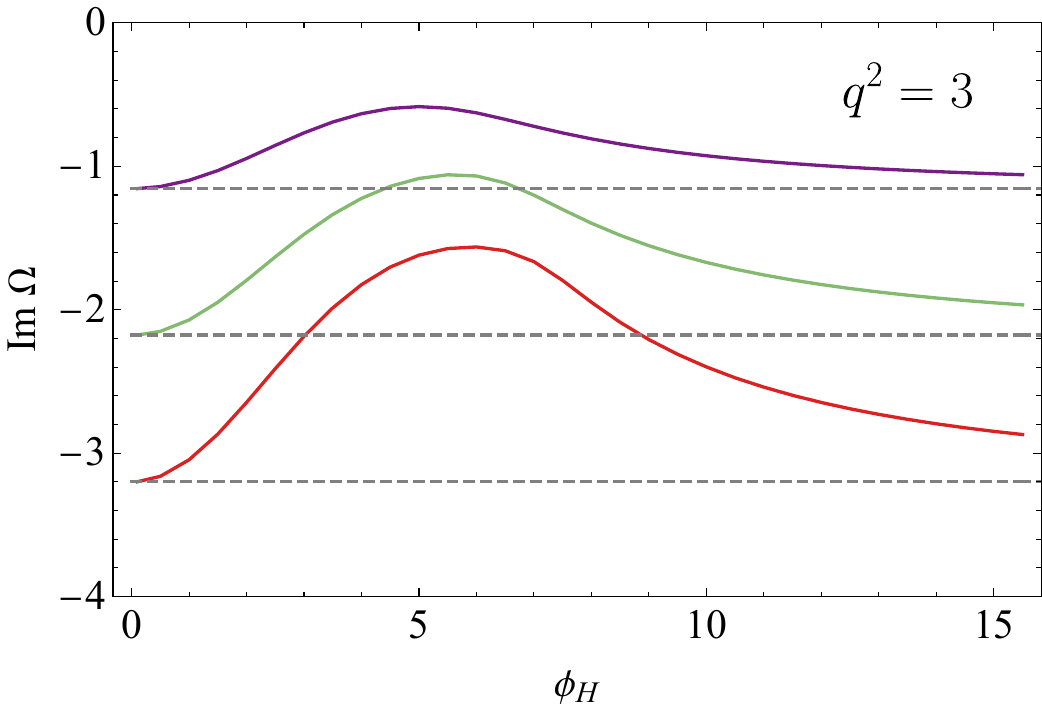}
    \caption{The real part (first row) and the imaginary part (second row) of the lowest three QNMs in Spin-2 sector as functions of $\phi_H$ at $q^2=(0, 3)$ for the crossover EoS with $B_4=0$. The Dashed lines show the results for the 5D-AdS-Schwarzschild black hole.}
    \label{fig: CO-Real-Momenta-EoS-Im-Re-Omega}
\end{figure}

In Fig.~\ref{fig: CO-Real-Momenta-EoS-Im-Re-Omega},  we show the real and imaginary part of the lowest QNMs for the crossover EoS with $B_4=0$ and real $q^2$. Note that, in the spin-2 sector, there are no hydro modes. From these figures, we see that each mode asymptotically approaches its known value for the 5D-AdS-Schwarzschild black hole in large (small) $\phi_H$ or small (large) temperature.  Moreover, it turns out that the dependence of these frequencies on the $q^2$ is very mild, especially in the imaginary part, and can be neglected in a first approximation. This is due to the soft nature of the crossover transition, which leads to a certain “ultralocality” of the dynamics of the nonequilibrium modes on top of a hydrodynamic flow \cite{Janik:2015waa}. Near the phase transition point, $\phi_c \approx 3.77$, there is a bump in the real and imaginary parts that reflects the deviations from the 5D-AdS-Schwarzschild results. Additionally, the modes of high and low temperatures do not mix, which is a unique feature for crossover transition. 

\begin{figure}[htb]
    \centering
    \includegraphics[width=0.55\textwidth]{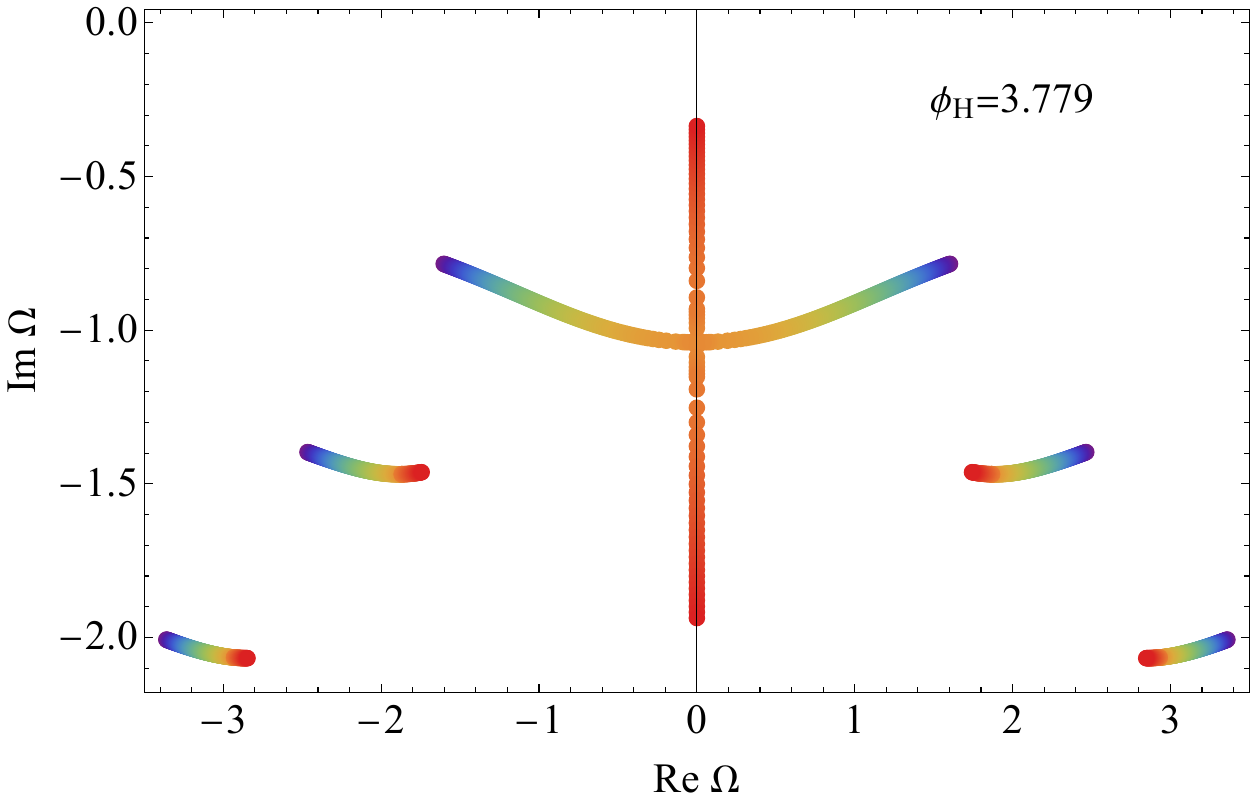}
    \caption{The collision of the lowest non-hydrodynamic modes at the critical point of the crossover EoS with $B_4=0$. The imaginary momentum goes from $q=0$ (Blue) to $q=i \sqrt{2}$ (red) and the collision occurs at $q^2=-1.7080$.}
    \label{fig: B40Imq}
\end{figure}

In Fig.~\ref{fig: B40Imq}, we show the collision between the non-hydro modes in the spin-2 sector at the critical temperature of the crossover EoS. Let us emphasize that in this sector, the collisions always occur in imaginary momenta, $q^2<0$.  We consider $-2 \leq q^2 \leq 0$ and use the rainbow style.
We repeat this calculation for each $\phi_H$ in the complex momenta plane and observe that collision always happens at $\theta = \pi$. 

\begin{figure}
    \centering
    \includegraphics[width=0.435\textwidth, valign=t]{rcScalarCO.pdf}
    \hspace{0.35cm}
    \includegraphics[width=0.425\textwidth, valign=t]{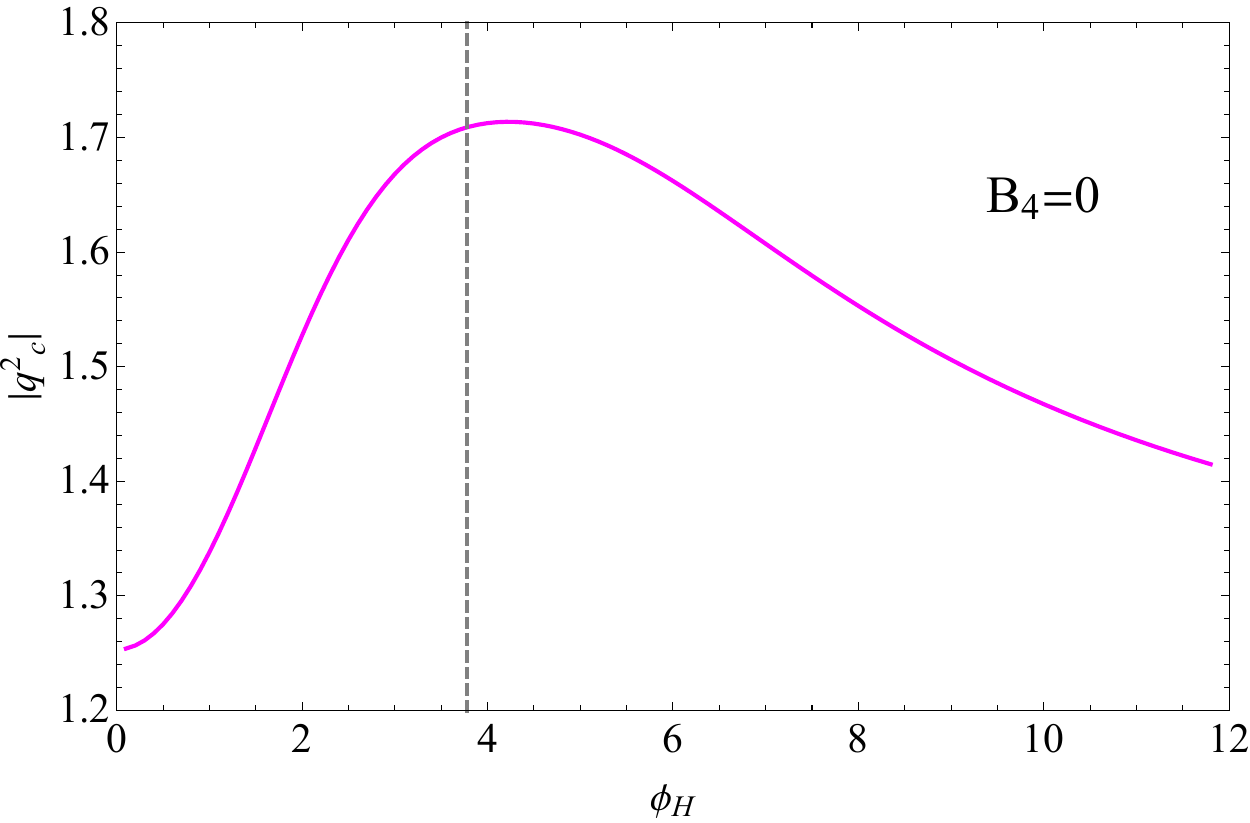}
    \caption{Plots of the radius of convergence for the lowest non-hydrodynamic modes in the spin-2 sector for the crossover phase transition. The collision always occurs at $\theta = \pi$. The dashed line indicates the critical point, $\phi_c = 3.779$. The left(right) plot shows the dependency of $q^2(T) (q^2(\phi_H))$. }
    \label{fig: convergence-radius-phiH-CO}
\end{figure}

In Fig. \ref{fig: convergence-radius-phiH-CO},  we show the $\vert q^2_c \vert$ in terms of $\phi_H$ and $T/T_c$, where the dashed line represents the transition point, $\phi_c = 3.779$. Near the transition point, $\vert q^2_c \vert$ increases, which can be interpreted as an improvement of the hydrodynamic series for this kind of phase transition. An interesting fact is that the high-temperature and low-temperature limits of $\vert q^2_c \vert$ seem to be equal, and they show the same validity limit for using the hydro expansion in these two regimes.

\subsection{Spin-1 sector}

\begin{figure}[htb]
    \centering
    \includegraphics[width=0.44\textwidth, valign=t]{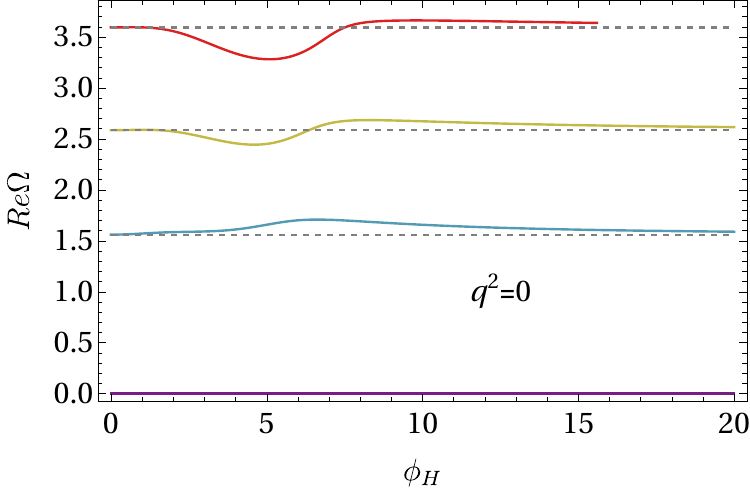}
    \hspace{0.2cm}
    \includegraphics[width=0.44\textwidth, valign=t]{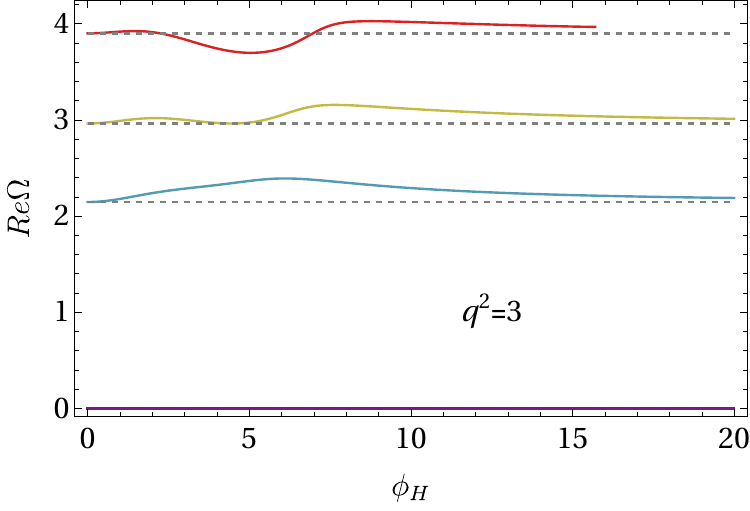}\\
    \vspace{0.2cm}
    \includegraphics[width=0.45\textwidth, valign=t]{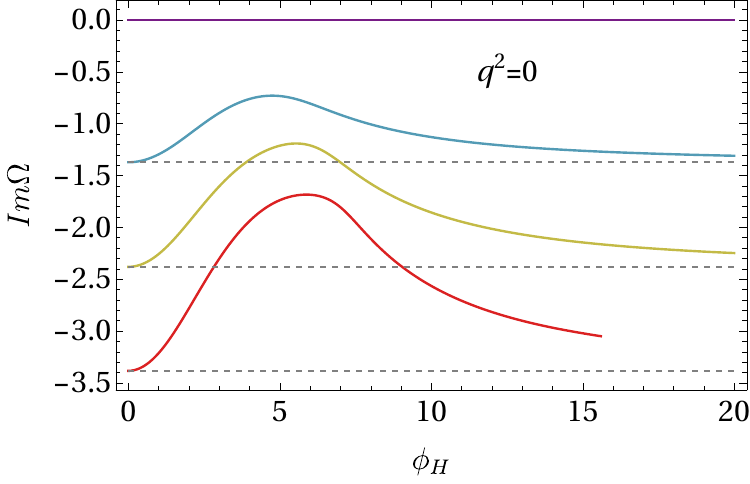}
\hspace{0.1cm}
    \includegraphics[width=0.45\textwidth, valign=t]{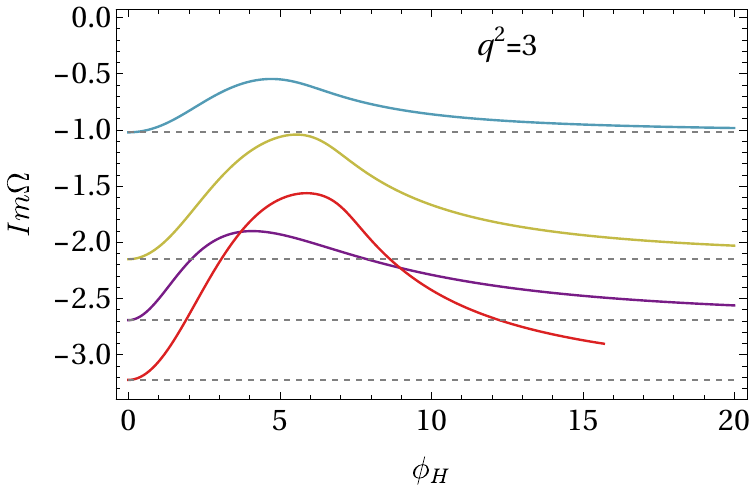}
    \caption{The real part (first row) and the imaginary part (second row) of the lowest three QNMs in the spin-1 sector as functions of $\phi_H$ at real  $q^2=(0, 3)$ for crossover EoS with $B_4=0$. The dashed lines show the results for the 5D-AdS-Schwarzschild black hole. The Magenta line in each plot represents the hydro modes, while the other lines show non-hydro modes.}
    \label{fig: CO-Real-Momenta-EoS-Im-Re-Omega-spin1}
\end{figure}

Similar to the previous subsection, in Fig. \ref{fig: CO-Real-Momenta-EoS-Im-Re-Omega-spin1}, we show the real and imaginary parts of the lowest modes for real $q^2= (0, 3)$ in the spin-1 sector for  $B_4=0$. Unlike the spin-2 case, in the spin-1 sector, we get a hydro mode for each $q^2$,  which is purely imaginary. This is the well-known shear mode that according to Eq.  \eqref{eq:hydromode} at small momenta defines the ratio $\eta/s$. The magenta lines represent the hydro mode at small momenta. Evolving the $q^2$ results in small modifications of the real parts, while the imaginary parts of the modes start to go through each other. This is due to the purely imaginary hydro modes and the collision always must occur in imaginary parts. Another observation is that non-hydro modes do not mix at very high or very low temperatures.   

\begin{figure}
    \centering
        \includegraphics[width=0.45\textwidth, valign=t]{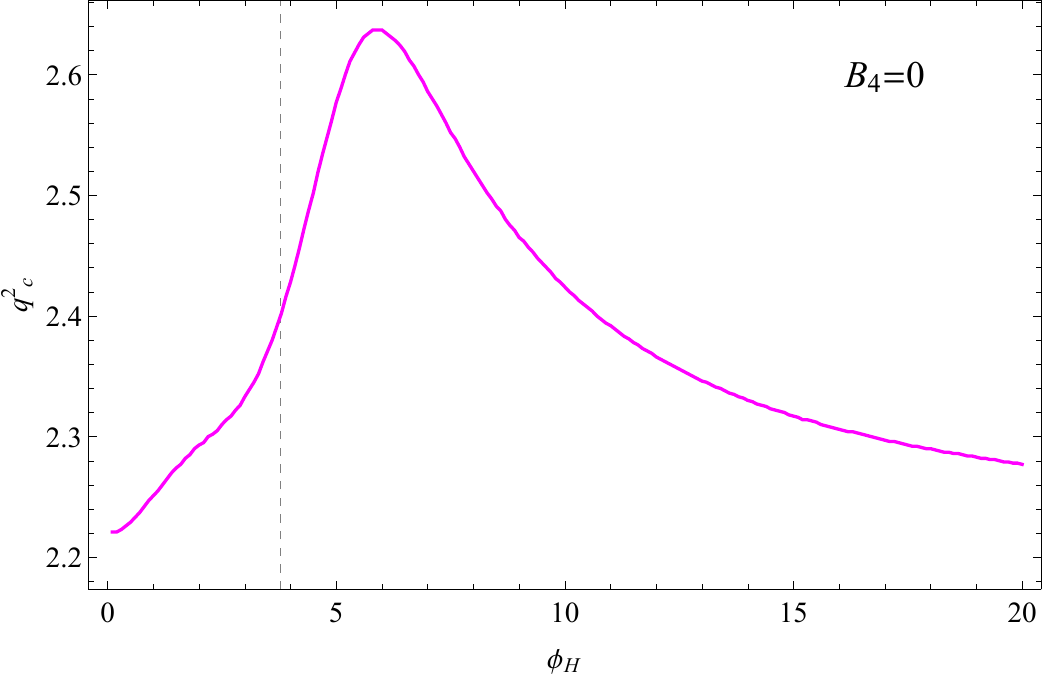}
        \includegraphics[width=0.45\textwidth, valign=t]{plot1-CO.pdf}
    \caption{Left panel: the  radius of convergence in spin-1 sector in terms of $\phi_H$,  Right panel: the same in terms of $T/T_c$. Plots correspond to the cross-over phase transition with $B_4=0$. The gray dashed line represents the location of the transition point, $\phi_H^c \approx 3.779$.}
    \label{fig: rc-shear-co}
\end{figure}

The radius of convergence or $q^2_c \geq 0$ in this sector is shown in Figure~\ref{fig: rc-shear-co}.  At very high or very low temperatures, it reaches the 5D-ADS values, $\vert q^2_c \vert \approx 2.224$.  The $q^2_c$ increases near the transition point because the crossover phase transition is not a true transition. Therefore,  the paradigm {"\it{breakdown of the hydrodynamics near the transition point"}} loses its validity. 

\begin{figure}
    \centering
    \includegraphics[scale=0.34,valign=t]{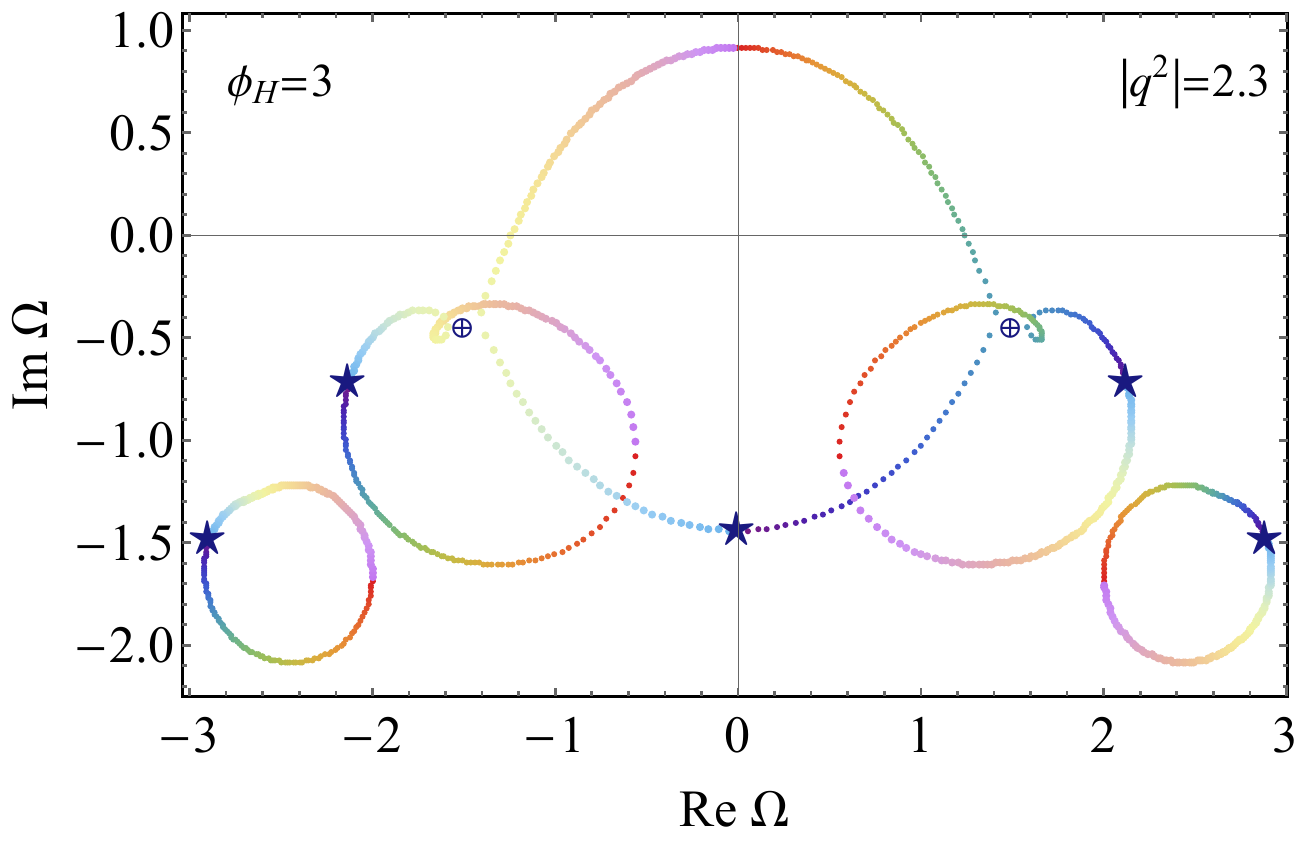}
    \includegraphics[scale=0.34,valign=t]{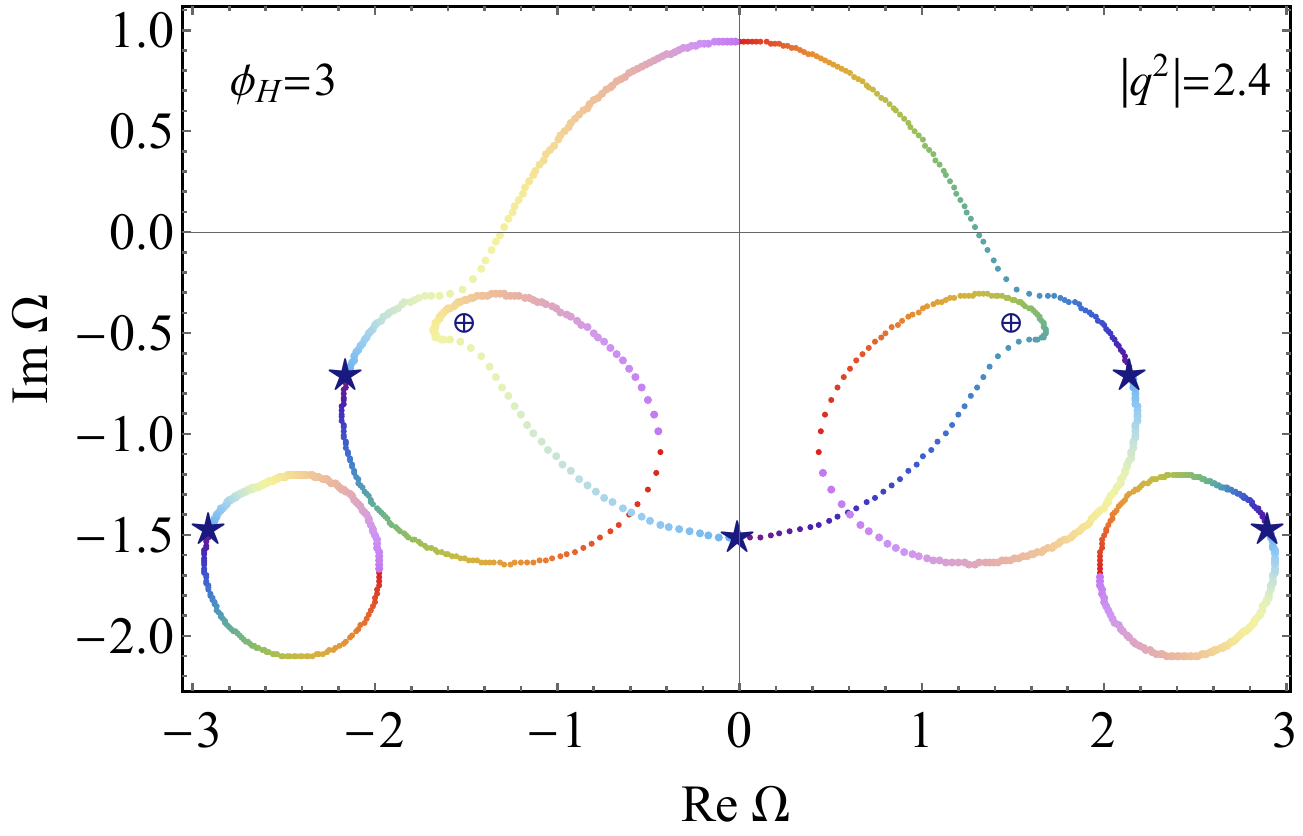}\\
    \vspace{0.3cm}
    \includegraphics[scale=0.34,valign=t]{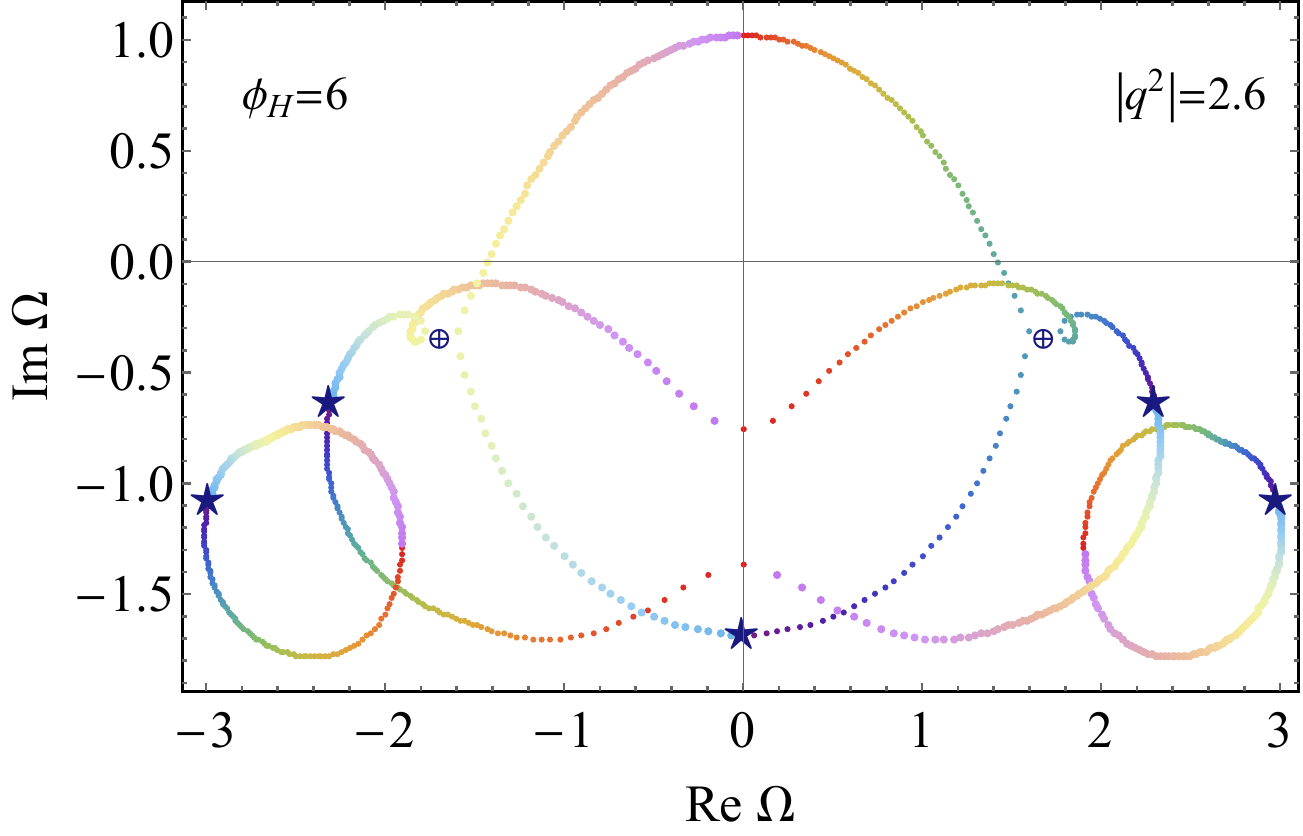}
    \includegraphics[scale=0.34,valign=t]{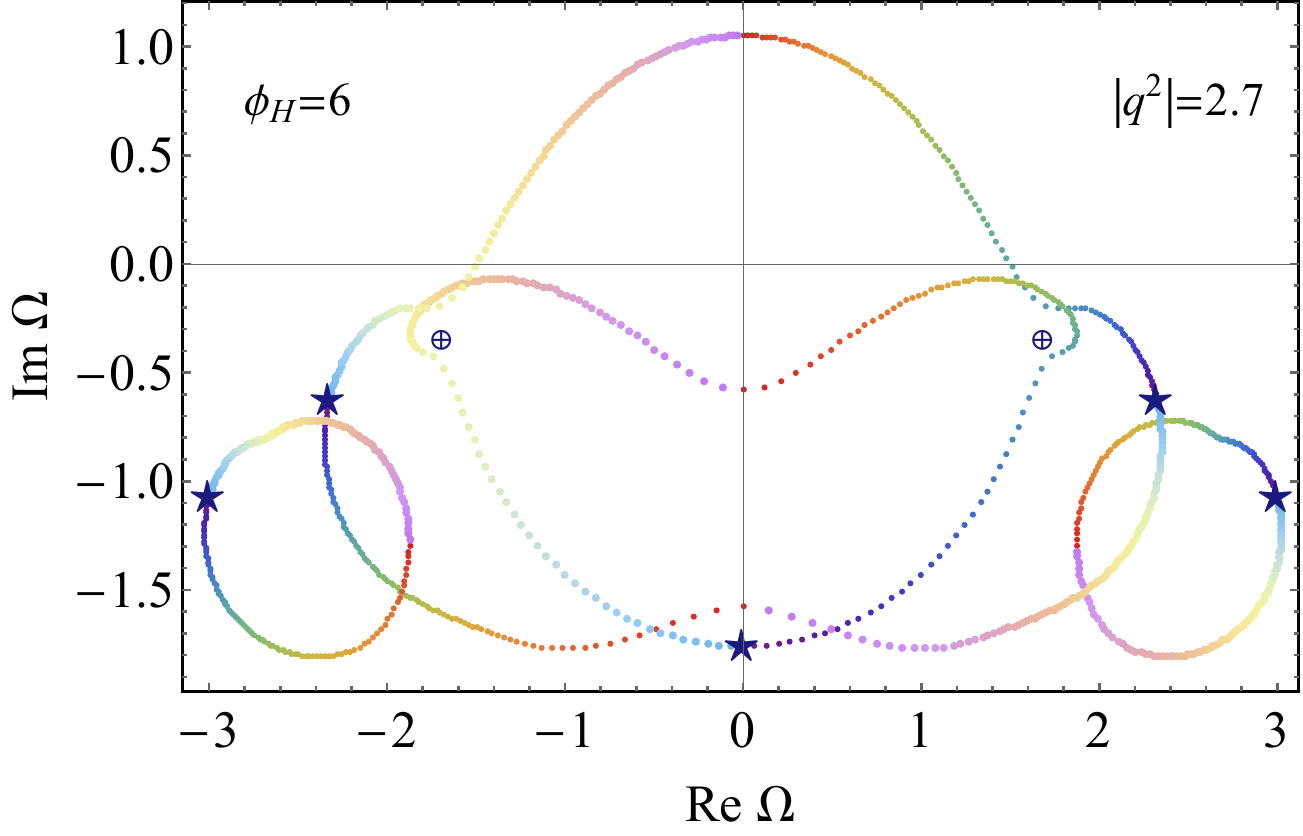}
    \caption{Mode collision for  $B_4=0$ in spin-1 sector. The top (bottom) rows indicate $\phi_H=3$ ($\phi_H=6$) data, and the left (right) panels demonstrate the modes before (after) the collision. At $\phi_H=3$, the collision occurs at $\vert q^2 \vert = 2.333$ with $\Omega_\star=\pm 1.498-0.437 i$, and for $\phi_H=6$, the collision occurs at $\vert q^2 \vert=2.637$ with $\Omega_\star=\pm 1.687-0.3344 i$. The marked circles in each plot indicate the location of $\Omega_\star$.}
    \label{fig: collision-mode-spin1-CO}
\end{figure}

\begin{figure}
    \centering
    \includegraphics[width=0.53\textwidth, valign=t]{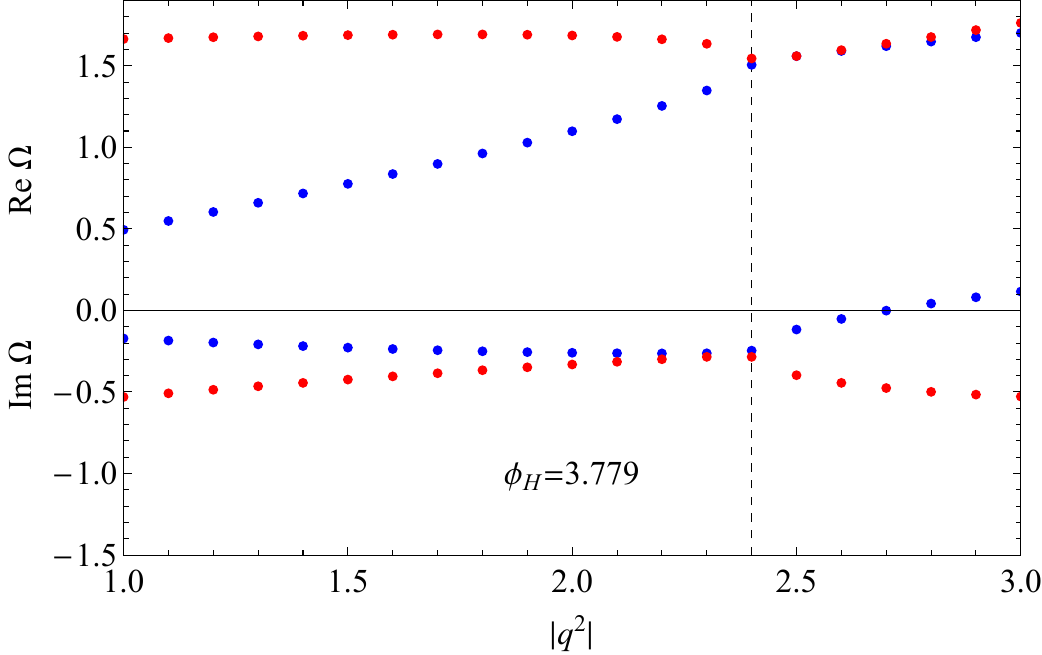}
    \caption{The plot of collision between the lowest hydro mode and a non-hydro mode at $\phi_H = 3.779$ for $B_4=0$ in the spin-1 sector. This collision occurs at $\theta=0.364 \pi$ and $\vert q^2 \vert = 2.4$ which is shown by the gray dashed line.  The blue and red dots correspond to the hydro and closest non-hydro modes, respectively. }
    \label{fig: collision-CO}
\end{figure}

To obtain the collision point at each $\phi_H$, we need to find the location where the imaginary and real parts of the hydro and closest non-hydro mode are equal to each other.  In the spin-1 sector, we were only involved with the metric perturbations (look at Eq. \eqref{eq: spin-1-p}), and the collision of the hydro modes and first gravity non-hydro modes define the radius of convergence. In Fig. \ref{fig: collision-mode-spin1-CO}, we show this collision for $\phi_H = (3, 6)$. The left (right) panels are before  (after) collision.  Before the collision, the modes have their path, but the collisions lead to path sharing in which the hydro modes change their way from a circle round to a more complex one.  Our criterion for the radius of convergence is the smallest value at which the hydro modes and smallest non-hydro modes collide with each other.  
For instance, in Fig. \ref{fig: collision-CO}, we show the approach of the hydro mode and the lowest gravity non-hydro modes at $\phi_H=3.779$. The collision occurs at $\theta=0.364 \pi$ and $\vert q^2 \vert = 2.4$, where real and imaginary parts are equal.
\subsection{Spin-0 sector}
\begin{figure}[htb]
    \centering
    \includegraphics[width=0.45\textwidth]{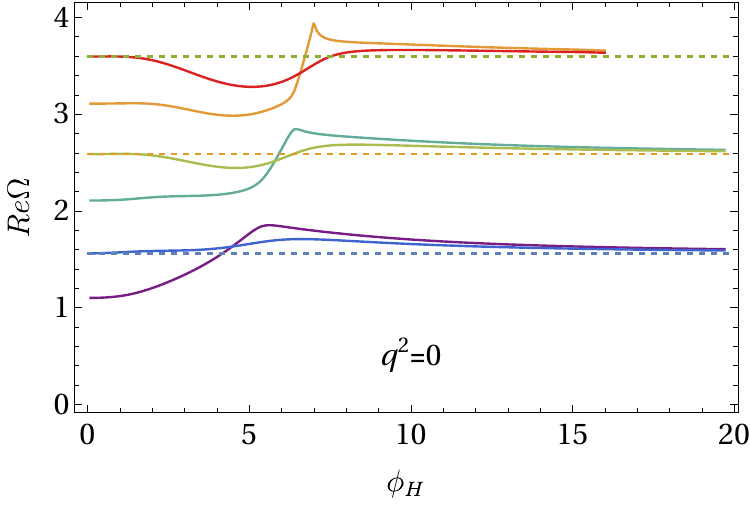}
    \hspace{0.15cm}
    \includegraphics[width=0.45\textwidth]{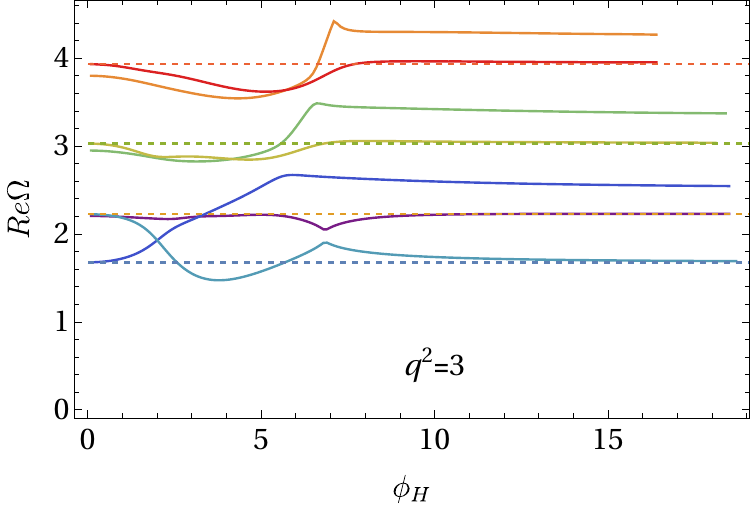}
    \\
    \includegraphics[width=0.45\textwidth]{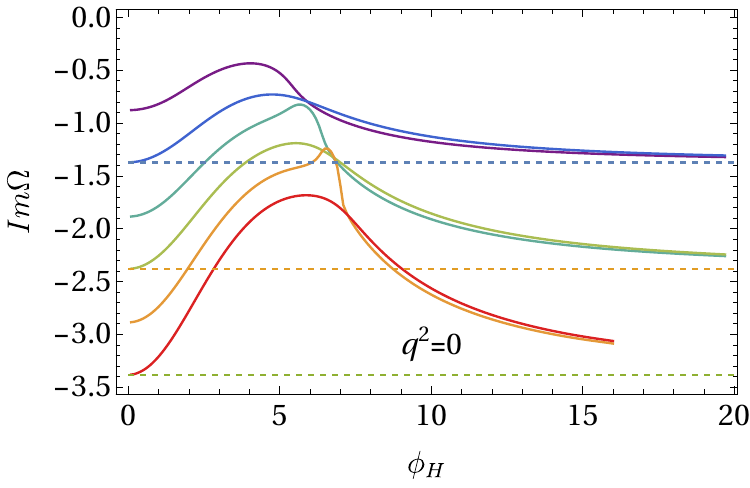}
    \hspace{0.15cm}
     \includegraphics[width=0.45\textwidth]{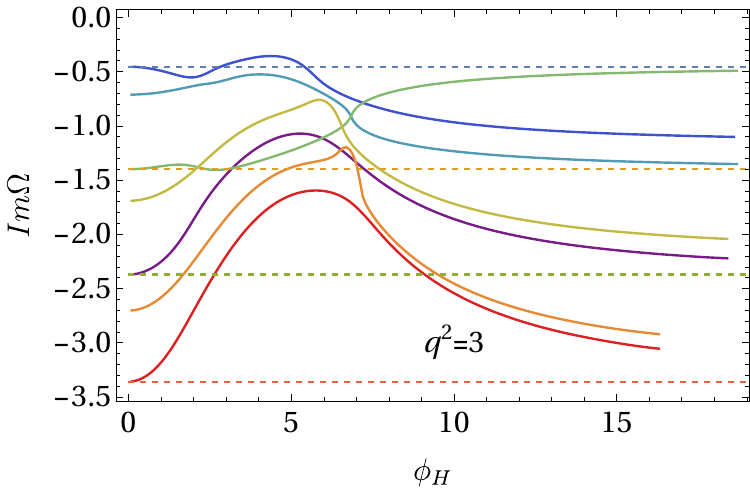}
    \caption{The real part (first row) and the imaginary part (second row) of the lowest QNMs in the spin-0 sector as functions of $\phi_H$ at $q^2=(0, 3)$ for $B_4=0$. The magenta line denotes the smallest mode value, while the gray dashed lines represent the results of the 5D-AdS-Schwarzschild black hole.}
    \label{fig:CO-spin0}
\end{figure}

In Fig. \ref{fig:CO-spin0},  we show the real and imaginary parts of the lowest QNMs for $B_4=0$ EoS at $q^2 = (0,3)$.  According to Eq. \eqref{eq:soundmaster}, there are two coupled perturbations including the gravity and scalar perturbations. Due to this doubling, we expect a more complex pattern. For instance, in Fig. \ref{fig:CO-spin0}, the real and imaginary parts of the lowest modes in the spin-0 sector are sketched for real momenta $q^2 = (0,3)$.  This figure is very different from analogous Figs. \ref{fig: CO-Real-Momenta-EoS-Im-Re-Omega} and \ref{fig: CO-Real-Momenta-EoS-Im-Re-Omega-spin1}. There lie two modes on top of each gray line, one corresponds to gravity, and the other belongs to the scalar field perturbations. Besides this doubling, there is an extra interesting feature concerning the value of $q^2$. At $q^2=0$, equations of gravity and scalar field decouple due to Eqs. \eqref{eq:soundmaster} and  \eqref{eq:Wsound}, and we get an even number of "non-hydro" modes. However, a non-vanishing $q^2$ leads to coupled equations, which apart from doubling at higher non-hydro modes, we get one hydro mode (smallest modes), and so are the odd modes. Patterns of collision are also different for $q^2=0$ and $q^2\neq 0$. At $q^2=0$, the first collision occurs between two higher modes around $\phi_H \approx 5$, while at $q^2 \neq 0$ this collision happens frequently between the smallest mode and higher non-hydro modes. At very high or very low temperatures this touch is among gravity and scalar field perturbations, while at intermediate points it occurs between the hydro and first gravity non-hydro modes.  This feature is seen in the real parts but in the imaginary parts, the collisions happen always between two similar types of modes. \footnote{By collision, we mean a collision that determines the radius of convergence, that is the collision between the hydro-more and first non-hydro modes. Other non-hydro mode collisions have nothing to do with hydrodynamic series.} Also the high temperature and low temperature limits of the hydro modes are the same.

\begin{figure}[htb]
    \centering
        \includegraphics[width=0.43\textwidth, valign=t]{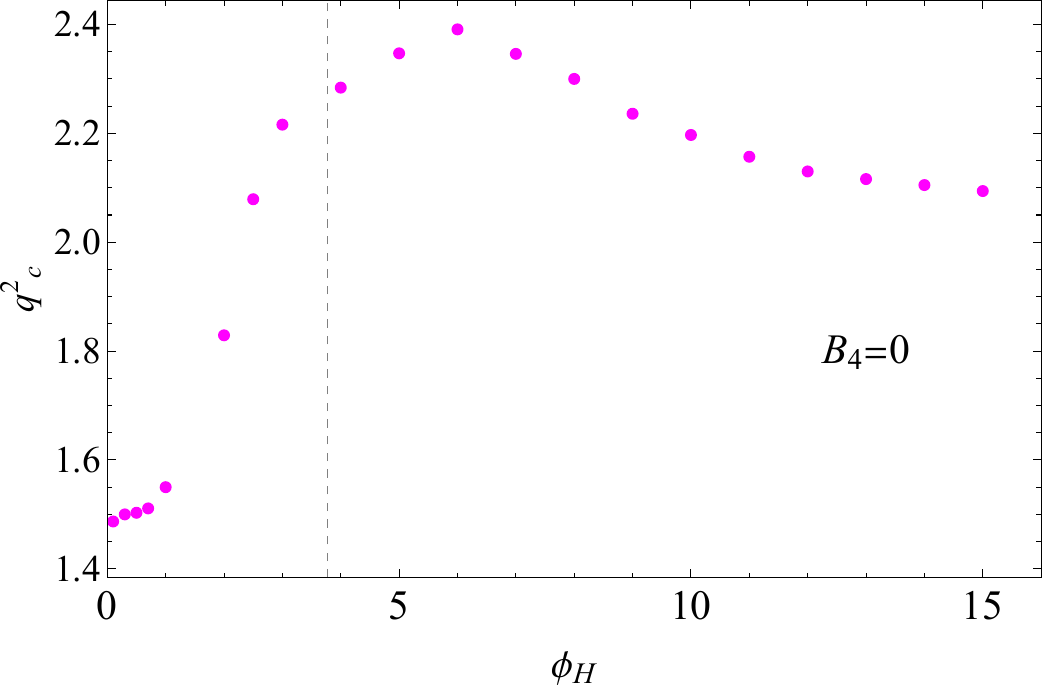}
        \hspace{0.4cm}
        \includegraphics[width=0.43\textwidth, valign=t]{q2c-T-CO-spin0.pdf}
    \caption{Plots of the radius of convergence in terms of $\phi_H$ (left panel) and $T/T_c$ (right panel) in spin-0 sector for crossover EoS. The Red dashed line represents the location of the critical point.}
    \label{fig: rc-sound-co}
\end{figure}

Finally, Fig. \ref{fig: rc-sound-co} demonstrates the radius of convergence in the $B_4 = 0 $ EoS for the spin-0 sector. Similar to the Figs. \ref{fig: convergence-radius-phiH-CO} and \ref{fig: rc-shear-co}, at low and high temperatures, the radius of convergence approaches an identical value $\vert q^2 \vert_c = 1. 486$ that reflects a kind of duality. Furthermore, it seems that the following relation holds between different spin sectors
\begin{align}\label{eq:compare-rc}
    (\text{Max} \vert q^2 \vert_c )_{\text{spin-2}} < (\text{Max} \vert q^2 \vert_c )_{\text{spin-0}} < (\text{Max} \vert q^2 \vert_c )_{\text{spin-1}}.
\end{align}
The term $(\text{Max} \vert q^2 \vert_c )$ refers to the maximum value of $\vert q^2 \vert_c$ in the $\phi_H$ or temperature plane. We will observe that relation \eqref{eq:compare-rc} holds for both second and first-order phase transitions.
\section{Second-order  phase transition}\label{sec: secondorder-results}
In this section, we review the numerical results of the second-order phase transition with  $B_4 = - 0. 0098$. Similar to the previous section, each spin sector is discussed in separate parts.
\subsection{Spin-2 sector}
\begin{figure}[htb]
    \centering
  \includegraphics[width=0.45\textwidth, valign=t]{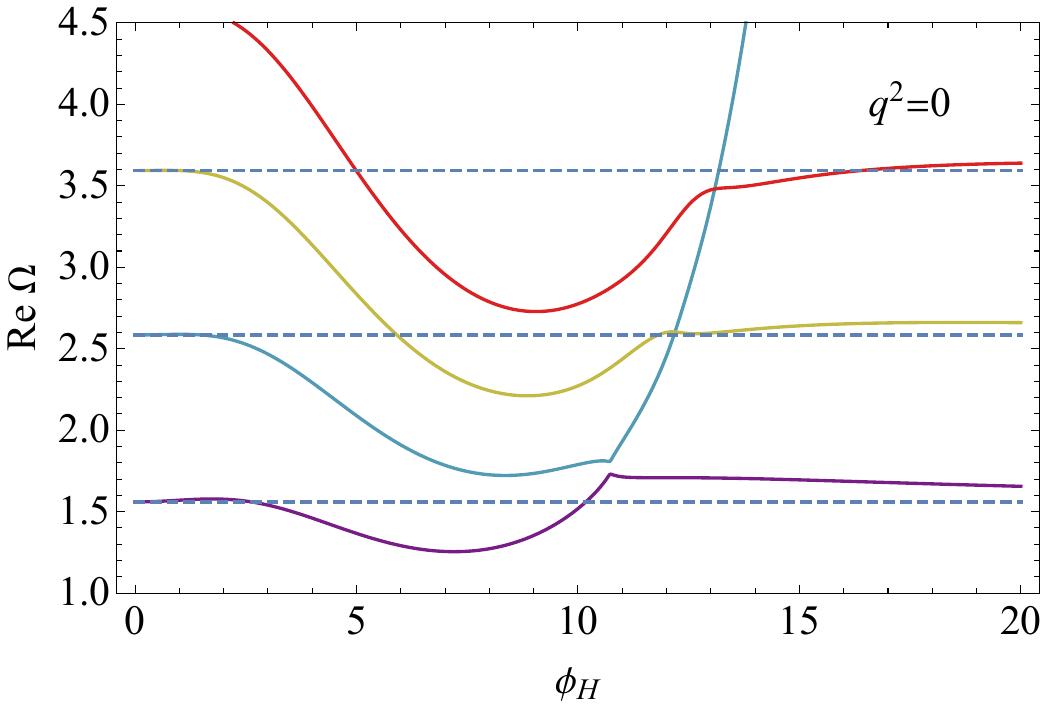}
  \hspace{0.5cm}
    \includegraphics[width=0.45\textwidth,valign=t]{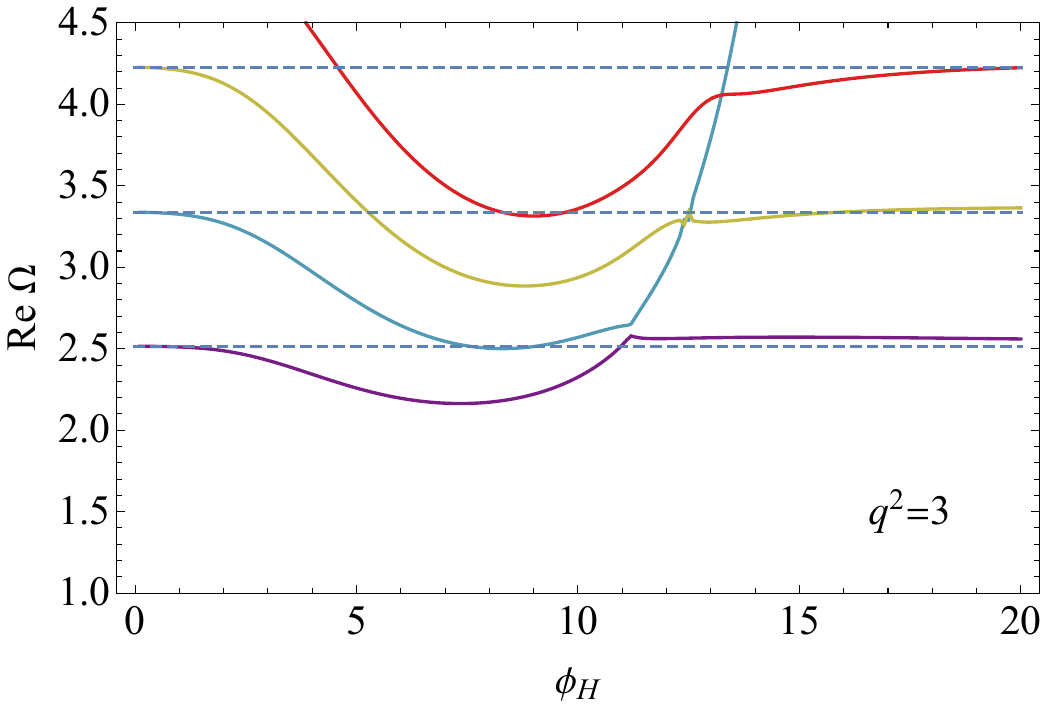}\\
    \includegraphics[width=0.45\textwidth, valign=t]{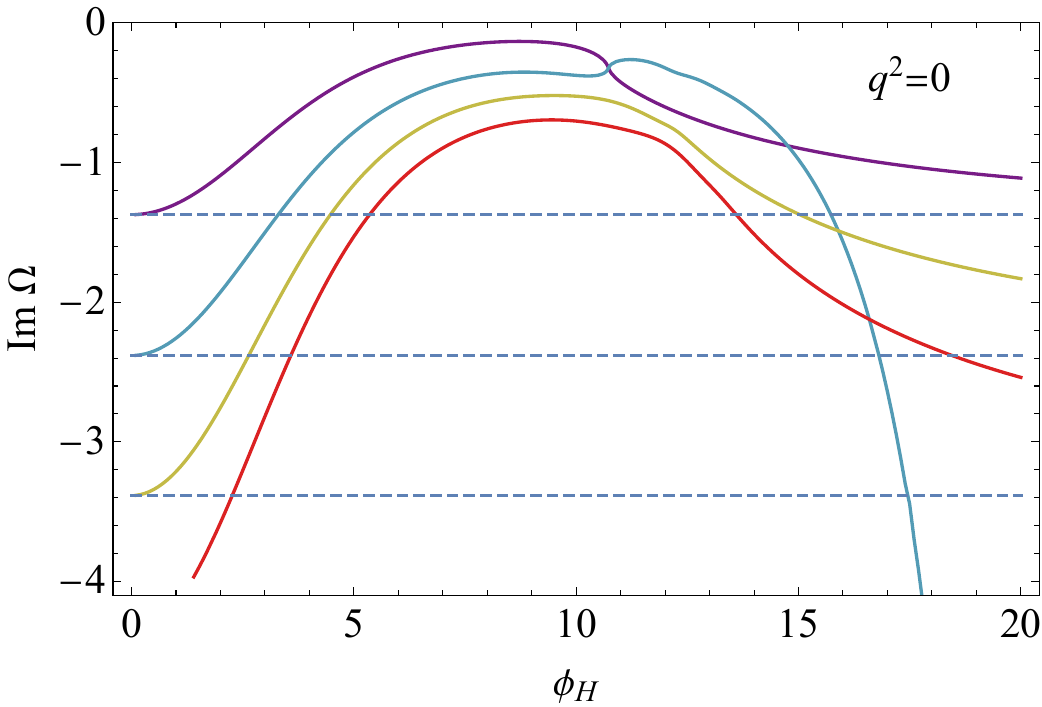}
    \hspace{0.5cm}
    \includegraphics[width=0.45\textwidth, valign=t]{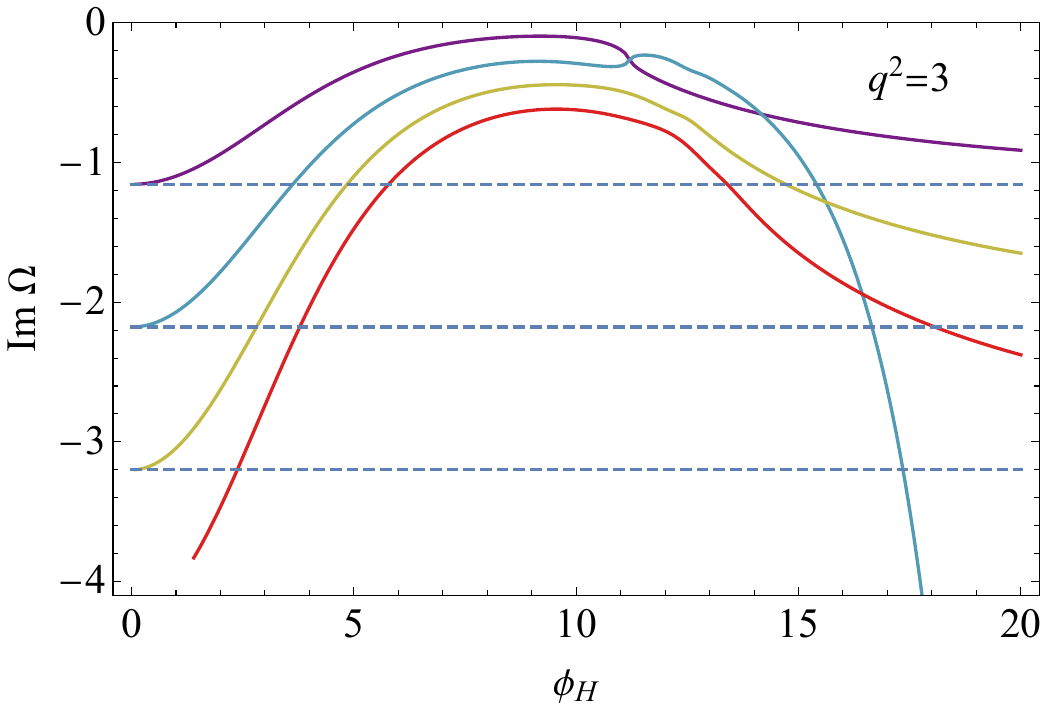}
    \caption{The real part (first row) and the imaginary part (second row) of the lowest three QNMs in the spin-2 sector as functions of $\phi_H$ at $q^2=(0,3)$ for the second-order EoS with $B_4=-0.0098$. The magenta line refers to the lowest non-hydro mode.}
    \label{fig: SO-Real-Momenta-EoS-Im-Re-Omega}
\end{figure}

In Fig.~\ref{fig: SO-Real-Momenta-EoS-Im-Re-Omega},  we show the real and imaginary parts of the lowest QNMs for the second-order phase transition by choosing real $q^2$. We observe that for $q^2\leq 3$, the second lowest mode in the high-temperature regime diverges to infinity in the low-temperature limit, while the QNM structure of the lowest frequencies approaches the 5D-AdS-Schwarzschild black hole. This happens such that the $i+1$-th mode at high temperature will be the $i$-th mode in low temperature, $\Omega_{i+1}(T\gg T_c )\rightarrow\Omega_{i}(T\ll T_c)$ for $i>2$. Furthermore, there is no collision between the modes.

\begin{figure}[htb]
    \centering
    \includegraphics[width=0.55\textwidth]{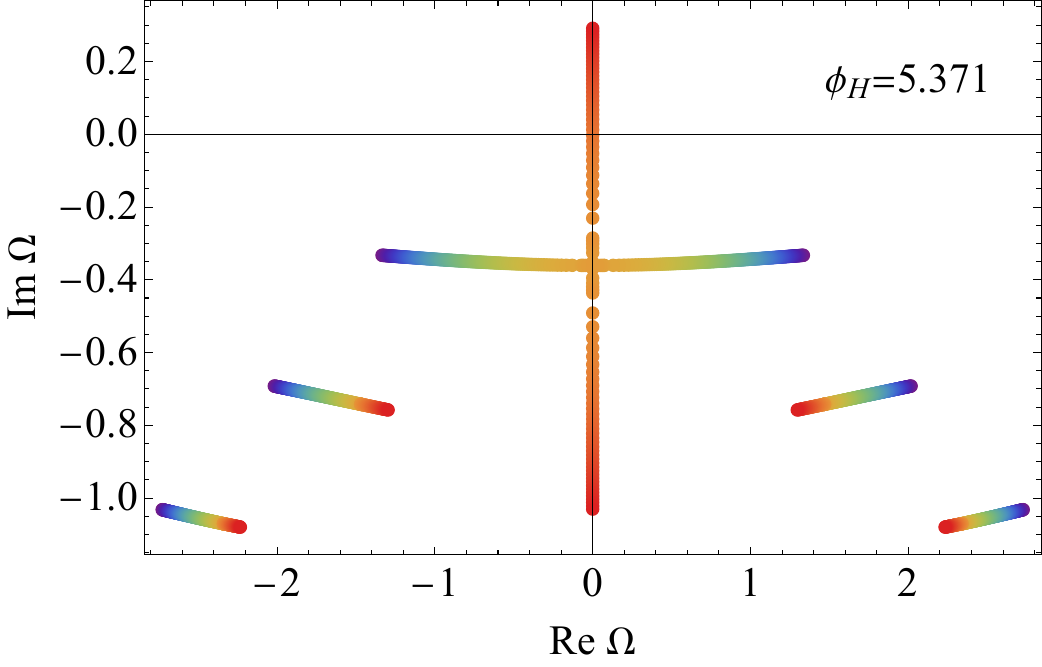}
    \caption{The plots of collision of the lowest non-hydro modes at the transition point of the second-order phase transition. The imaginary momentum goes from $q=0$ (purple) to $q=i \sqrt{2}$ (red) and the collision occurs at $q^2= -1.6148$.}
    \label{fig: 2ndImq}
\end{figure}

In Fig.~\ref{fig: 2ndImq}, we show the collisions that occur at the critical temperature of the second-order phase transition.  In this sector, the collisions always occur in imaginary momenta or $q^2<0$. We consider $ -2 \leq q^2 \leq 0$ and use the rainbow style. The collision at $\phi_c$ occurs at $q^2= -1.6148$. In the spin-2 sector, the first collision always occurs at negative $q^2$ and is between the two closest gravity non-hydro modes.

\begin{figure}[htb]
    \centering
    \includegraphics[width=0.4\textwidth, valign=t]{rcScalar2ndPT.pdf}
    \hspace{0.4cm}
    \includegraphics[width=0.4\textwidth, valign=t]{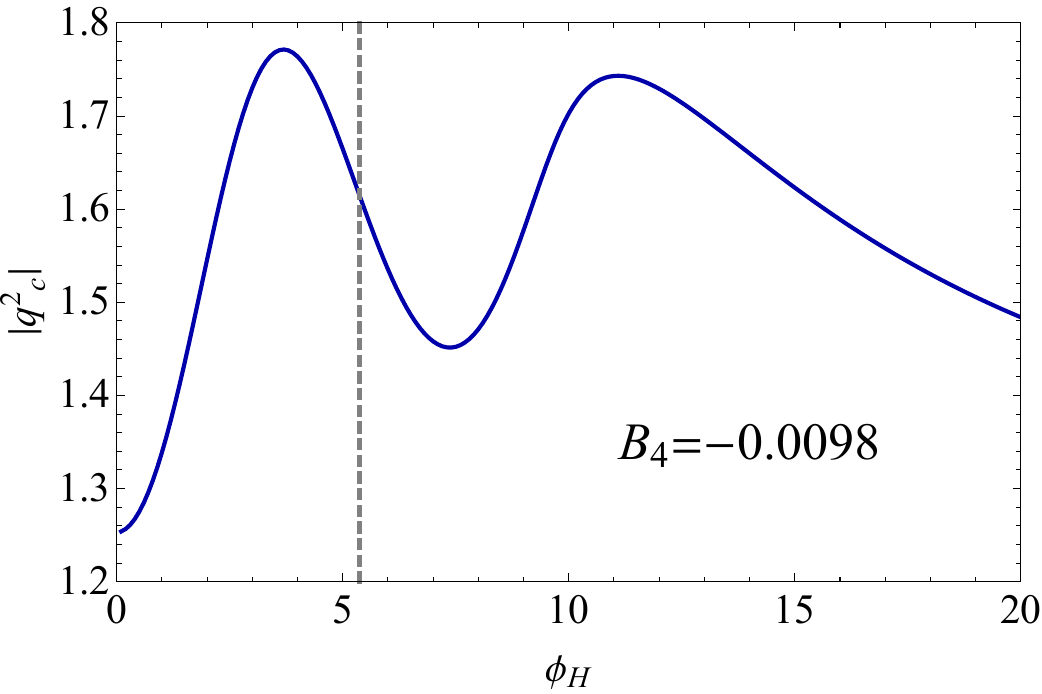}
    \caption{Plots of the radius of convergence for the lowest non-hydrodynamic modes in the spin-2 sector for second-order phase transition. The left (right) panel shows the dependency of $T/T_c$ ($\phi_H$).}
    \label{fig: convergence-radius-phiH-2nd}
\end{figure}

Fig. \ref{fig: convergence-radius-phiH-2nd} shows the radius of convergence in terms of $T/T_c$ and $\phi_H$. Unlike the crossover transition, we observe a decrease near the transition point, indicating less validity of hydro expansion. 

\subsection{Spin-1 sector}
\begin{figure}[htb]
    \centering
    \includegraphics[width=0.4\textwidth, valign=t]{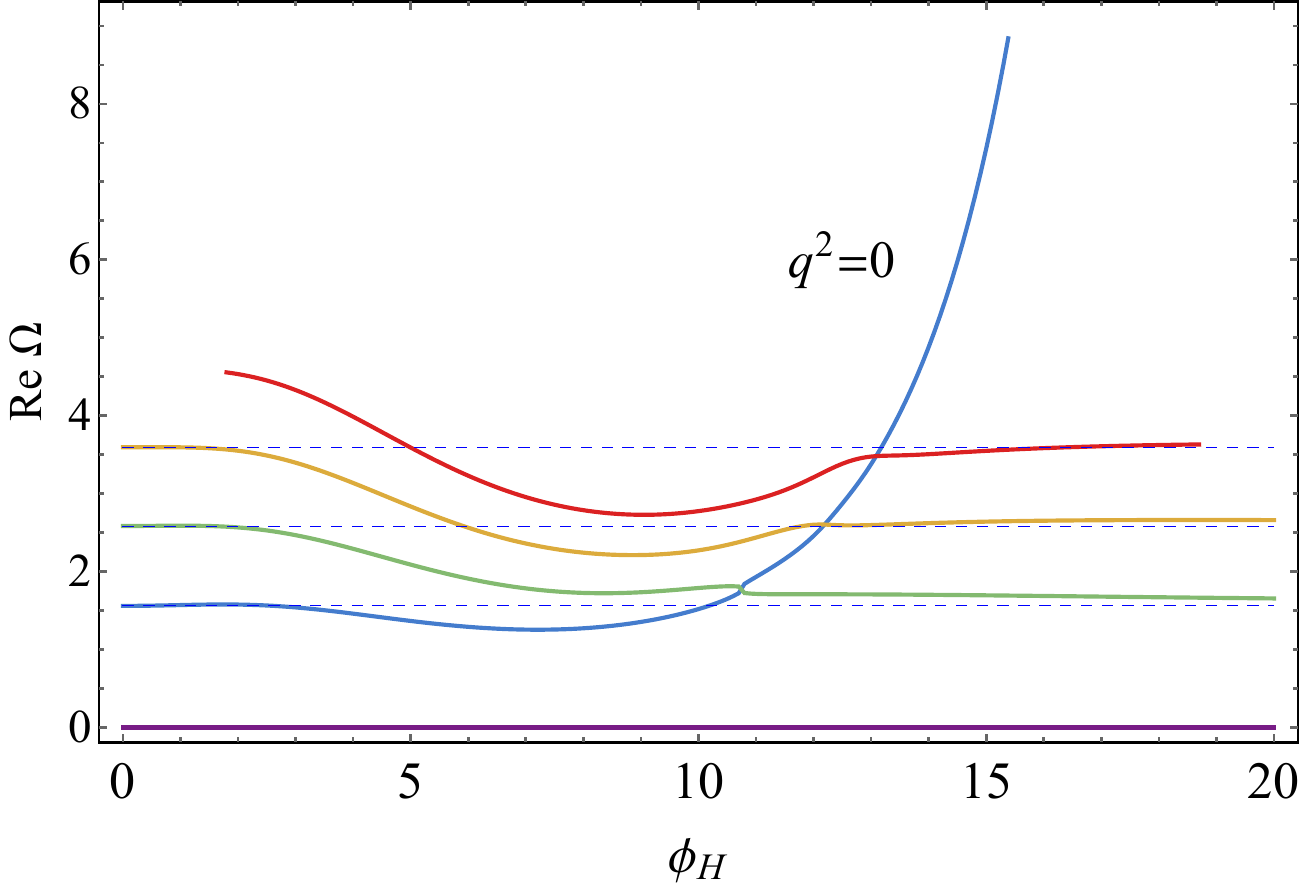}
    \hspace{0.5cm}
    \includegraphics[width=0.4\textwidth, valign=t]{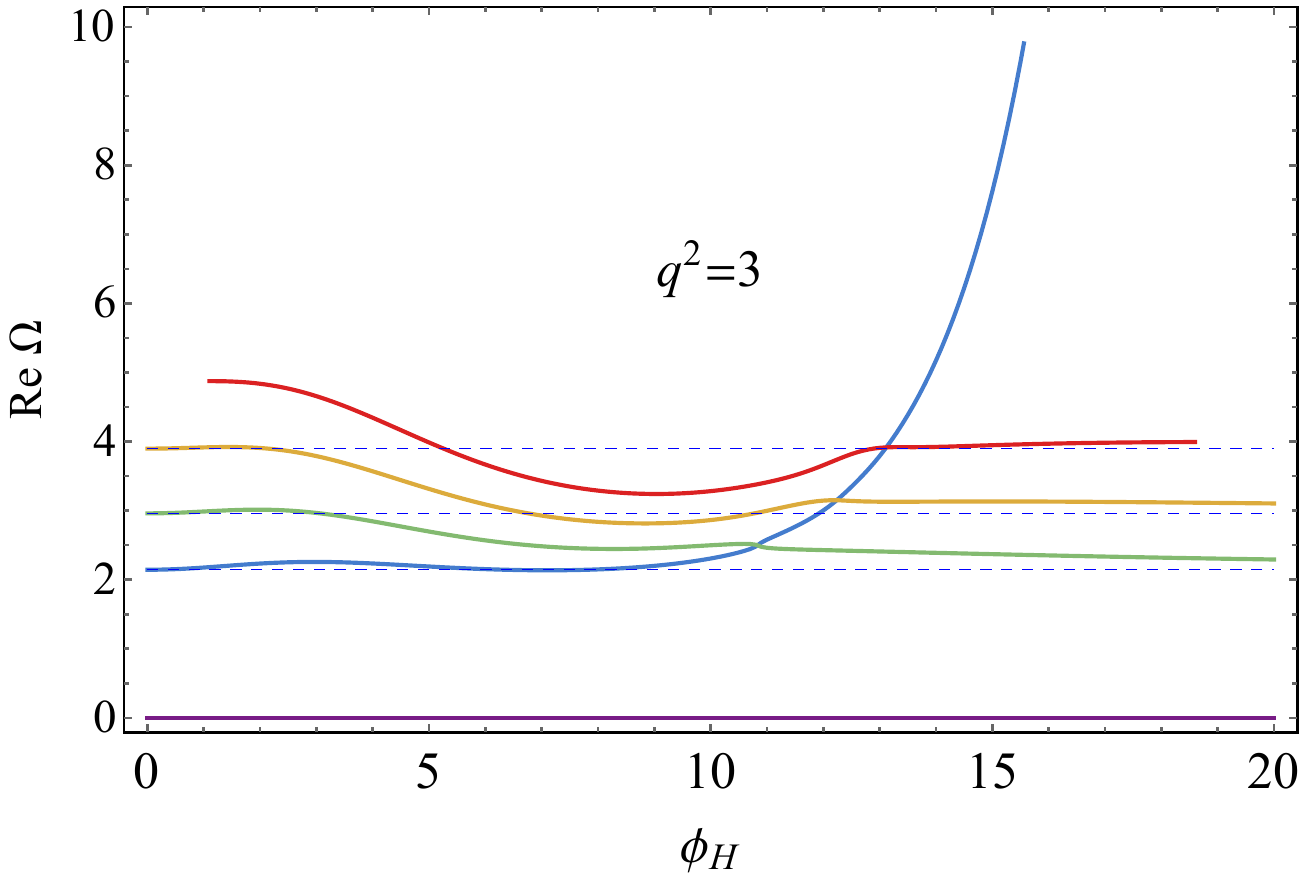}\\
    \includegraphics[width=0.4\textwidth,  valign=t]{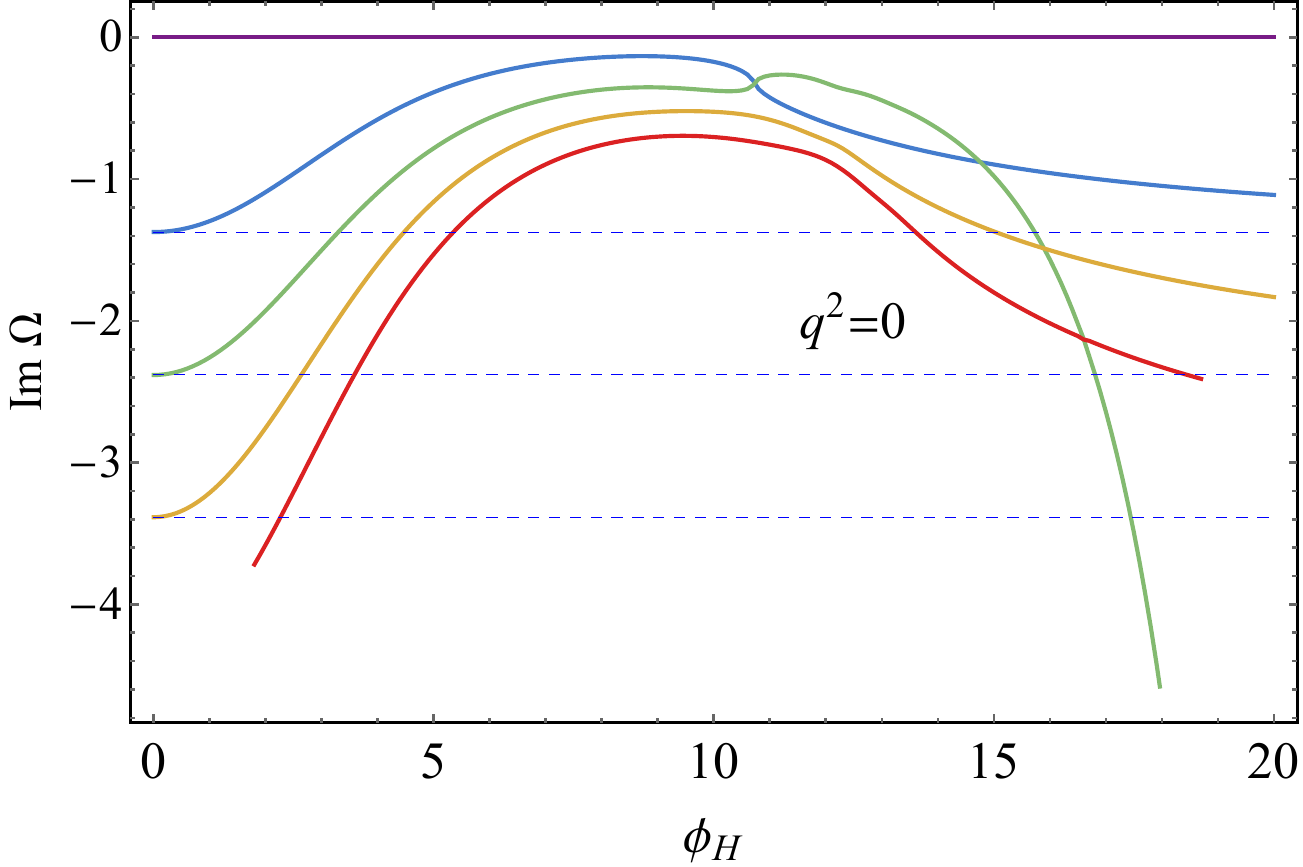}
    \hspace{0.5cm}
    \includegraphics[width=0.4\textwidth,  valign=t]{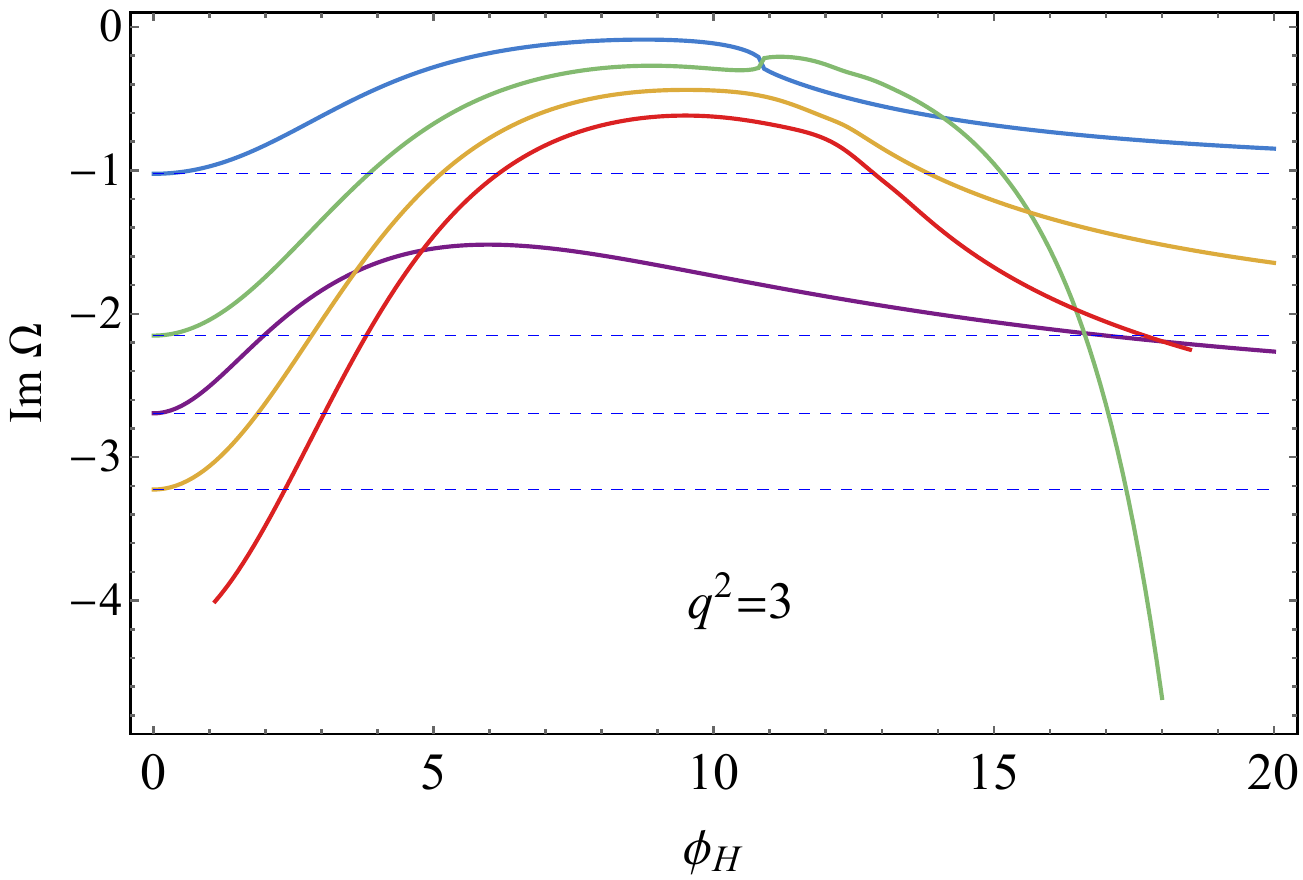}
    \caption{The real part (first row) and the imaginary part (second row) of the lowest three QNMs in the spin-1 sector as functions of $\phi_H$ at $q^2=(0, 3)$ for the second-order EoS with  $B_4=-0.0098$. The magenta line represents the hydro modes, while the other lines show non-hydro modes.}
    \label{fig: SO-Real-Momenta-EoS-Im-Re-Omega-spin1}
\end{figure}

In Fig. \ref{fig: SO-Real-Momenta-EoS-Im-Re-Omega-spin1}, we show the real and imaginary parts of the lowest three modes for real $q^2=(0, 3)$ in the second-order EoS. We observe the mixing of the modes between high temperature and low temperature.  However, there are some differences between Figs.  \ref{fig: SO-Real-Momenta-EoS-Im-Re-Omega-spin1} and \ref{fig: SO-Real-Momenta-EoS-Im-Re-Omega}. In Fig. \ref{fig: SO-Real-Momenta-EoS-Im-Re-Omega}, only the gravity non-hydro modes play the role, and the lowest non-hydro mode remains the same at low and temperatures. But in Fig. \ref{fig: SO-Real-Momenta-EoS-Im-Re-Omega-spin1}, the hydro modes come into play so that at $q^2 = 0$, the lowest mode doesn't change, and the second (third) mode goes to infinity in real (imaginary) parts. At $q^2 \neq 0$, the latter keeps going in addition to the hydro mode (the third mode in imaginary parts at high temperatures), which collides with others in between and becomes the second mode at low temperatures.  

\begin{figure}[htb]
    \centering
    \includegraphics[width=0.44\textwidth, valign=t]{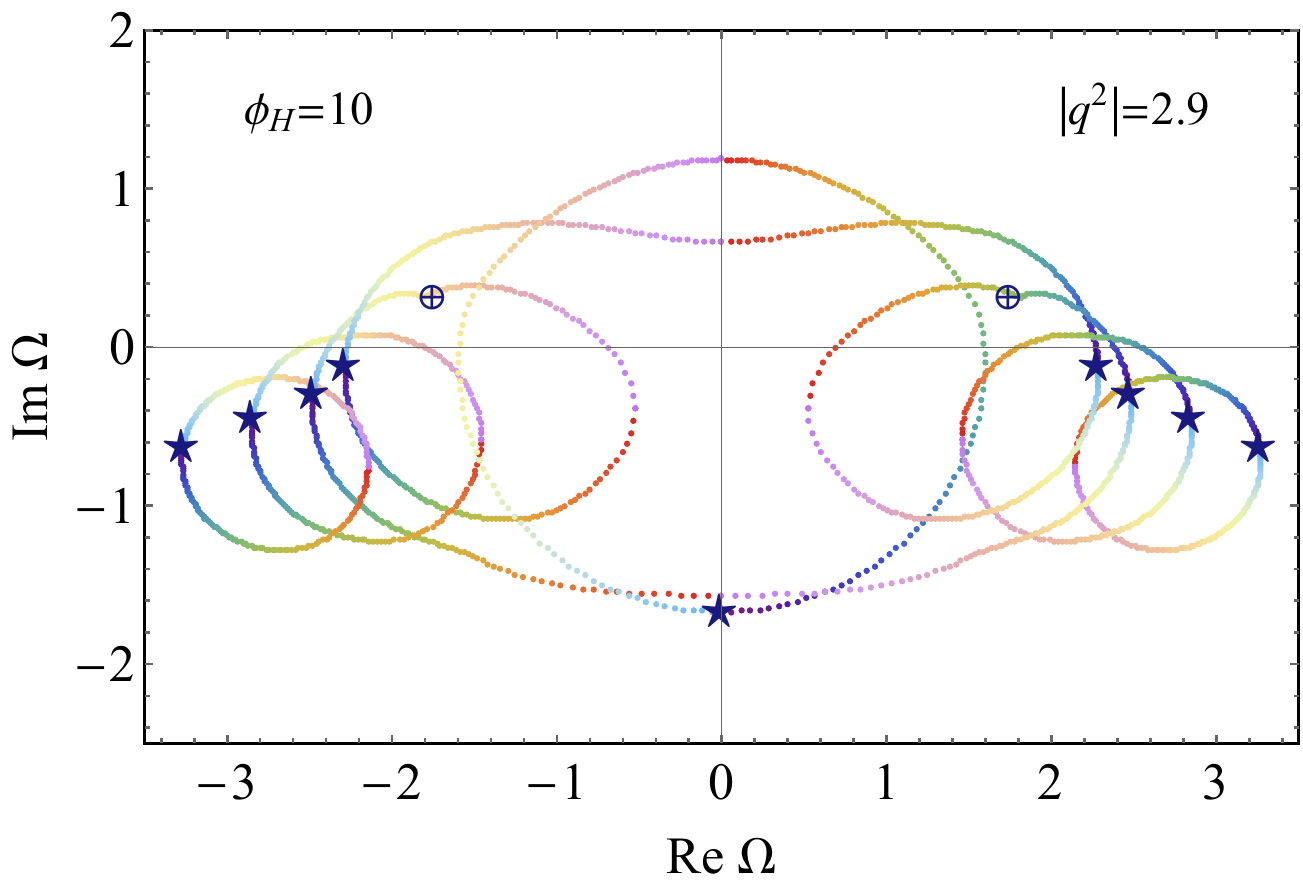}
    \hspace{0.5cm}
    \includegraphics[width=0.44\textwidth, valign=t]{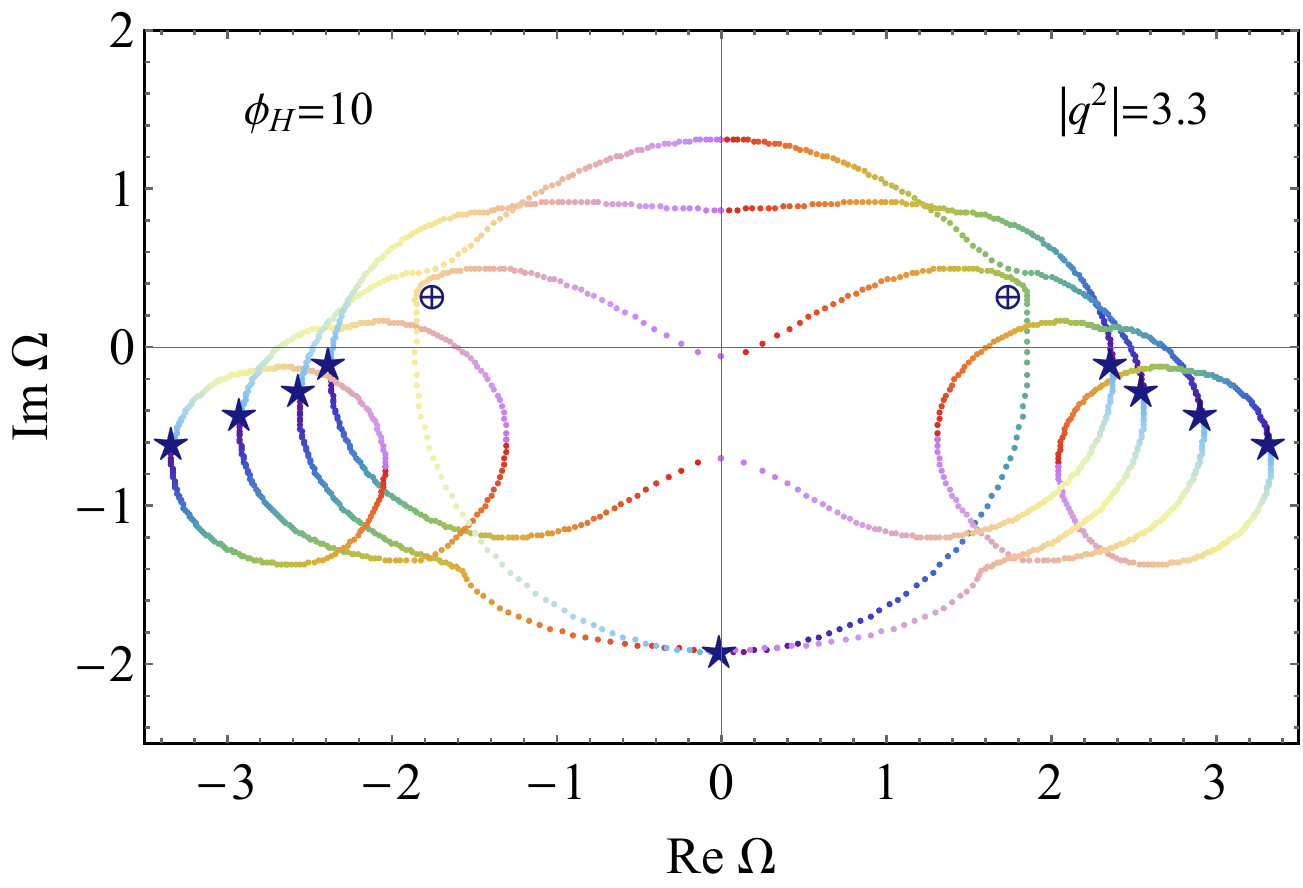}\\
    \includegraphics[width=0.44\textwidth, valign=t]{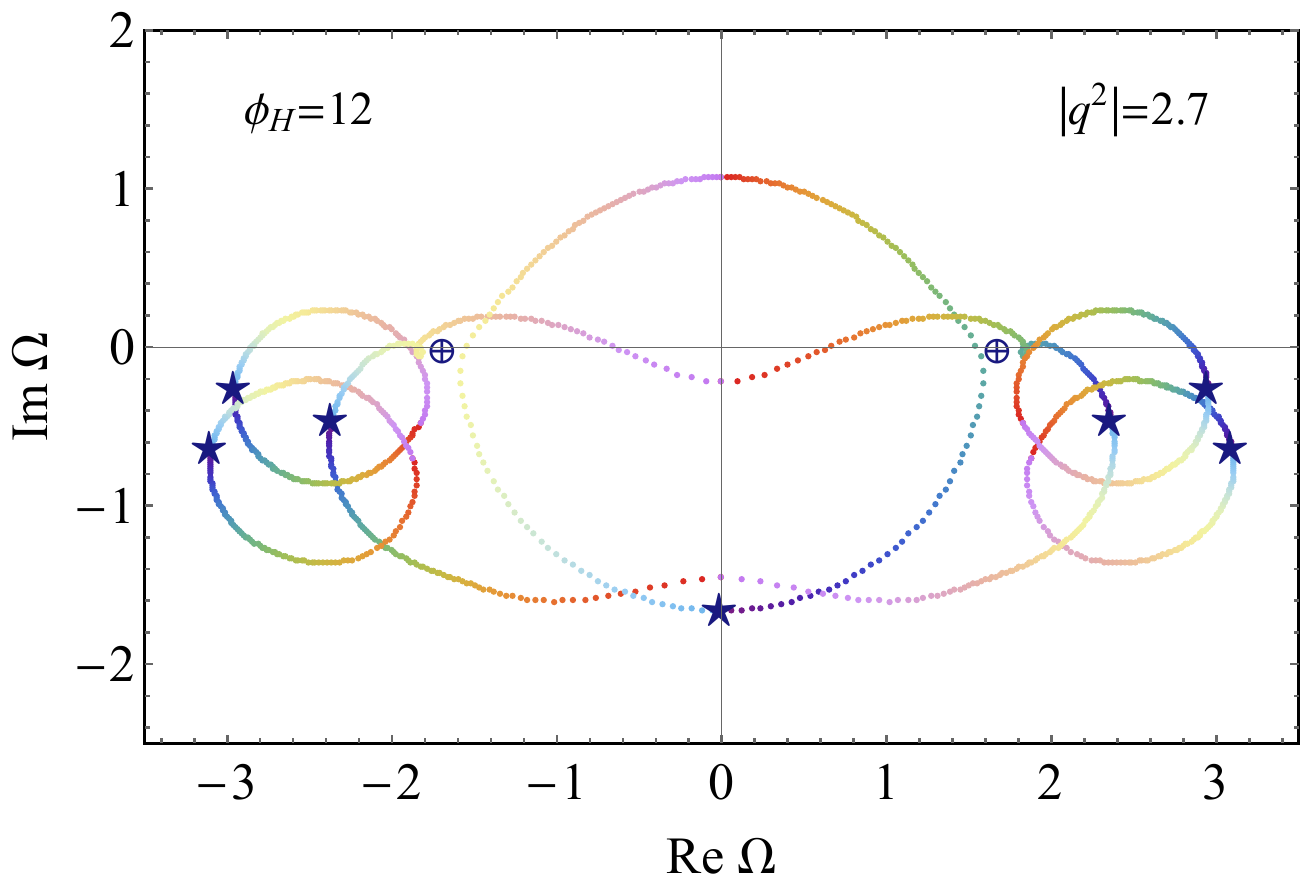}
    \hspace{0.5cm}
    \includegraphics[width=0.44\textwidth, valign=t]{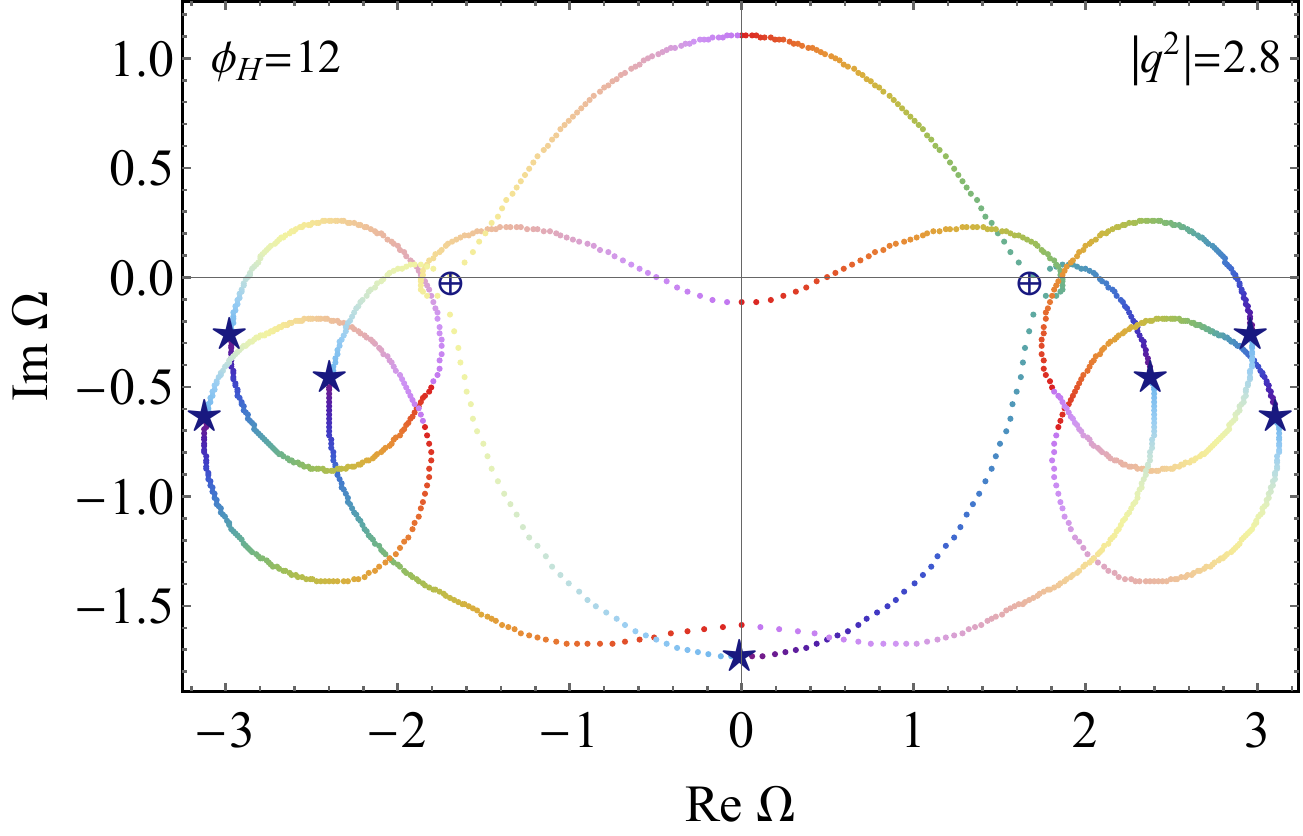}
    \caption{Plots of the mode collision for second-order EoS in spin-1 sector. Top (bottom) row plots indicate $\phi_H=10$ ($\phi_H=12$) data, and the left (right) panels demonstrate the modes before (after) the collision. At $\phi_H=10$, collision occurs at $\vert q^2 \vert=3.105$ with $\Omega_\star=\pm 1.747+0.328 i$, and for $\phi_H=12$ collision occurs at $\vert q^2\vert=2.784$ with $\Omega_\star=\pm 1.682-0.011 i$. The dashed circles in each plot show the location of $\Omega_\star$.}
    \label{fig: collision-mode-spin1-SO}
\end{figure}

In Fig.~\ref{fig: collision-mode-spin1-SO}, we present the mode trajectories before and after the collision for $\phi_H = 10$ and $\phi_H = 12$ for complex momenta choices. Before the collision, the path of the hydro and non-hydro modes are separate, but after that, they share their path. Accordingly, following the path after the collision brings us to the next Riemann surface \cite{Withers:2018srf}.

\begin{figure}
    \centering
    \includegraphics[width=0.44\textwidth,valign=t]{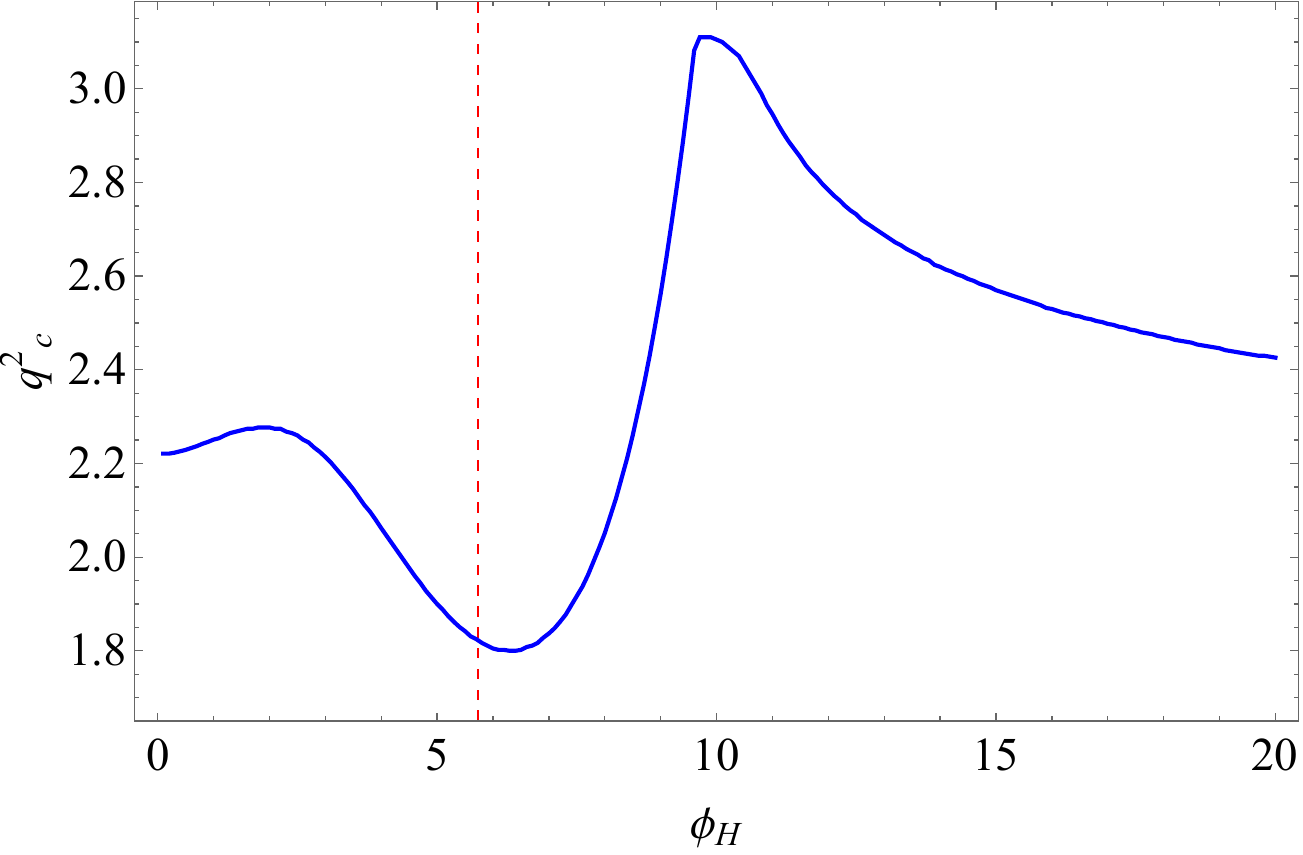}
        \hspace{0.1cm}
    \includegraphics[width=0.44\textwidth,valign=t]{rcvsT-spin1-SO.pdf} 
    \caption{Plots of the radius of convergence in terms of $\phi_H$ (left panel), and $T/T_c$(right panel) for second-order phase transition in the spin-1 sector. The Red dashed line represents the location of the critical point.}
    \label{fig: rc-shear-so}
\end{figure}

\begin{figure}[htb]
    \centering
    \includegraphics[width=0.48\textwidth,valign=t]{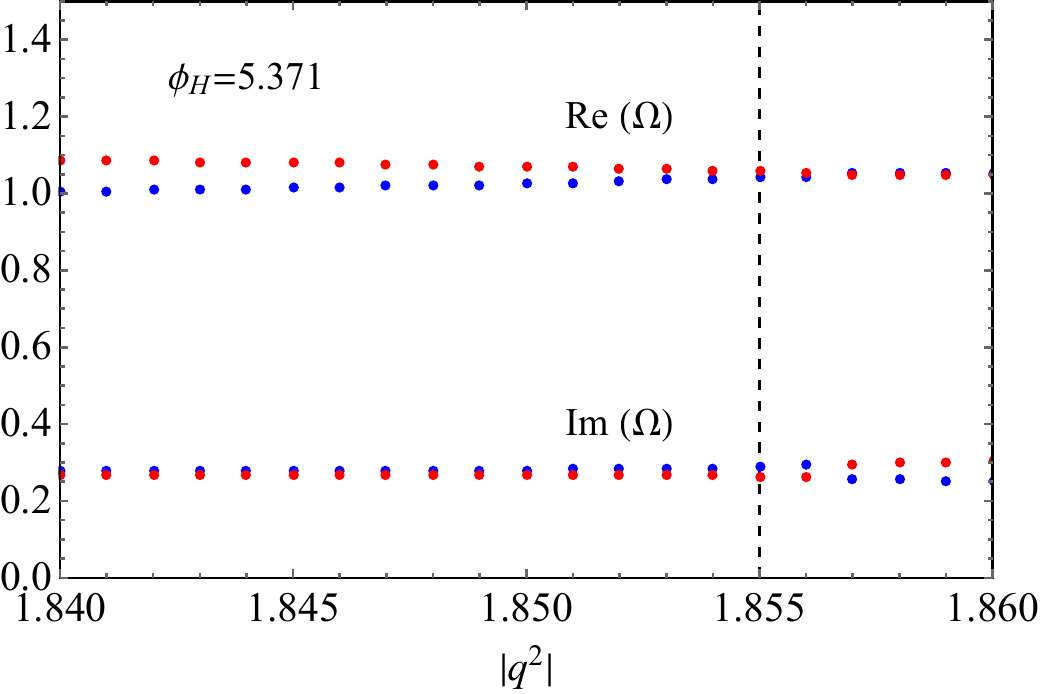}
    \caption{Plots of the collision of lowest non-hydro mode with the hydro mode at $\phi_H=5.371$ for the second-order phase transition in the spin-1 sector. It occurs at $\theta= 0.575 \pi$ and $\vert q^2\vert = 1.855$.}
    \label{fig: collision-SO}
\end{figure}

In Fig.~\ref{fig: rc-shear-so}, we show the radius of convergence in the spin-1 sector for the second-order EoS. Similar to the crossover transition, spin-1 has the greatest $q^2_c$ that occurs around $\phi_H = 11$. This is the point where the pattern of mode collision changes from hydro with the second non-hydro to hydro with the first non-hydro. Likewise, near the transition point, $q^2_c$ decreases which can be a sign of {\it{“breakdown of the hydrodynamic series near the transition point.”}}
Moreover, in Fig.~\ref{fig: collision-SO}, we demonstrate the approach of the hydro and the closest non-hydro mode towards each other at $\phi_H = 5.371$. The collision occurs at $\vert q^2\vert=1.855$ and $\theta = 0.575 \pi$.

\subsection{Spin-0 sector}
\begin{figure}[htb]
    \centering
    \includegraphics[width=0.42\textwidth, valign=t]{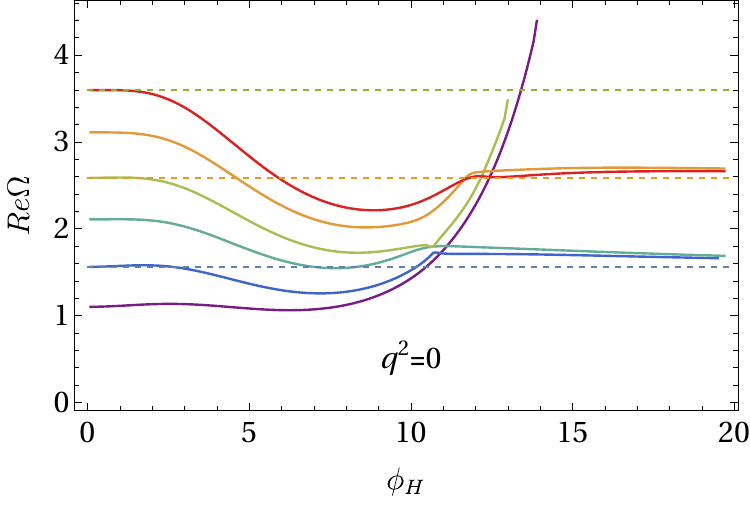}
    \hspace{0.3cm}
      \includegraphics[width=0.42\textwidth, valign=t]{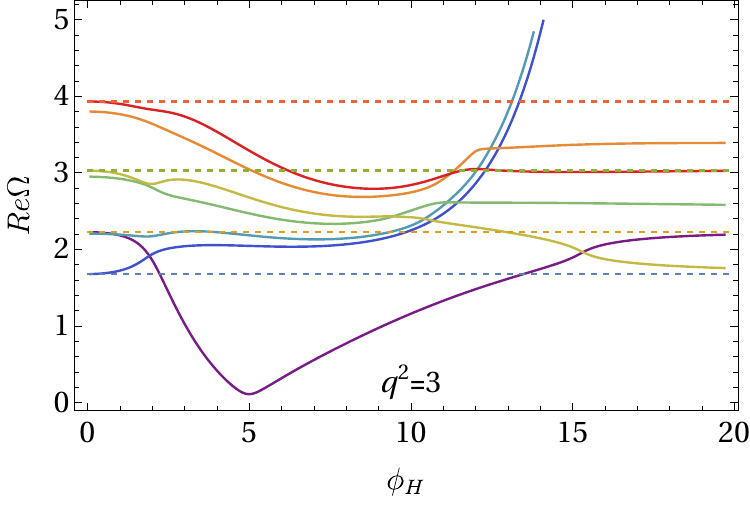}
      \vspace{0.4cm}
      \includegraphics[width=0.42\textwidth, valign=t]{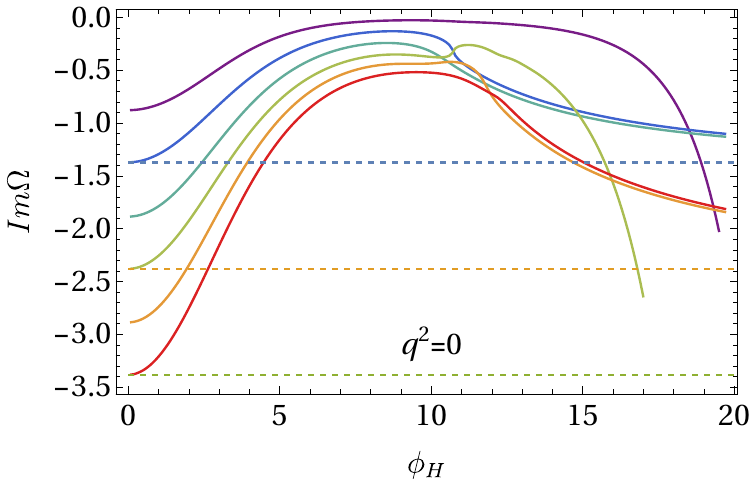}
      \hspace{0.3cm}
     \includegraphics[width=0.43\textwidth, valign=t]{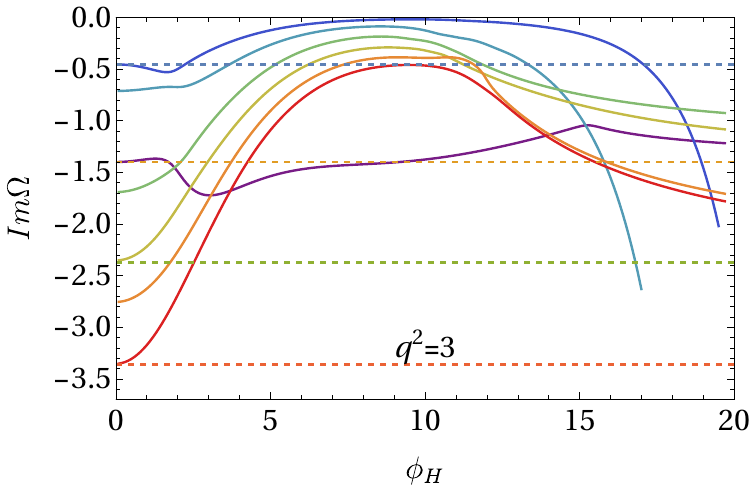}
       \caption{The real part (first row) and the imaginary part (second row) of the lowest three QNMs in the spin-0 sector as functions of $\phi_H$ at $q^2=(0, 3)$ for the second order EoS.}
    \label{fig: reim-SO-spin0-q03}
\end{figure}

\begin{figure}[htb]
    \centering
        \includegraphics[width=0.45\textwidth, valign=t]{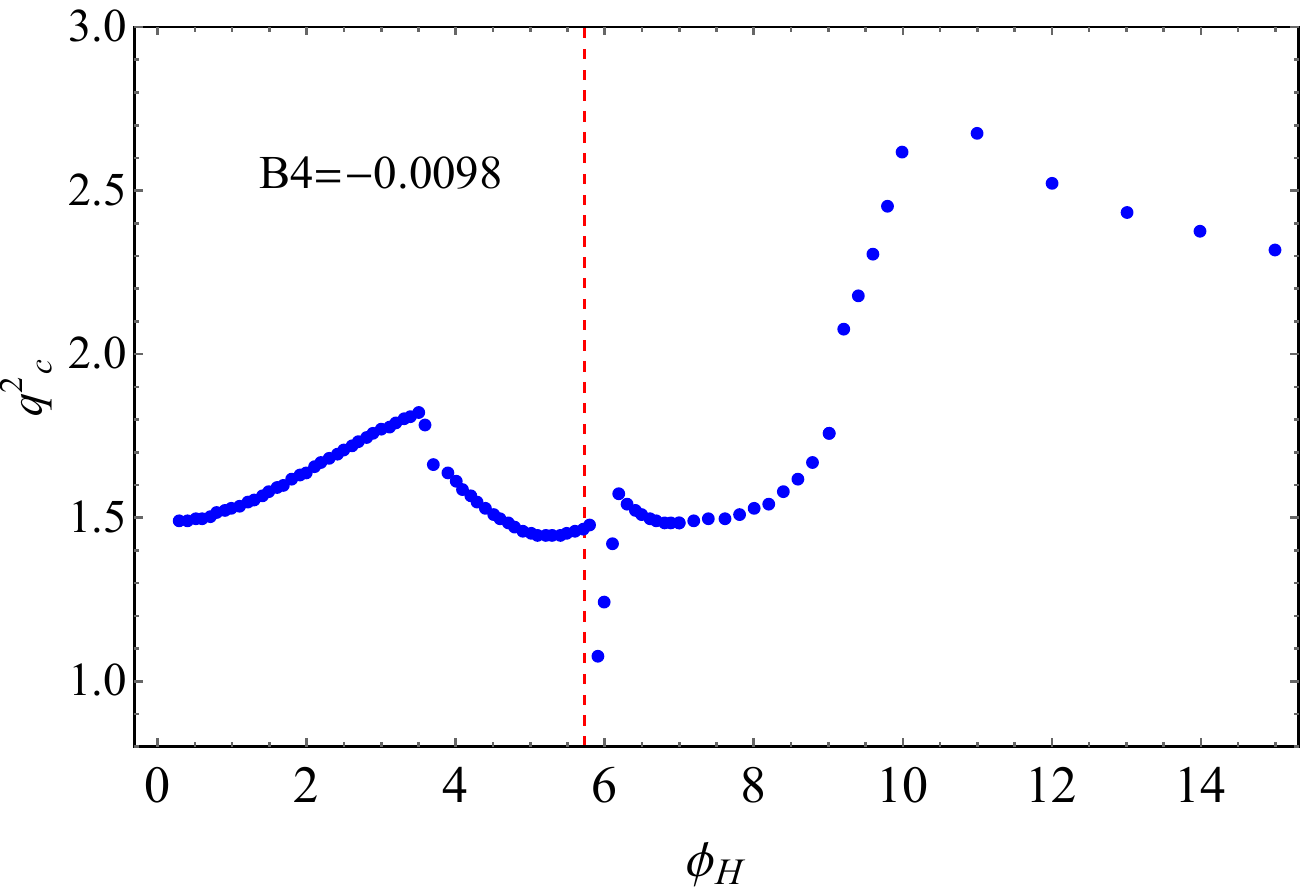}
        \hspace{0.4cm}
        \includegraphics[width=0.45\textwidth, valign=t]{plot-q2c-T.pdf}
    \caption{Plots of the radius of convergence in terms of $\phi_H$ (left panel), and $T/T_c$ (right panel)  for second-order EoS with $B_4=-0.0098$ in the spin-0 sector. The Red dashed line represents the location of the transition point.}
    \label{fig: rc-spin0-So}
\end{figure}

In each section, the spin-0 parts exhibit intricate structures. For example, Fig. \ref{fig: reim-SO-spin0-q03} displays the real and imaginary parts of the lowest QNMs for real $q^2 = (0,3)$. The real and imaginary parts of the QNMs have a similar general structure to that of Fig. \ref{fig: SO-Real-Momenta-EoS-Im-Re-Omega-spin1}, except for the mode doubling structure that arises from the coupling of scalar and gravity perturbations.
The radius of convergence for second-order EoS in the spin-0 sector is illustrated in Fig. \ref{fig: rc-spin0-So}. The dip near the transition point indicates a reduced validity for the hydrodynamics series. We observe that the $q^2_c$ approaches the same value at high and low temperatures. We also observe that Eq. \eqref{eq:compare-rc} holds for the second-order results.

\section{First-order phase transition}\label{sec: firstorder-results}
The first-order results were obtained with $B_4 = - 0.02$. Due to the complex nature of first-order transitions compared to other types of transitions, we expect the curves to be more intricate and involved.
\subsection{Spin-2 sector}

\begin{figure}[htb]
    \centering
    \includegraphics[width=0.46\textwidth, valign=t]{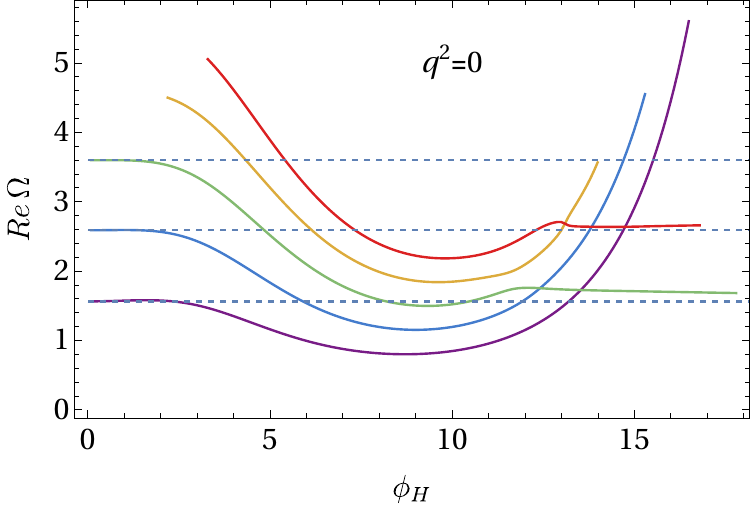}
    \hspace{0.3cm}
    \includegraphics[width=0.46\textwidth, valign=t]{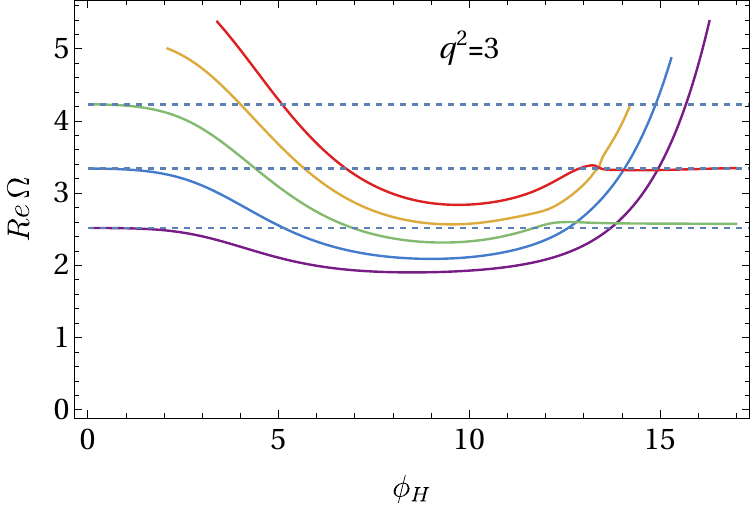}\\
    \includegraphics[width=0.48\textwidth, valign=t]{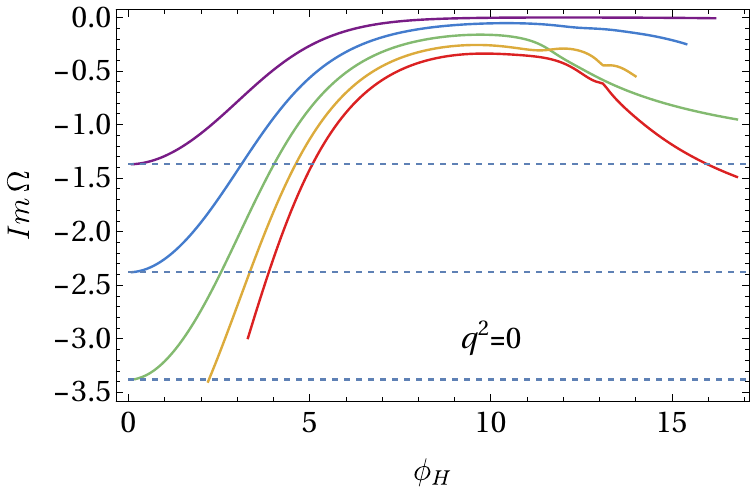}
    \hspace{0.3cm}
    \includegraphics[width=0.48\textwidth, valign=t]{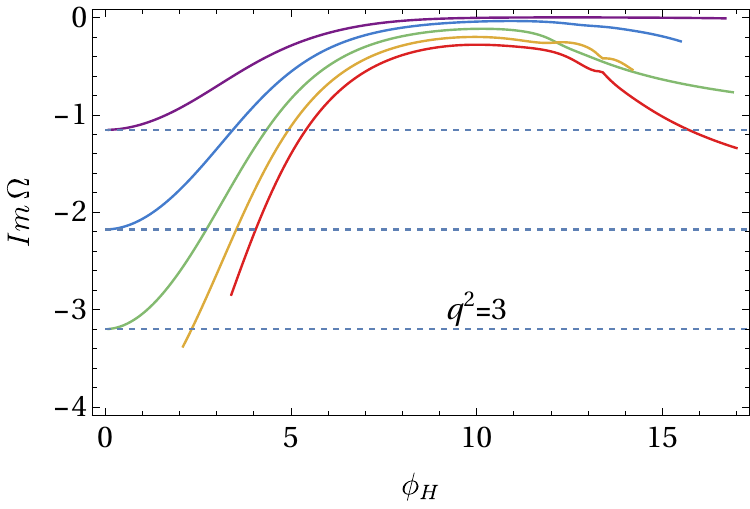}
    \caption{The real part (first row) and the imaginary part (second row) of the lowest three QNMs in Spin-2 sector as functions of $\phi_H$ at real $q^2=(0, 3)$ for the first-order phase transition with $B_4=-0.02$.}
    \label{fig: FO-Real-Momenta-EoS-Im-Re-Omega}
\end{figure}

In Fig.~\ref{fig: FO-Real-Momenta-EoS-Im-Re-Omega}, we show the real and imaginary parts of the lowest QNMs for the first-order phase transition and real $q^2$. We observe some similarities compared to Fig. \ref{fig: SO-Real-Momenta-EoS-Im-Re-Omega}. For $\vert q^2\vert<3$, the first three non-hydro modes diverge to infinity at low temperatures. Moreover,  we note that $\Omega_{2k+1}(T \gg T_c) \to \Omega_k(T \ll T_c)$ for $k \geq 1$. We also observe no mode collisions (i.e., the same real and imaginary parts at a given $\phi_H$). These observations suggest that the collisions must occur for negative $q^2$, which is indeed the case.

\begin{figure}[htb]
    \centering
    \includegraphics[width=0.5\textwidth]{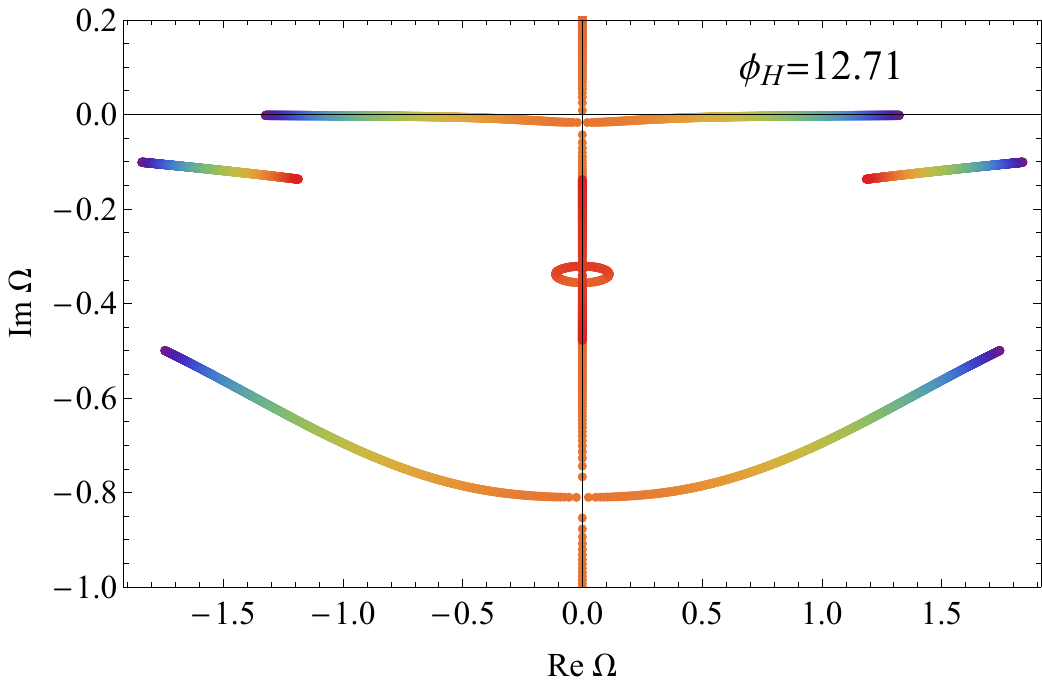}
    \caption{The plot of collision of the lowest non-hydro modes at $\phi_H=12.71$ of the first-order EoS in the spin-2 sector. The collision occurs at $q^2=-1.740$ at which two sets of modes collide at different frequencies $\Omega=(-0.0167i, -0.8100 i)$. For larger temperatures (smaller $\phi_H$), the modes with the lowest imaginary frequency collide at the smallest $\vert q^2\vert$ (similar to Fig.~\ref{fig: B40Imq}). For lower temperatures (larger $\phi_H$), the other set of modes collide at smaller $\vert q^2\vert$ (see right panel in  Fig.~\ref{fig: convergence-radius-phiH-1st}). }
    \label{fig: B4-0.2Imq}
\end{figure}

Fig.~\ref{fig: B4-0.2Imq} displays the mode patterns at $\phi_H = 12.71$ for purely negative $\vert q^2\vert$ to illustrate the collisions between non-hydrodynamic modes in the spin-2 sector. As you can see, the collision occurs at two $\Omega = (- 0.0167i,- 0.8100i)$. It is a universal feature that collisions always occur for negative $q^2$ in the spin-2 sector.

\begin{figure}[htb]
    \centering
    \includegraphics[width=0.41\textwidth, valign=t]{rcScalar1stPT.pdf}
    \hspace{0.4cm}
    \includegraphics[width=0.41\textwidth, valign=t]{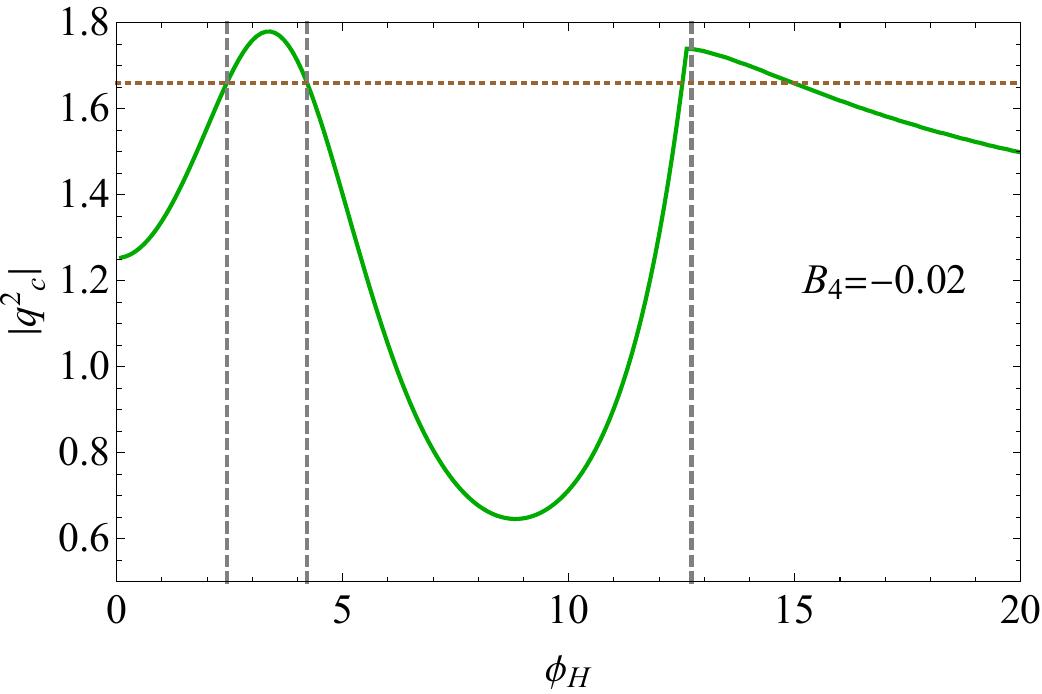}
    \caption{The radius of convergence for the lowest non-hydrodynamic modes in the spin-2 sector for first-order phase transitions. The left (right) panel shows the dependency of $T/T_c$ ($\phi_H$). The inside plot focuses on near-transition points.}
    \label{fig: convergence-radius-phiH-1st}
\end{figure}

To conclude this subsection, we present the radius of convergence for non-hydro modes collisions in the spin-2 sector in Fig.  \ref{fig: convergence-radius-phiH-1st}. Due to the coupling between gravity and scalar fields, the collision occurs between the gravity and scalar non-hydro modes. In the first-order transition, there are stable and unstable branches between the transition lines. We take the stable branches to discuss the radius of convergence. As shown in Fig.  \ref{fig: convergence-radius-phiH-1st}, the radius of convergence increases near the transition point in the stable branches. This indicates that the validity of using the hydro expansion near the transition point is becoming larger.

\subsection{Spin-1 sector}
\begin{figure}[htb]
    \centering
    \includegraphics[width=0.42\textwidth, valign=t]{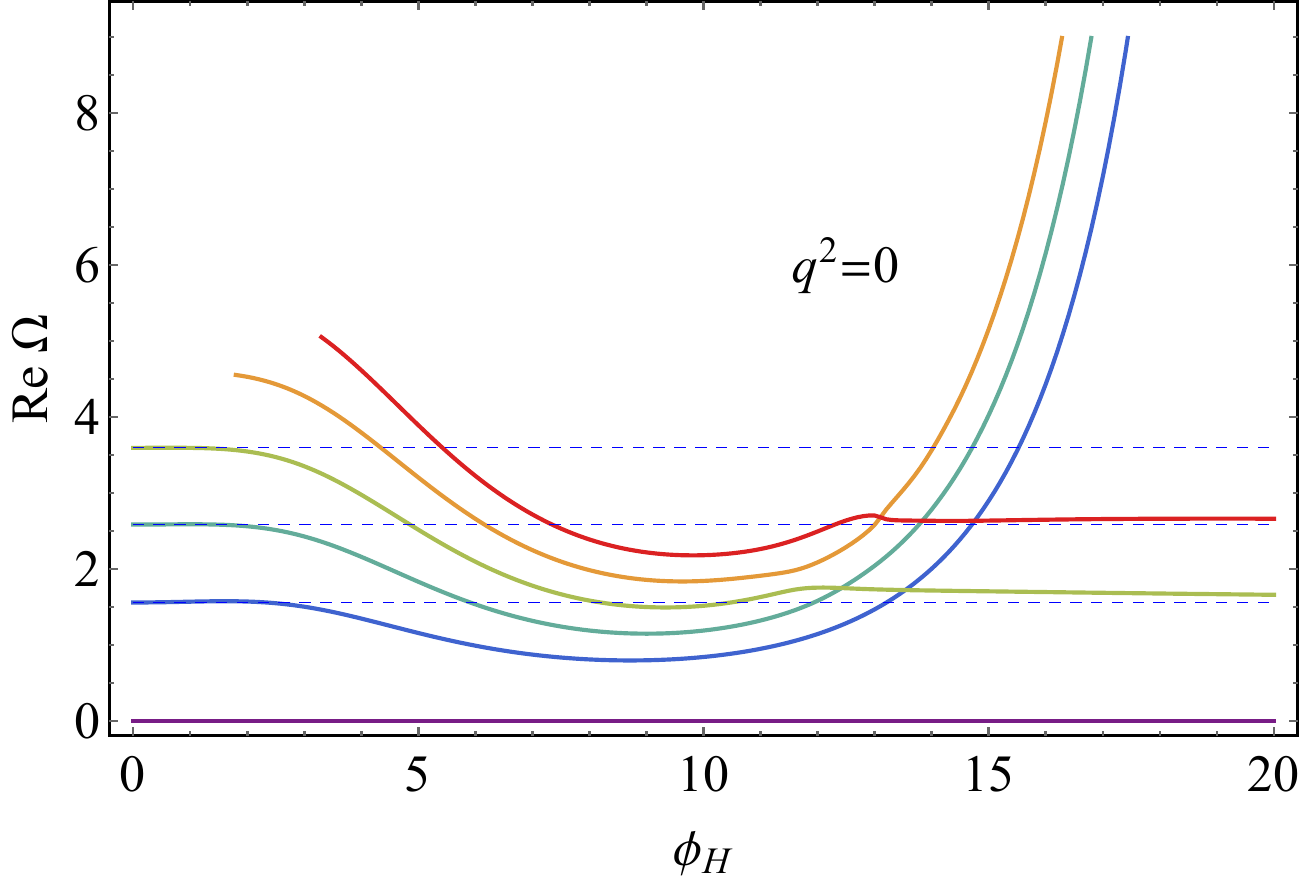}
    \hspace{0.4cm}
    \includegraphics[width=0.42\textwidth, valign=t]{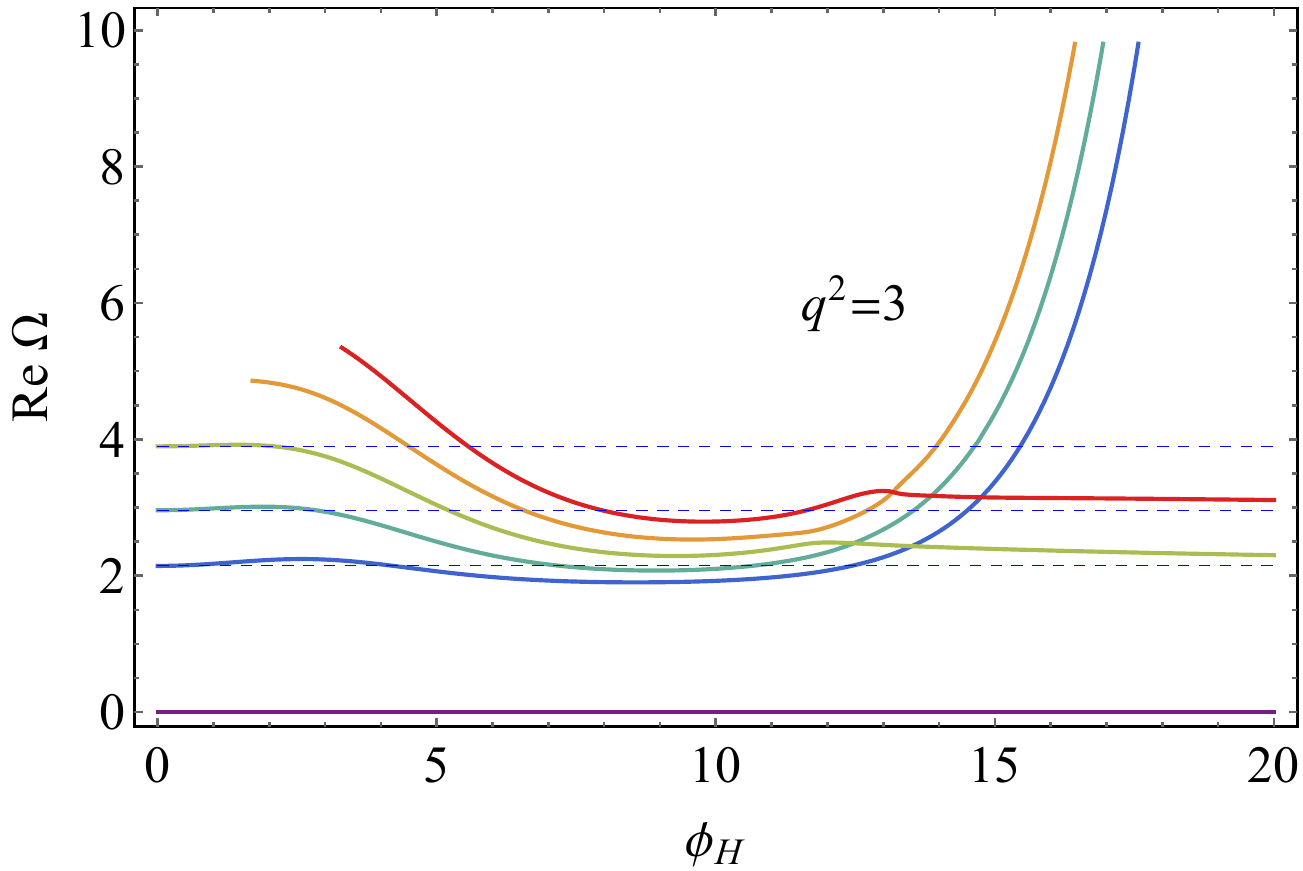}\\
    \includegraphics[width=0.42\textwidth, valign=t]{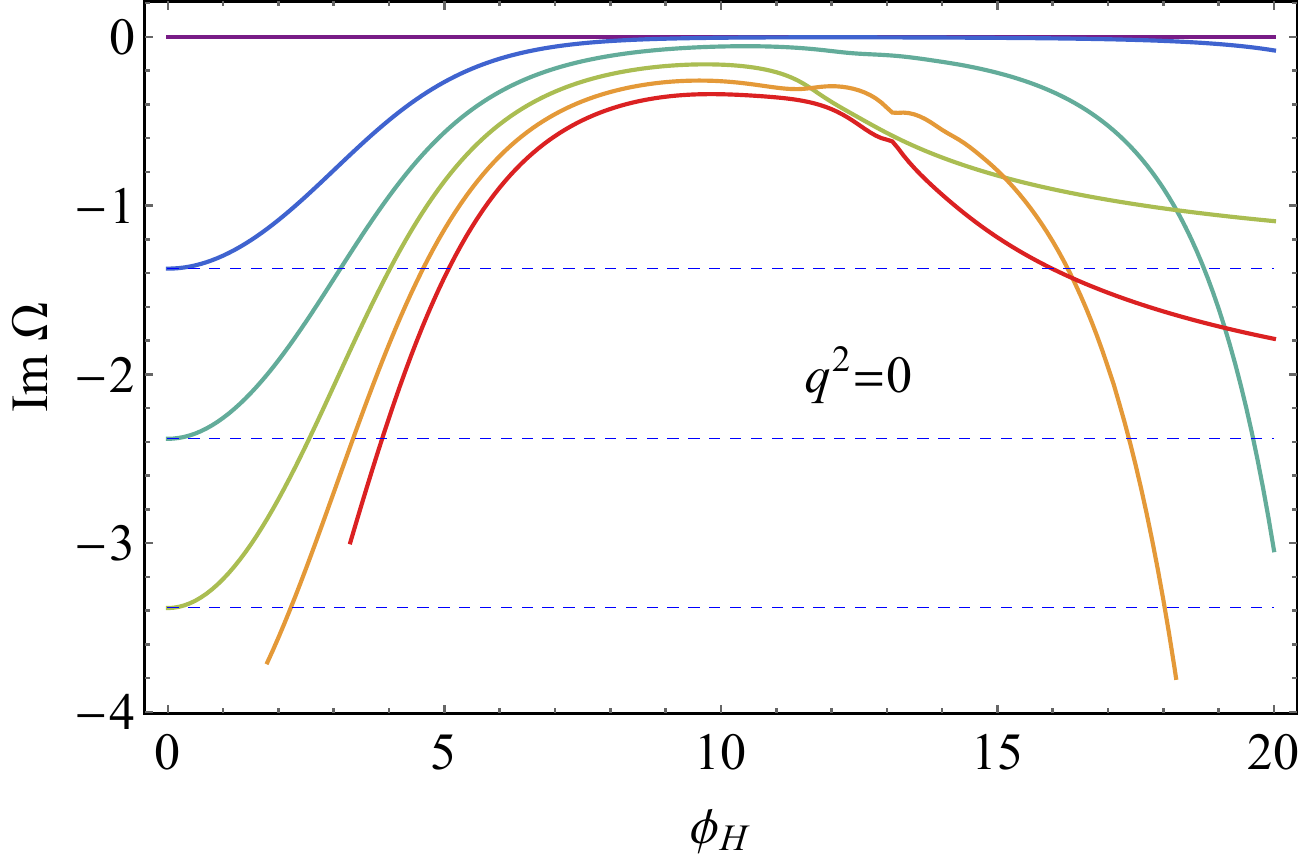}
    \hspace{0.4cm}
    \includegraphics[width=0.42\textwidth, valign=t]{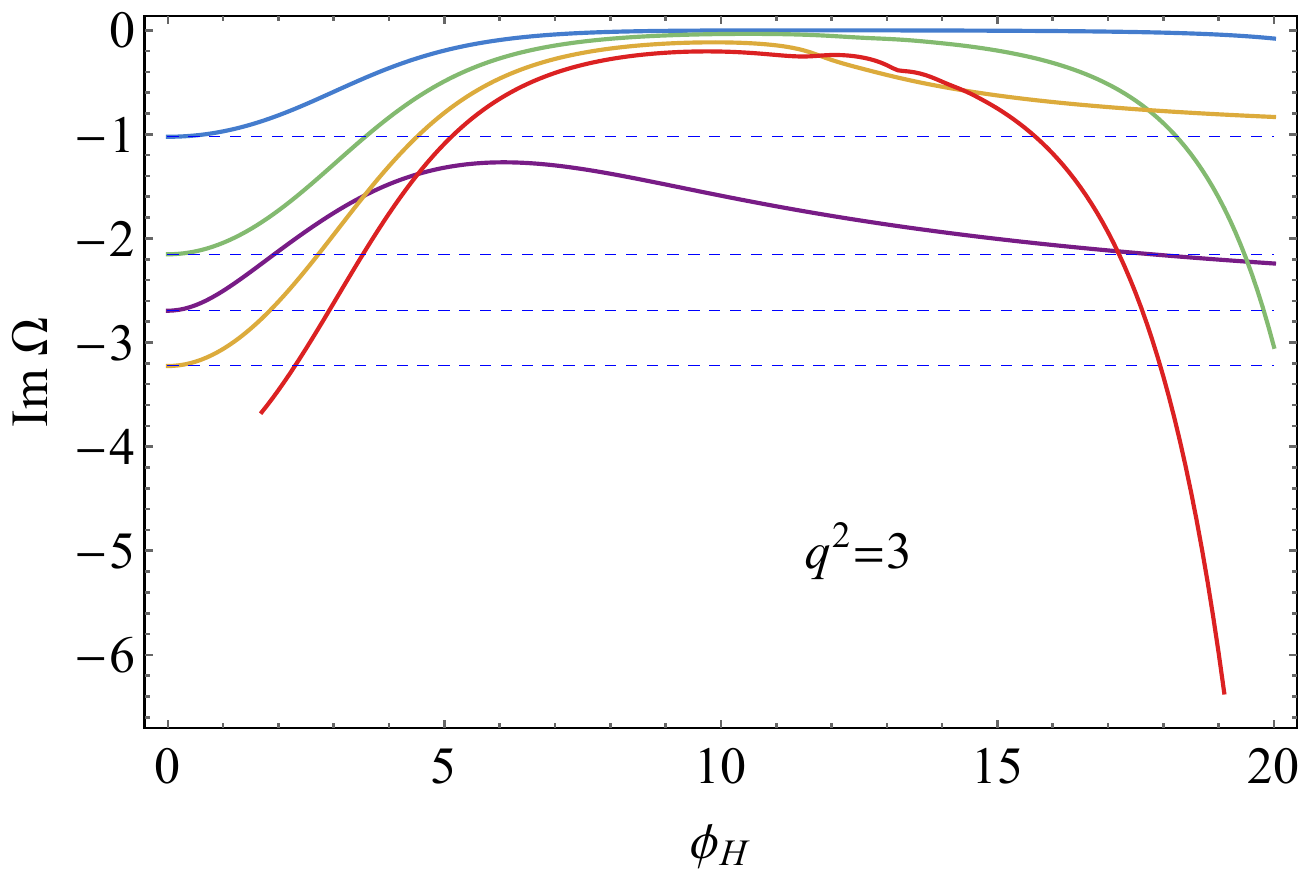}
    \caption{The real (first row) and the imaginary part (second row) of the lowest three QNMs in the spin-1 sector of first-order transition as functions of $\phi_H$ for real  $q^2=(0, 3)$. The magenta lines represent the lowest hydro modes, while the other lines show non-hydro modes.}
    \label{fig: FO-Real-Momenta-EoS-Im-Re-Omega-spin1}
\end{figure}

\begin{figure}[htb]
    \centering
    \includegraphics[width=0.44\textwidth, valign=t]{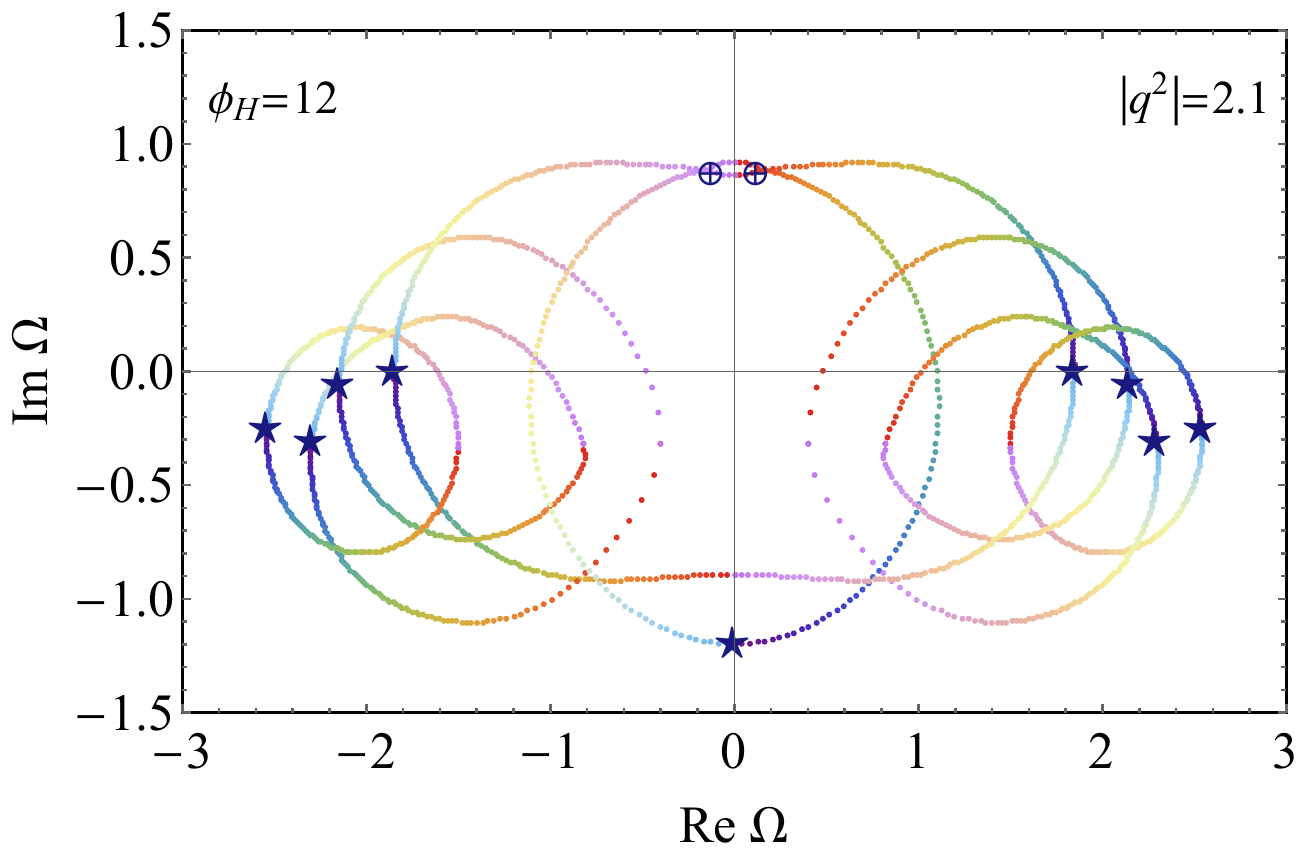}
    \hspace{0.3cm}
    \includegraphics[width=0.44\textwidth, valign=t]{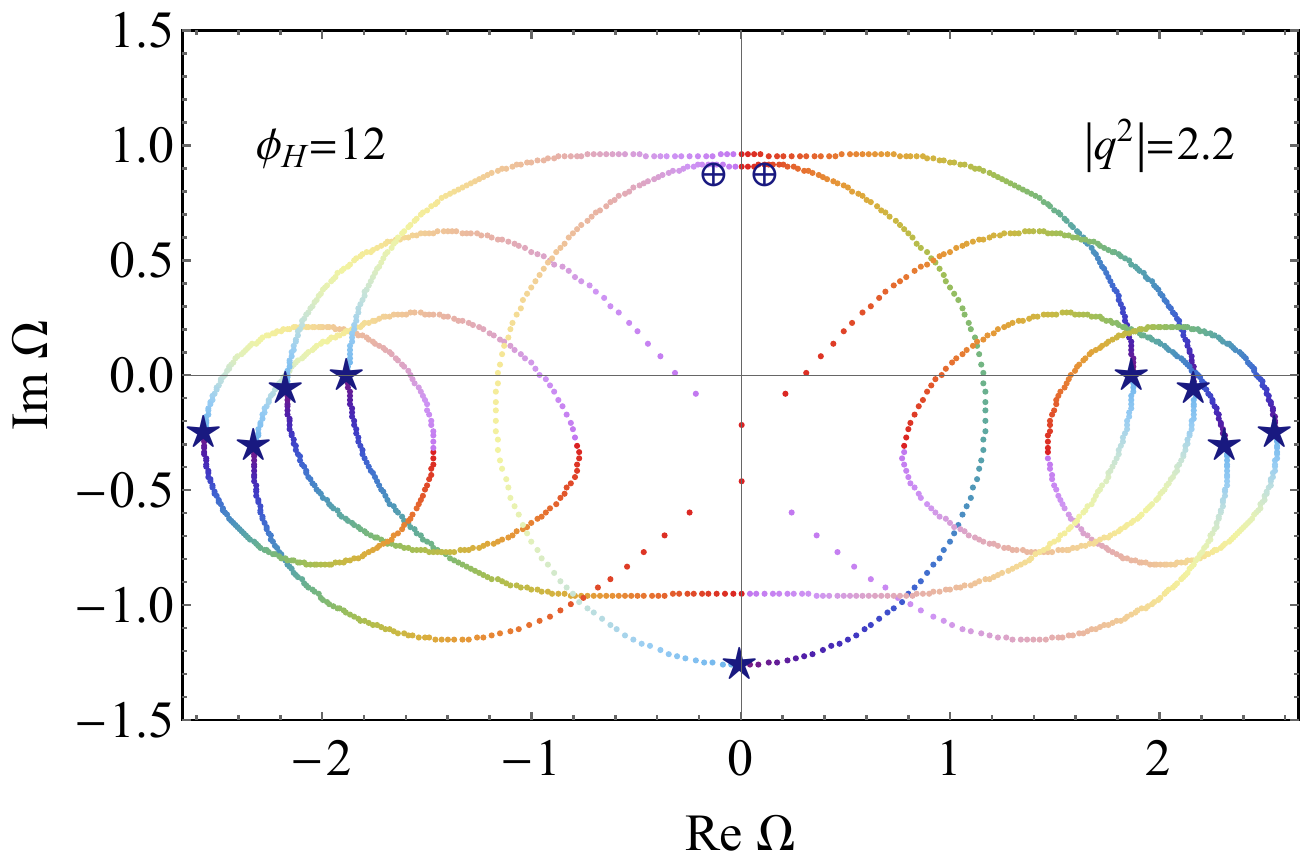}\\
    \includegraphics[width=0.44\textwidth, valign=t]{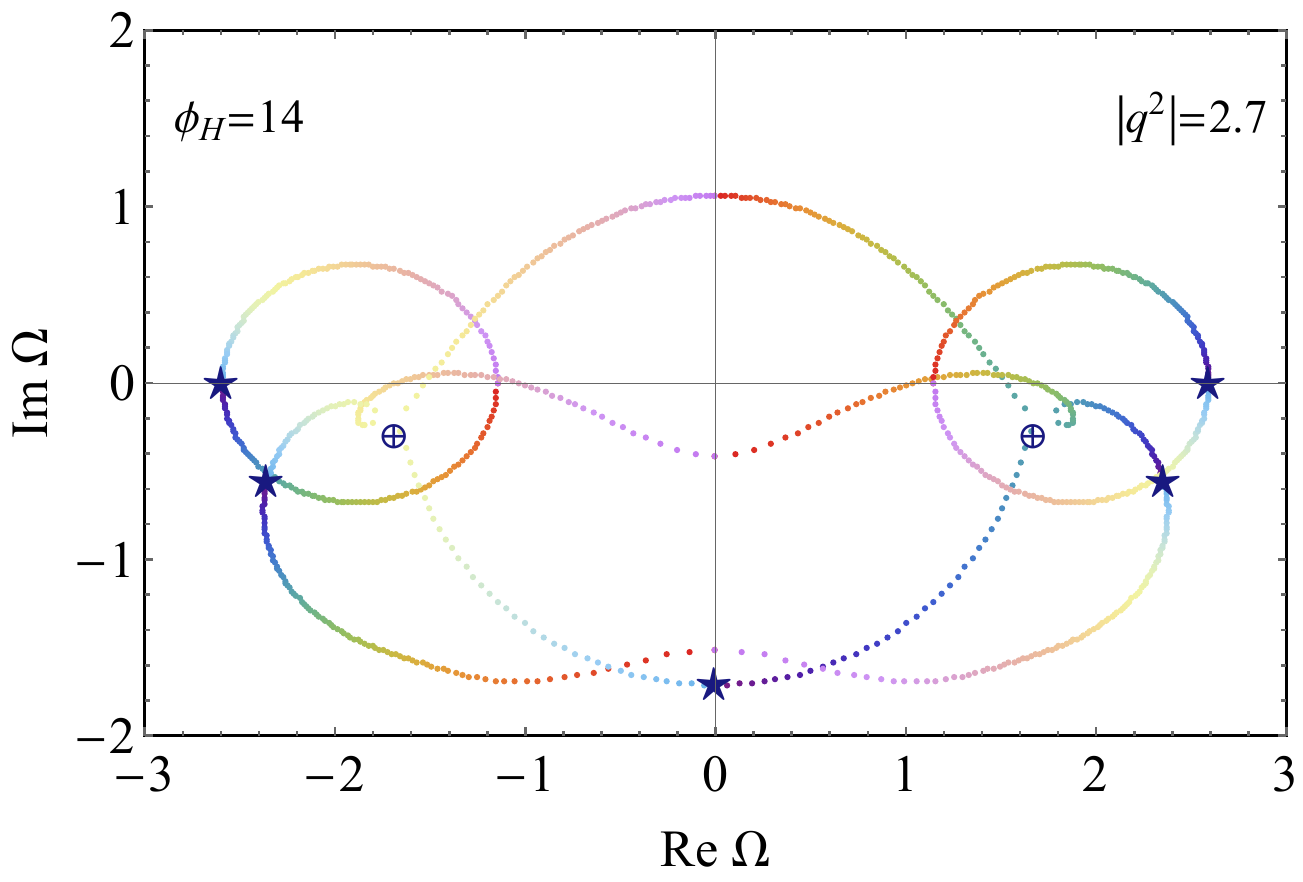}
    \hspace{0.3cm}
    \includegraphics[width=0.44\textwidth, valign=t]{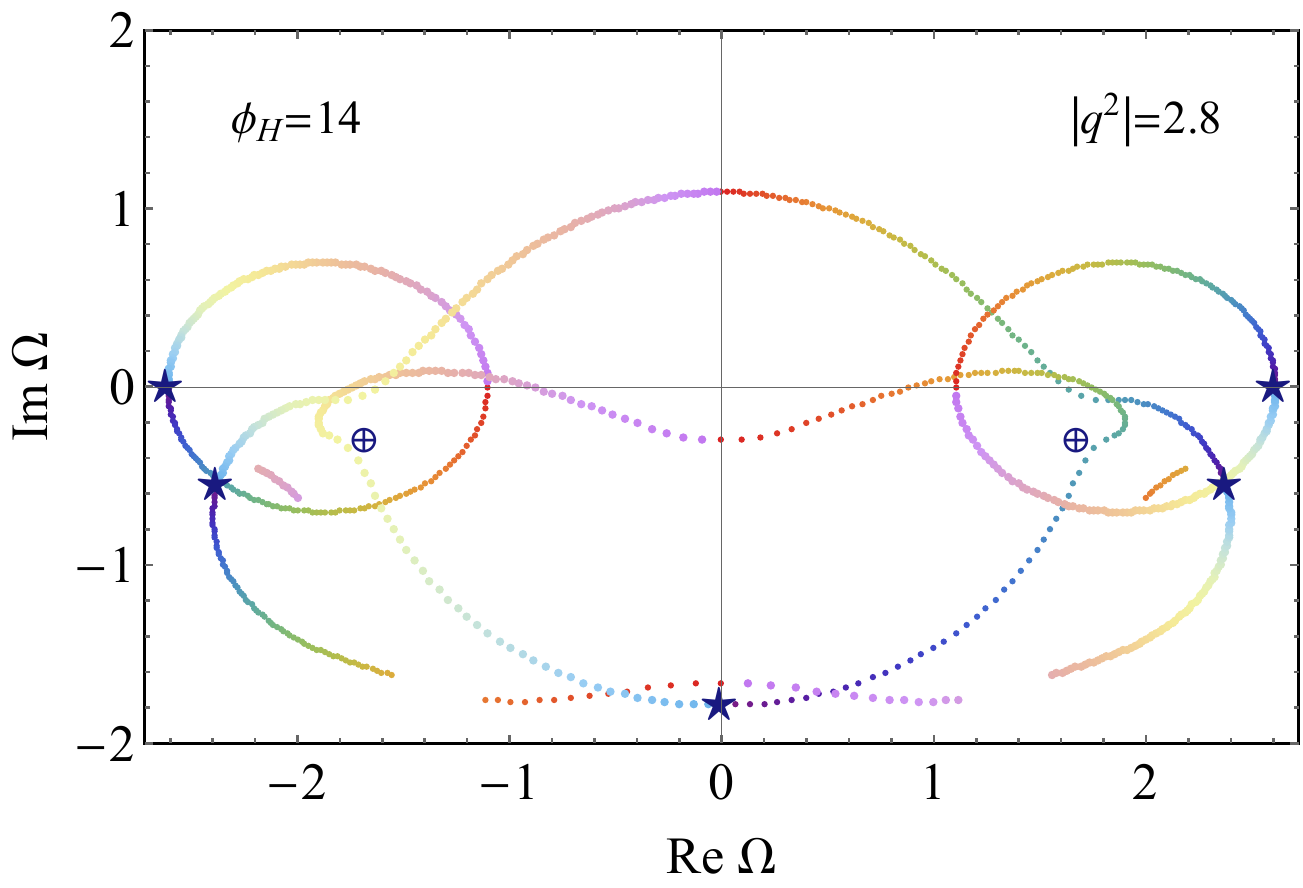}
    \caption{The Mode collision for the first-order transition in spin-1 sector. The top (bottom) row plots show $\phi_H=12$ ($\phi_H=14$) data, and the left (right) panels demonstrate the modes before (after) the collision. At $\phi_H=12$, collision occurs at $\vert q^2\vert=2.117$ with $\Omega_\star=\pm 0.122+0.887 i$, and for $\phi_H=14$ collision occurs at $\vert q^2\vert=2.714$ with $\Omega_\star=\pm 1.678-0.286 i$.}
    \label{fig: collision-mode-spin1-FO}
\end{figure}

\begin{figure}[htb]
    \centering
    \includegraphics[width=0.45\textwidth, valign=t]{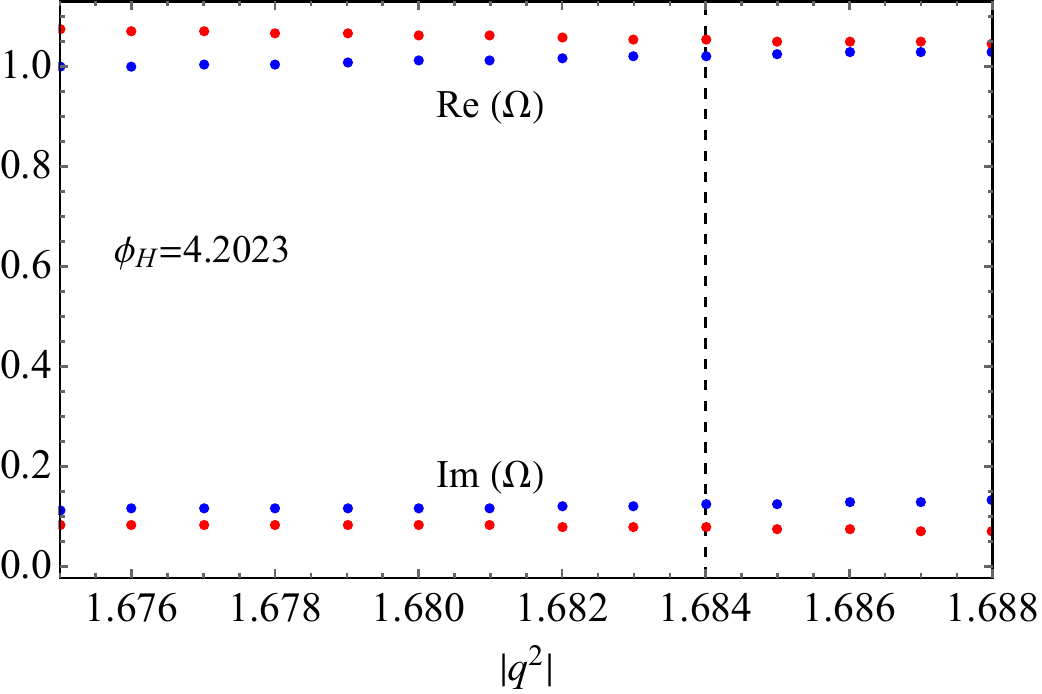}
    \caption{Collision of the lowest non-hydro mode with the hydro mode at $\phi_H=4.2023$ for the first-order phase transition in spin-1 sector. This collision occurs at $\theta=0.53 \pi$.}
    \label{fig: collision-1st}
\end{figure}

Fig. \ref{fig: FO-Real-Momenta-EoS-Im-Re-Omega-spin1} displays the real and imaginary parts of the lowest three modes for real $q^2 = (0,3)$ for the first-order EoS and the spin-1 sector. In the real parts, besides the lowest mode, the $(2^{\text{nd}}, 3^{\text{rd}}, 5^{\text{th}})$ modes diverge to infinity at low temperatures, and the other modes converge to their 5D-AdS limit. In the imaginary parts,  by increasing the $\vert q^2\vert$, the hydro mode evolves with $\phi_H$ and returns to its place at low temperatures. The first non-hydro mode goes to zero at low temperatures, while some of the $\Omega_{k}(T \gg T_c) \to -\infty(T\ll T_c)$ for $k \geq 3$.
Fig. \ref{fig: collision-mode-spin1-FO} shows the mode collision for two complex $q^2$ at $\phi_H = 12$ and $\phi_H = 14$. The left (right) parts correspond to before (after) the collision. In the spin-1 sector, the collision happens between the gravity hydro and non-hydro modes. At $\phi_H = 12$, the collision occurs for purely imaginary momenta that are very close to $\theta = \pi$, and $\text{Im} \Omega > 0$. At $\phi_H = 14$, it happens for $\text{Im} \Omega < 0$ around $\theta = 0.4 \pi$.
To further demonstrate, Fig. \ref{fig: collision-1st} displays the real and imaginary parts of the gravity hydro and first non-hydro mode at $\phi_H = 4.2023$ in terms of $\vert q^2\vert$ near the collision point. The collision occurs at $\vert q^2\vert = 1.684$ and $\theta = 0.53 \pi$.

\begin{figure}[htb]
    \centering
        \includegraphics[width=0.44\textwidth, valign=t]{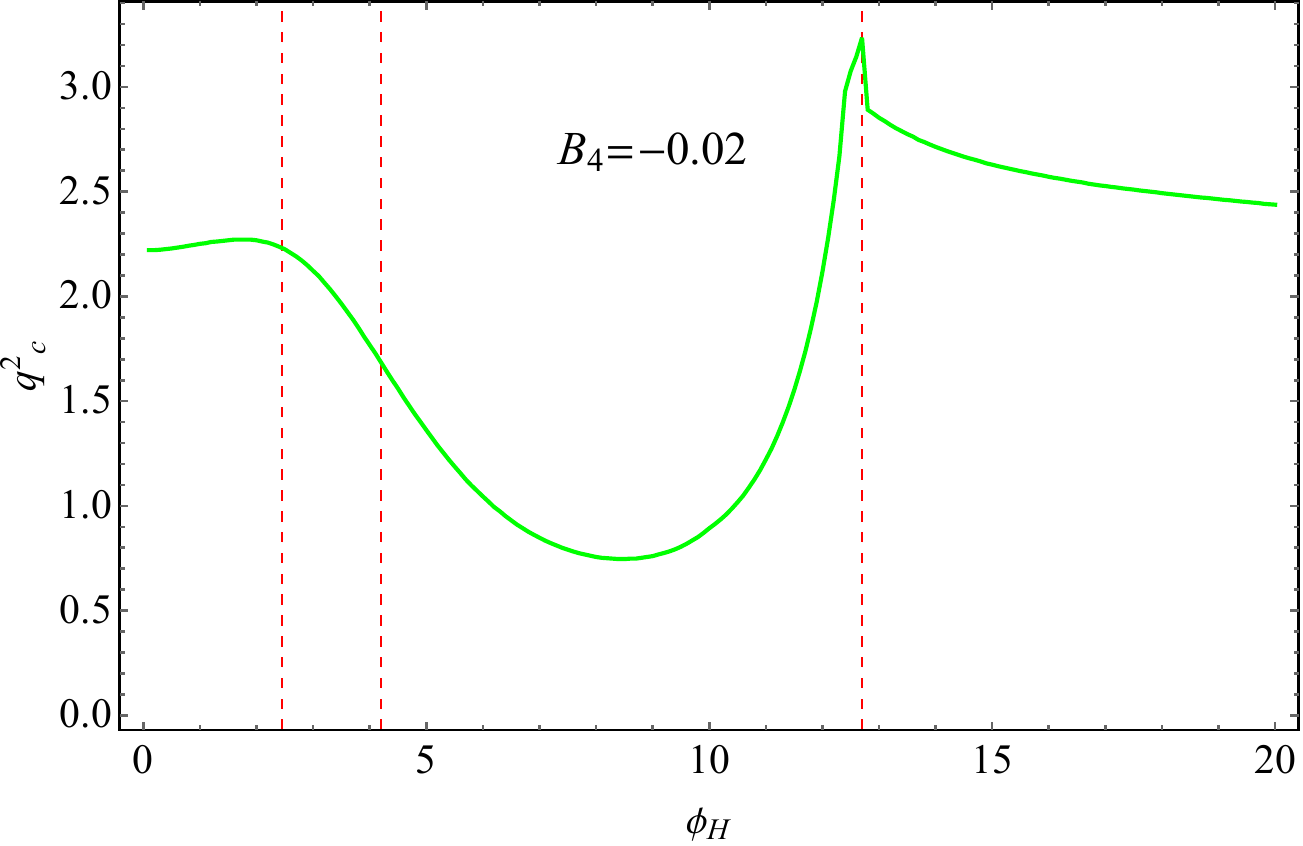}
        \hspace{0.4cm}
        \includegraphics[width=0.44\textwidth, valign=t]{plot1-FO.pdf}
    \caption{Plots of the radius of convergence in spin-1 sector for the first-order EoS. The left (right) panel shows the dependency of $\phi_H$ ($T/T_c$).  The dashed lines represent the location of critical points.}
    \label{fig: rc-shear-fo}
\end{figure}

Additionally, in Fig. \ref{fig: rc-shear-fo} we represent the radius of convergence in the spin-1 sector in terms of $\phi_H$ and $T/T_c$. This is similar to what happens in the spin-0 sector, where the $q^2_c$ increases on the stable branches, as seen in Fig. \ref{fig: convergence-radius-phiH-1st}.
\subsection{Spin-0 sector}
\begin{figure}[htb]
    \centering
    \includegraphics[width=0.42\textwidth, valign=t]{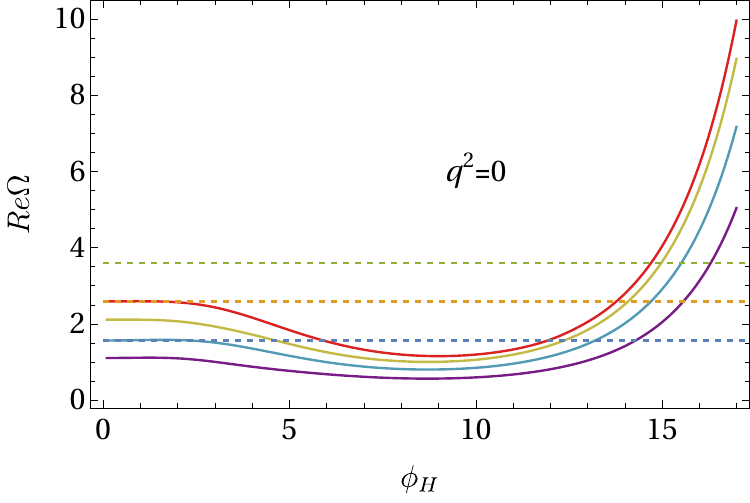}
    \hspace{0.3cm}
     \includegraphics[width=0.42\textwidth, valign=t]{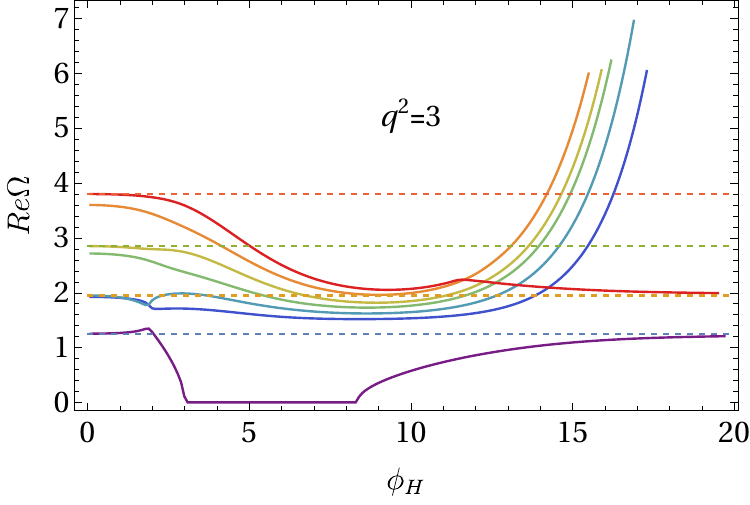}
     \includegraphics[width=0.42\textwidth, valign=t]{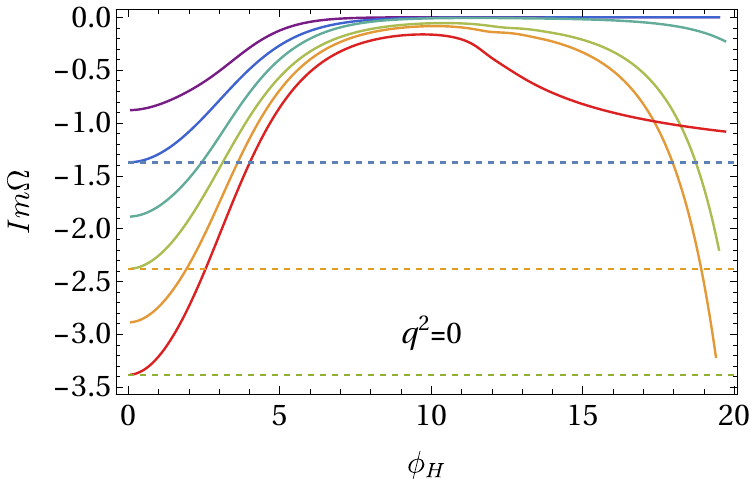}
     \hspace{0.3cm}
      \includegraphics[width=0.42\textwidth, valign=t]{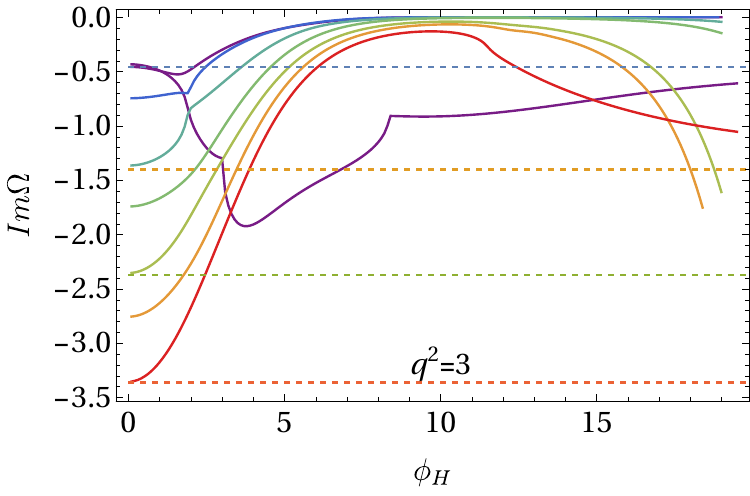}
    \caption{The real part (first row) and the imaginary part (second row) of the lowest three QNMs in the spin-0 sector of first-order EoS as functions of $\phi_H$ at $q^2=(0, 3)$.}
    \label{fig: FO1-Real-Momenta-EoS-Im-Re-Omega-spin0}
\end{figure}

In Fig. \ref{fig: FO1-Real-Momenta-EoS-Im-Re-Omega-spin0},
we show the real and imaginary parts of the modes for real $q^2 = (0,3)$ in the spin-0 sector. For the real parts, at $q^2=0$, all modes diverge to infinity at low temperatures except the lowest mode, while in the imaginary part, this happens for a few modes. By increasing the $q^2$, the hydro modes come into play, and in $3 \lesssim \phi_H \lesssim 8$, the real part of the hydro modes vanishes, and the imaginary part is split. This is a sign of mode collision, which is a well-known phenomenon in hydrodynamics. Apart from the hydro modes, the behavior of the real and imaginary parts of the other modes is similar to that of $q^2 = 0$.  

\begin{figure}[htb]
    \centering
    \includegraphics[width=0.44\textwidth, valign=t]{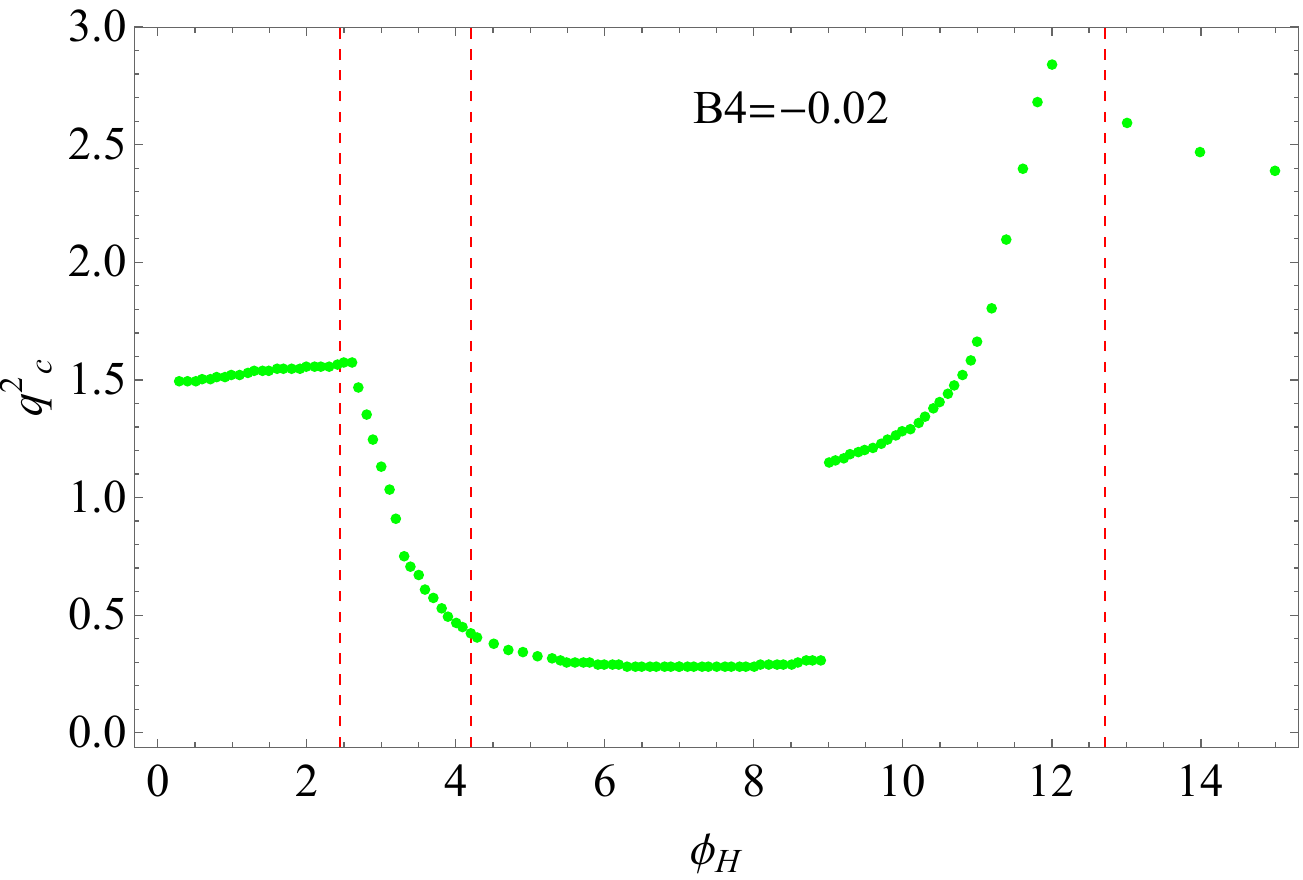}
    \hspace{0.4cm}
    \includegraphics[width=0.44\textwidth, valign=t]{plot-q2-T-FO-spin0.pdf}
    \caption{Radius of convergence in the spin-0 sector for the first-order phase transition. The left(right) panel shows the dependency of $\phi_H$ ($T/T_c$).}
    \label{fig: rc-FO-spin-0}
\end{figure}

Finally, in Fig. \ref{fig: rc-FO-spin-0}, we present the radius of convergence of the spin-0 sector in the first-order phase transition. Similar to the spin-1 and spin-2 sectors, there is a rise near the transition point of the stable branch, which seems to be a unique feature of the first-order transitions. Likewise, Eq. \eqref{eq:compare-rc} holds in the first-order EoS. 
\section{Pole-skipping}\label{sec: pole-skipping}
It has been confirmed that pole-skipping is a general feature of every quantum field theory that has a dual gravity interpretation \cite{Grozdanov:2017ajz, Blake:2018leo, Grozdanov:2023txs, Amrahi:2023xso}. It is the other side of quantum chaos that has been manifested in the linearized equations of motion. It works as follows: In the linearized equations of motion for gravity perturbations in the spin-0 sector, the $"vv"$ component of Einstein’s equations becomes trivial on the chaos point, namely $\omega = i \lambda_L = 2 \pi T i$ and $ k = i k_0 = i 2\pi T /v_B$, where $v_B$ is the butterfly velocity or velocity of information propagation   \cite{Grozdanov:2017ajz, Blake:2018leo}. The result is that the hydro poles skip around this point, and the corresponding Greens function becomes multivalued in such a way that different trajectories with various slopes converge to this point. This is very interesting because the out-of-equilibrium properties of a many-body quantum system can be probed even at the level of near-equilibrium situations.

Motivated by this, we study the chaotic nature of linearized equations in the spin-0 sector solutions for each type of phase transition. To do this, we have to modify Eq. \eqref{eq:soundmaster} and take the fluctuations in the Eddington-Finkelstein coordinates into account, as follows \cite{Amrahi:2023xso}
\begin{align}
\delta \Phi = \bigg(\delta g_{v v}, \delta g_{r v}, \delta g_{rr}, \delta g_{r z}, \delta g_{v z}, \delta g_{x x}, \delta g_{z z}, \delta \phi \bigg).
\end{align}
Solving the linearized equations is a very hard task. However, a series solution exists for every single point in the bulk. Among them, we scrutinize the near horizon and assume a series ansatz for each physical perturbation
\begin{align}\label{expansion}
	&\delta g_{M N}(u) = \sum_{n = 0}^{\infty} \delta g_{M N}^{(n)}(u_H)\, (u -u_H)^n,\nn\\
	&\delta \phi(u) = \sum_{n = 0}^{\infty} \delta\phi^{(n)}(u_H)\, (u -u_H)^n,
	\end{align}
 and insert these into the equations. We observe that the "$v v$" component of Eq. \eqref{eq: EoMs} at the lowest non-vanishing order takes the following form
 \begin{align}\label{eqgvv}
	\delta g_{v v}^{(0)}(u_H) \left( k^2 + \frac{i \omega e^{2 A(u_H)} V(u_H)}{4 \pi T} \right)+ (\omega -2 i \pi  T) (2 k \delta g_{z v}^{(0)}(u_H)+ \omega  \delta g_{x^i x^i}^{(0)}(u_H)) = 0,
	\end{align} 
where $i = x, y, z$. In the linearized equations of motion for gravity perturbations in the spin-0 sector, the $"vv"$ component of Eq. \eqref{eqgvv} imposes a non-trivial constraint on the near-horizon components $\delta g_{v v}^{(0)}(u_H)$, $\delta g_{z v}^{(0)}(u_H)$, $\delta g_{x^i x^i}^{(0)}(u_H)$ for general $\omega$ and $k$.  However,  at the point  $\omega_{*}=i\lambda_{L}=2 \pi Ti$ the metric component $\delta g_{v v}^{(0)}(u_H)$ decouples from the others and additionally at $k = \sqrt{e^{2 A(u_H)} V(u_H)/2}$ or $k = i k_0 = i \sqrt{6 \pi T e^{A(u_H) - B(u_H)} A'(u_H)}$ the Eq. \eqref{eqgvv} becomes identically zero \cite{Amrahi:2023xso}. Therefore, it doesn't imply any constraint on the near-horizon components. Its message is that there exists one extra ingoing mode at this point, which leads to the pole-skipping in the retarded energy density correlation function $G^{R}_{T^{00}T^{00}}(\omega,k)$  at the chaos point. In other words, slightly away from the chaos point with $\omega=i\lambda+\epsilon\,\delta\omega$ and $k=i k_{0} + \epsilon\, \delta k$ where $\vert \epsilon \vert\ll 1$ and at leading order in $\epsilon$,  we have a  family of different ingoing modes by different slopes $\frac{\delta \omega}{\delta k}$. This slope can be chosen such that it corresponds to the resulting ingoing mode near the chaos point with different asymptotic solutions at the boundary. If one chooses $\frac{\delta \omega}{\delta k}$ as follows
	\begin{align}\label{eqslope}
	\frac{\delta \omega}{\delta k} = \frac{2 k_{0} \delta g^{(0)}_{v v}(u_H)}{\frac{k_{0}^2}{2 \pi T}\delta g^{(0)}_{v v}(u_H) - 2 k_{0} \delta g^{(0)}_{z v}(u_H) - 2 \pi  T \delta g^{(0)}_{x^i x^i}(u_H)},
	\end{align}
 we will obtain an ingoing mode that matches continuously onto the normalizable solution at the boundary. All these lines pass through the chaos point, and we see different lines with slopes \eqref{eqslope} away from that point in the $T_{0 0}$ correlation functions.

 The multivaluedness of the boundary retarded Green's function has another manifestation. Recently, it has been reported that at higher Matsubara frequencies, i.e. $\omega = \omega_n = - 2 i n \pi T$  the equations of motion of scalar field perturbations exhibit a pole-skipping property\cite{Grozdanov:2023txs, Blake:2019otz}. This is because at these points, the equations give no constraints on $\delta \phi^{(n)}(u_H)$,  and these unknown coefficients reflect many hydrodynamics poles around the $\omega_n$ with special slopes \cite{Blake:2019otz}. We would like to explore these features in our model. To do this, we expand the equations of motion for scalar perturbation around the horizon. The result is as follows
\begin{align}\label{pole}
	\mathcal{I}_{1} & = M_{1 1}(\omega, k^2) \delta \phi^{(0)}(u_H) + (2 \pi T - i \omega) \delta \phi^{(1)}(u_H),\nn\\
	\mathcal{I}_{2} & = M_{2 1}(\omega, k^2) \delta \phi^{(0)}(u_H) + M_{2 2}(\omega, k^2) \delta \phi^{(1)}(u_H) + (4 \pi T - i \omega) \delta \phi^{(2)}(u_H),\\
	\mathcal{I}_{3} & = M_{3 1}(\omega, k^2) \delta \phi^{(0)}(u_H) + M_{3 2}(\omega, k^2) \delta \phi^{(1)}(u_H) + M_{3 3}(\omega, k^2) \delta \phi^{(2)}(u_H) +  (6 \pi T - i \omega) \delta \phi^{(3)}(u_H),\nonumber
\end{align}
where the coefficients $M_{i j}(\omega, k^2)$ take the following form
\begin{align}
	M_{i j}(\omega, k^2) = i \omega a_{i j} + k^2 b_{i j} + c_{i j},
\end{align}
with $a_{i j}, b_{i j}, c_{i j}$ are determined by the background solutions in \eqref{eq: metric} and their derivatives on the horizon. The special form of these expressions is very complicated, and they have nothing to do with our goals. The Expressions $\mathcal{I}_{i}$ are combinations of gravity perturbations with specific coefficients. Eq. \eqref{pole} shows that at frequencies $\omega = \omega_n$, it is not possible to read the coefficients iteratively from $\delta \phi^{(0)}(r_H)$. It means that $\delta \phi^{(n)}(u_H)$ are free parameters near the horizon. Also, at the point $\omega = \omega_n$, the first $n$ equation is decoupled, and we can solve a simple matrix equation  as it follows
\begin{align}
	\mathcal{M}^{n}(\omega, k^2) \cdot \delta \tilde{\phi}= \mathcal{I},
\end{align} 
for $\delta \tilde{\phi} = \left(\delta \phi^{(0)}(u_H), \cdots, \delta \phi^{(n-1)}(u_H)\right)$.  However, we observe that at $k = i k_0$ and $\omega = \omega_n$, $\mbox{det}\, \mathcal{M}^n(\omega_n, k_0) = 0$. Therefore, solutions for the linear equations \eqref{pole} are labeled with two free parameters \cite{Grozdanov:2019uhi, Grozdanov:2023txs, Blake:2019otz}.

\begin{figure}[htb]
    \centering
    \includegraphics[width=0.54\textwidth, valign=t]{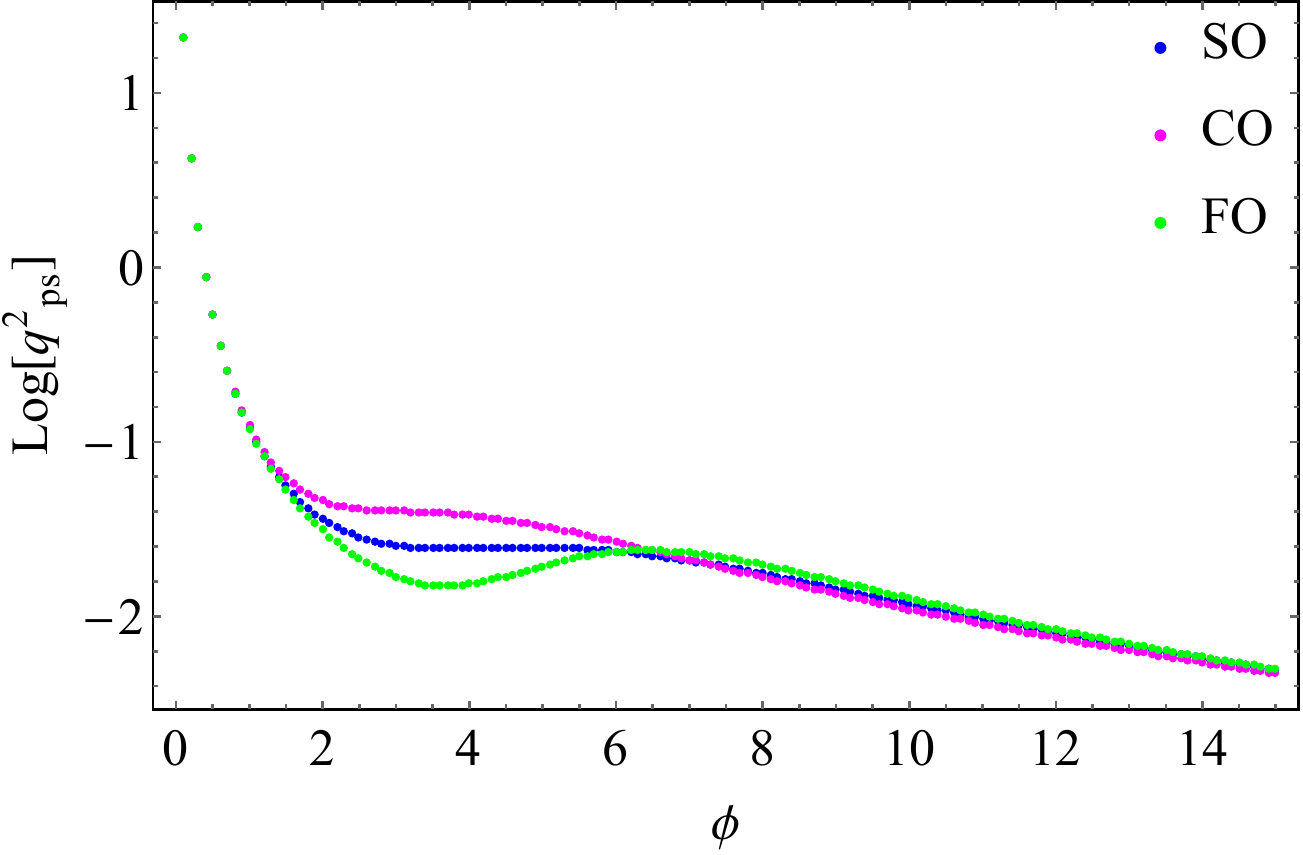}
    \caption{Logarithmic sketch of chaos momenta in terms of $\phi_H$ for different kinds of phase transition, second-order (SO), crossover (CO), and first-order (FO).}
    \label{fig: ps-spin-0}
\end{figure}

\begin{figure}[htb]
    \centering
    \includegraphics[width=0.46\textwidth, valign=t]{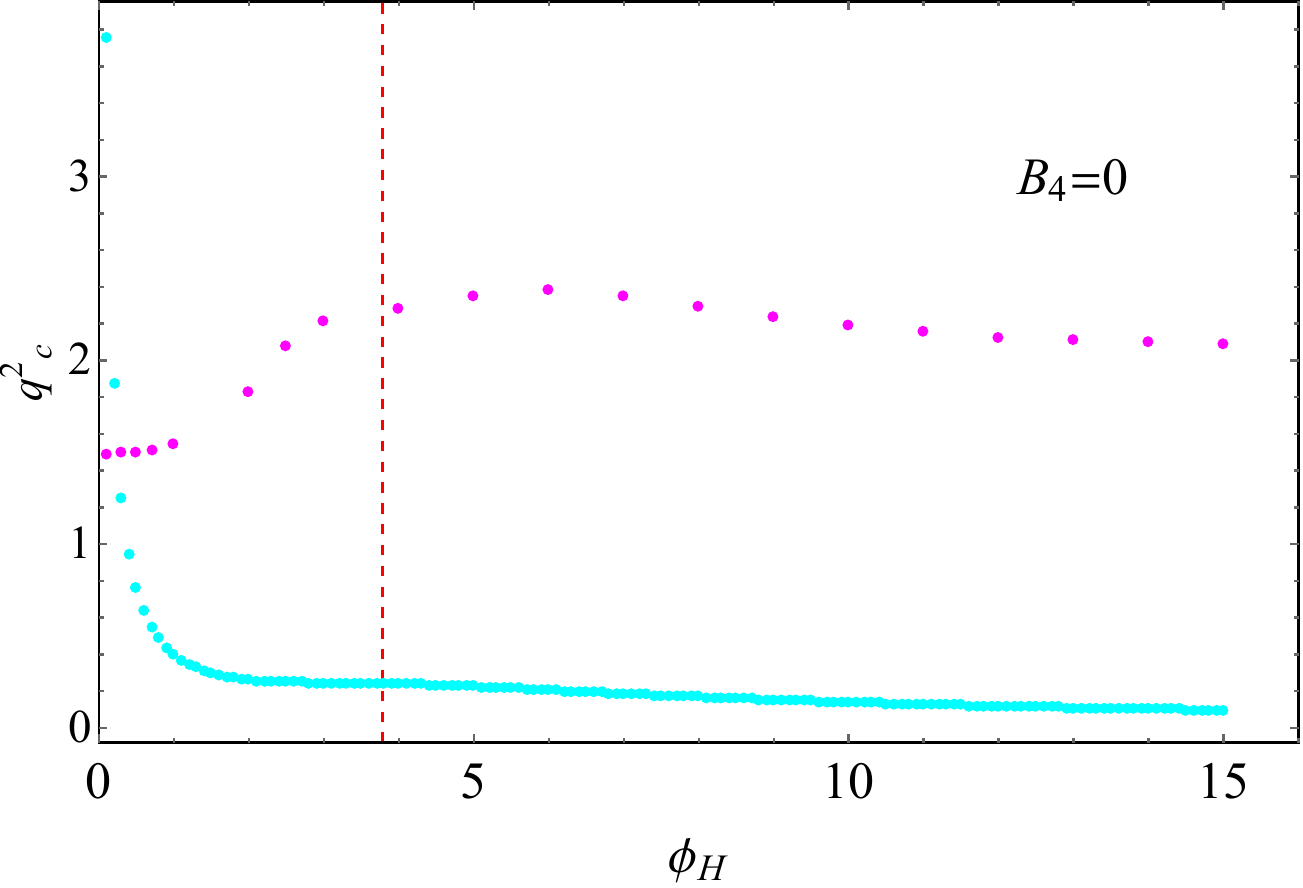}
    \hspace{0.4cm}
    \includegraphics[width=0.46\textwidth, valign=t]{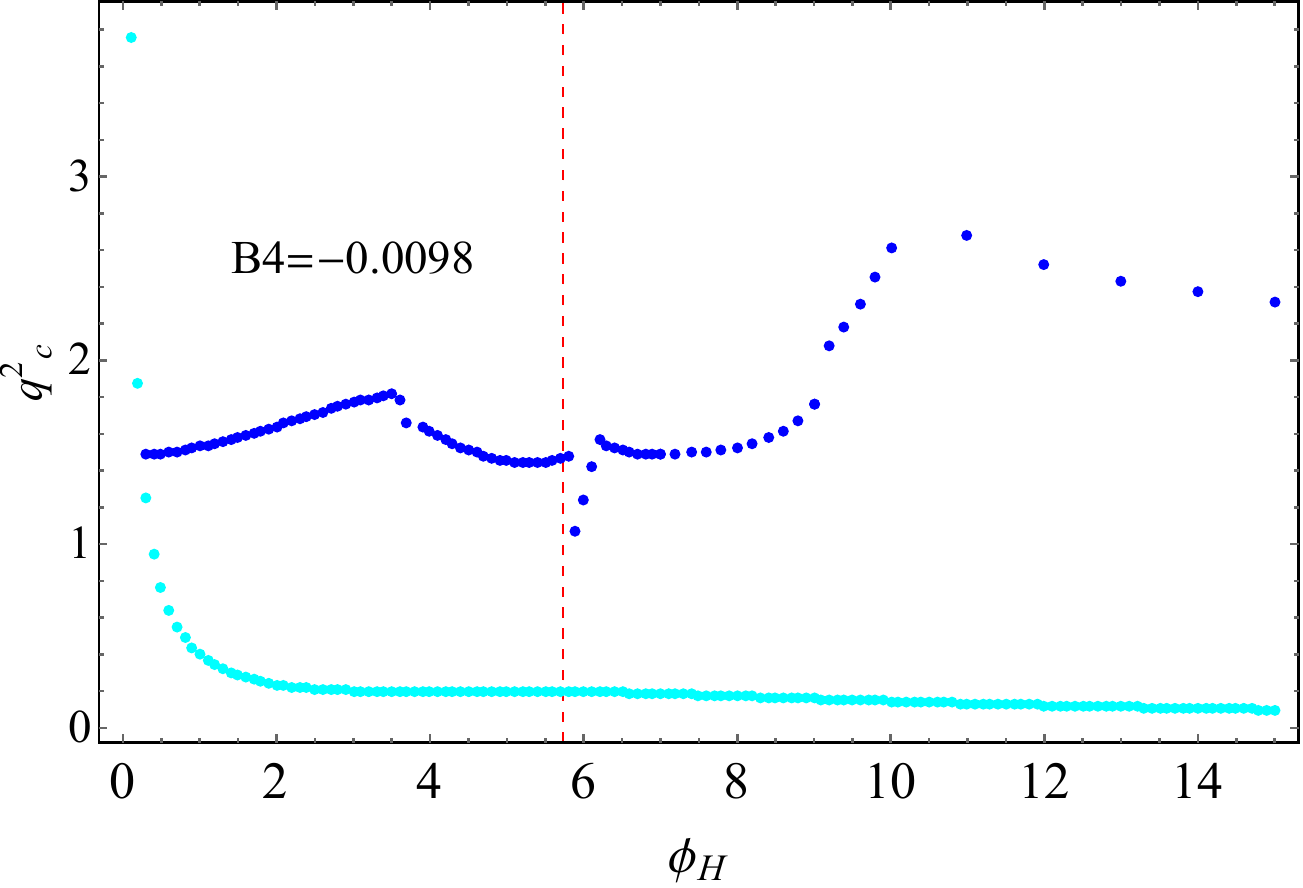}
    \includegraphics[width=0.46\textwidth, valign=t]{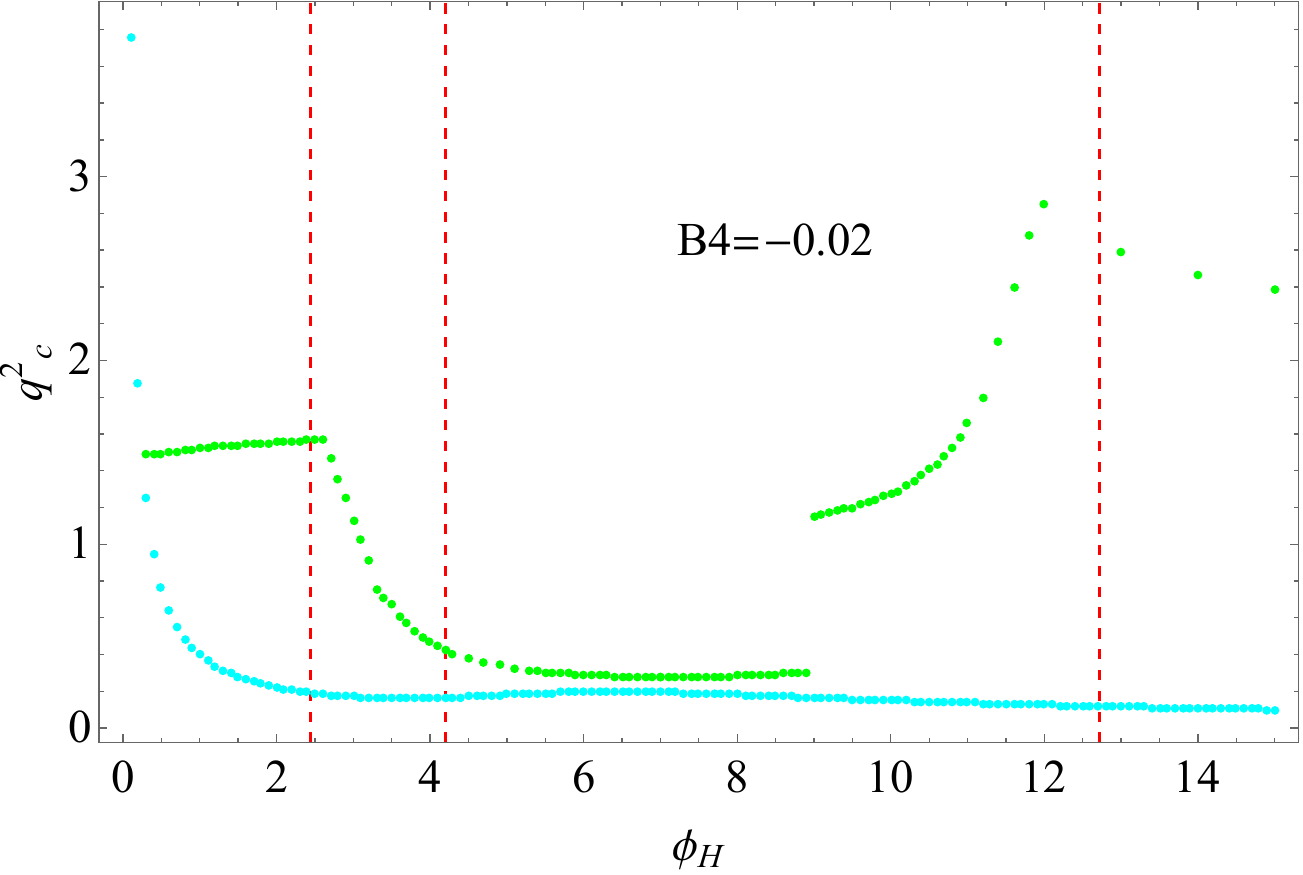}
    \caption{The chaos momenta and radius of convergence are compared for different types of phase transitions. The cyan plots represent the $q^2_{ps}$ points, while the dashed lines indicate the location of the phase transition.}
    \label{fig: ps-compare-spin-0}
\end{figure}

In Fig. \ref{fig: ps-spin-0}, we plot $\ln(q^2_{ps}) = \ln(k_0^2/(2 \pi T)^2)$ for different phase transition in terms of $\phi_H$. At very high temperatures, no remarkable difference is seen for $q^2_{ps}$ between various kinds of phase transitions. However, in the region $1 \lesssim \phi_H \lesssim 6$, it seems that $(q^2_{ps})_{\text{FO}} < (q^2_{ps})_{\text{SO}}< (q^2_{ps})_{\text{CO}}$.  We can infer that $q^2_{ps}$ is useful in identifying the location of phase transition.  This is because the relation $q^2_{ps} \sim V(\phi_H)$ results from $k = \sqrt{e^{2 A(u_H)} V(u_H)/2}$. From the physical point of view, it is interesting that such an out-off-equilibrium variable, i.e. $q^2_{ps}$ would enable us to see the critical points. It has been observed that in the 1RCBH model, the butterfly velocity as a dynamical and out-off-equilibrium probe, can show the location of the critical point, \cite{Amrahi:2023xso}. It is worthwhile to examine the relationship between the $q^2_{ps}$ and $V(\phi_H)$ in other holographic models with a critical point.  
In Fig. \ref{fig: ps-compare-spin-0}, we compare chaos momenta and radius of convergence for different types of phase transitions. The cyan points refer to the $q^2_{ps}$ values, and the others represent the $q^2_c$. It is seen that there is a possibility of collision between $q^2_c$ and $q^2_{ps}$ at high temperatures around $\phi_H \sim 0.27$, which is independent of the phase transition. However, $q^2_{ps} < q^2_c$ always, i.e., chaos momenta are in the range of hydro validity.
  
\section{Conclusion}\label{sec: conclusion}
One of the significant challenges in applying relativistic hydrodynamics (RH) is dealing with scenarios near critical points, which can occur in low-energy colliders. It is highly valuable to estimate the validity of the RH series during or at transition points. Within the AdS/CFT conjecture, we extensively study the many-body dynamics of a strongly coupled and critical field theory that is dual to a gravity model with a self-interacting scalar field in one higher dimension. This model, known as the Einstein-Klein-Gordon model, is a phenomenological string theory construction. The model effectively represents the critical strong field theory through the parameter $B_4$ in the superpotential function, allowing for crossover $(B_4=0)$, second-order $(B_4 = -0.0098)$, and first-order $(B_4<-0.0098)$ phase transitions. We run the first-order codes with $B_4=-0.02$. According to the fluid/gravity conjecture, small perturbations of bulk fields on the background are equivalent to studying the RH for the field theory on the boundary. With this in mind, we investigate the dynamics of linearized fluctuations in the spin-2, spin-1, and spin-0 sectors for each type of phase transition.

Our study found that $\eta/s = 1/(4\pi)$ for each phase transition and $\xi/s$ exhibits peaks around the phase transition. Regardless of the phase transition, the first collisions in the spin-2 sectors occur at negative $q^2$ between two non-hydrodynamic modes. In contrast, the determining collision for the radius of convergence in the spin-1 sector happens between the hydrodynamic mode and the closest gravity non-hydrodynamic mode at positive $q^2$. In the spin-0 sector, the metric and scalar perturbations are coupled. We find that at minimal and large temperatures, the collision occurs between the hydrodynamic and scalar non-hydrodynamic modes, while at intermediate temperatures, it happens between the hydrodynamic mode and the gravity non-hydrodynamic mode. At $q^2=0$, the spin-0 sector equations for the gravity and scalar fields are decoupled, leading to an even number of modes. However, at $q^2\neq 0$, the number of modes is odd. The high-temperature and low-temperature limits of the hydrodynamic modes in the spin-1 and spin-0 sectors remain the same for real $q^2$. For second-order and first-order phase transitions, some of the non-hydrodynamic modes go to infinity at low temperatures, leading to other modes taking their places. Our results have shown that the paradigm "{\it{breakdown of the hydrodynamic series near the transition points}}" seems to be (not true, true, not true) for the (crossover, second-order, first-order) phase transition. We have observed that the high-temperature and low-temperature limits of $q^2_c$ are equal, which may indicate the same equality for using the low-momentum expansion in these regimes. Furthermore, we find that Equation \eqref{eq:compare-rc} holds between the different spin sectors irrespective of the kinds of phase transition. Additionally, we obtain a closed-form result for $q^2_c$ in the spin-2 sector at high temperatures, regardless of the phase transition type. It is $q^2_c = 1.486\, e^{i \theta_c}$ with $\theta_c \approx \pi$. Our study has found that the lowest non-hydrodynamic modes of the scalar field for large $-\vert q^2\vert$ approach one of the hydrodynamic (gravity) modes at high temperatures. In this sector, we calculate the radius of convergence for scalar fields with different conformal weights. We investigate the pole-skipping feature for near-equilibrium modes of gravity and scalar fields, which is studied through the $"v v"$ component of Einstein's equations at the chaos point for higher Matsubara frequencies. From this exploration, we have observed that $q^2_{ps}$ is always less than $q^2_c$ at high temperatures, and $q^2_{ps}$ decreases as $\phi_H$ increases. This may indicate that within the hydrodynamic validity region, we have a chance to observe chaos.

This study can be extended to exploring hydrodynamic solutions on top of thermodynamics with critical EoS, having Gubser or Bjorken flow. These critical models, derived from Thermal Quantum Field Theories with auxiliary fields, could validate Eq. \eqref{eq:compare-rc} and compare results with ours. We anticipate that our findings should be consistent in these models. Another way is to generalize this work to more real systems, namely the holographic models that are very close to the QCD phase diagram \cite{DeWolfe:2010he}. These models include gauge fields, and it is necessary to complete and correct the master formula approach for the gauge+scalar+gravity fields. Furthermore, it is valuable to investigate the validity of our results in other gravity models, particularly the Eq. \eqref{eq:compare-rc}. 
\section*{Acknowledgement}
We thank cordially H. Soltanpanahi for his earlier contribution to this work, especially for providing the numerical codes and giving background numerical solutions.

\begin{appendix}
\section{Holographic Renormalization}\label{appendix-HRG}
Having a suitable counter-term part in the ADS/CFT paradigm is very important to derive finite results for one-point functions of the boundary theory \cite{Skenderis:2002wp}.  The Hamilton-Jacobi approach is a convenient way to obtain this term order by order in derivatives with respect to "r" which resembles Hamiltonian time, $\tau$ \cite{Papadimitriou:2016yit}. Moreover, it has a great advantage to use the superpotentials because, in the process of holographic renormalization, superpotentials arise naturally. In $d$ dimensions of boundary space, it is defined as 
\begin{align}
V(\phi) = 2 \left(\frac{\partial W(\phi)}{\partial \phi}\right)^2  - \frac{d}{d-1} W(\phi)^2.
\end{align}
Indeed, the superpotentials as counter-term fix the ambiguous coefficient of the finite $\phi^4$ term to the unique value that gives zero free energy for the ground state dual to the domain wall geometry. Therefore, by construction, counter-terms can be obtained in terms of the superpotentials. We follow the recipes of  \cite{Papadimitriou:2016yit} to derive the counter-terms and the results are shown below
\bea
S_{\textrm{ct}}&=&-\frac{1}{16\pi \GN}\int_{\partial M}\extd^4x\sqrt{\gamma}\,\Big[W(\phi) + R(\gamma) \left(I(\phi) + J(\phi)\right) \Big],
\eea
where $\gamma$ is the determinant of the boundary-induced metric and
\begin{align}
    &I(\phi) = \frac{1}{2} \int^\phi \, d\tilde{\phi} \frac{1}{W'(\tilde{\phi})} = \frac{1}{2} \ln \frac{\left(1 - 4 B_4 \phi^2\right)^{\frac{1}{2}}}{\phi},\nn\\
    &J(\phi) = - e^{- 2 A(\phi)} \int^\phi \, d\tilde{\phi} \frac{ e^{2 A(\tilde{\phi})}}{W'(\tilde{\phi})},\nn\\
    & A(\phi) = - \frac{1}{6} \int^\phi \, d\tilde{\phi} \frac{W(\tilde{\phi})}{W'(\tilde{\phi})} = - \ln \phi - \frac{\phi^2}{48} + \frac{1 + 96 B_4}{192 B_4} \ln \left(1 - 4 B_4 \phi^2 \right).
\end{align}
It is worth mentioning that these results are obtained in the Gubser gauge where $u \equiv \phi(u)$. The one-point functions are given as functional derivatives of the renormalized action with respect to boundary fields
\be
\langle {\mathcal{O}_{\phi}} \rangle = \lim_{u\rightarrow0} \frac{u^{-\frac{3}{2} }}{\sqrt{-\gamma}}\frac{\delta S_{\textrm{ren}}}{\delta\phi},\qquad\qquad
\langle T_{ij} \rangle =2 \lim_{u\rightarrow0} \frac{u^{-2}}{\sqrt{-\gamma}}\frac{\delta S_{\textrm{ren}}}{\delta\gamma^{ij}},\label{EMT}
\ee
with the holographically renormalized action consists of the bulk action $S_{\textrm{\tiny{bulk}}}$, the Gibbons--Hawking--York boundary term $S_{\textrm{\tiny{GHY}}}$ and the holographic counter-term $S_{\textrm{\tiny{ct}}}$
\be
S_{\textrm{ren}} = S_{\textrm{bulk}}+S_{\textrm{GH}}+S_{\textrm{ct}}.
\ee
For the metric in the Eq. \eqref{eq: metric}, we get $R(\gamma) = 0$. By ignoring the terms related to the equations of motion, the one-point functions are given as 
\begin{align}
    &\langle T_{ij} \rangle = \frac{1}{8 \pi G_5} \lim_{u\rightarrow0} u^{-2} \left(K_{i j} - (K + \frac{W}{2} )\gamma_{i j} \right),\nn\\
    & \langle {\mathcal{O}_{\phi}} \rangle = -\frac{1}{16 \pi G_5}\lim_{u\rightarrow0} u^{-\frac{3}{2}} e^{- B(u)} H(u)^\frac{1}{2}.
\end{align}
By carefully manipulating the equation, we can omit the divergent terms. We can then substitute the near boundary expansion of $G(\phi)$ from the numerical results of Eq. \eqref{eq: master}. This substitution yields the one-point functions
\begin{align}
\varepsilon = \langle T_{tt} \rangle &=\frac{1}{8\pi G_5}\,\Big(-\frac{1}{64}- \frac{9 V_6}{2} + \frac{a_2}{4} + 15 a_2^2 + 6 a_4 - \frac{B_4}{4}\Big)\nn,\\
p = \langle T_{xx} \rangle &=\frac{1}{8\pi G_5}\, \Big(\frac{1}{576}- \frac{3 V_6}{2} + \frac{a_2}{4} + 5 a_2^2 + 2 a_4 + \frac{B_4}{4}\Big),\nn\\
\langle \mathcal{O} \rangle= \langle {\mathcal{O}_{\phi}} \rangle &=\frac{1}{8\pi G_5}\,\Big(\frac{1}{48}   + \frac{a_2}{4}+ B_4 \Big),
\end{align}
where 
\begin{align}
    a_2 = \frac{\tilde{G}(0)}{2}, \qquad a_4 = \frac{\tilde{G}''(0)}{8}, \qquad V_6 = \frac{B_4(24 B_4 + 1)}{3}.
\end{align}
These one-point functions respect the expected Ward identity, which is given by 
\begin{align}
\langle T^{i}_{~i} \rangle = \langle \mathcal{O}_{\phi} \rangle.\label{ward}
\end{align}
The non-zero trace of the energy-momentum tensor is generally due to the breaking of conformal symmetry in the presence of a dimensionful source.

\section{Near-horizon expansions}\label{appendix-nearhor}
To study the thermodynamics of the system, it is crucial to expand the functions near the horizon point  ($\phi = \phi_h$). We can do this by using the following expansion
\begin{align}
    &A(\phi) = \sum_{n = 0}^{\infty} A_n(\phi_h) \frac{(\phi -\phi_h)^n}{n!}, \nn\\
    &B(\phi) = \sum_{n = 0}^{\infty} B_n(\phi_h) \frac{(\phi -\phi_h)^n}{n!}, \nn\\
     &H(\phi) = \sum_{n = 1}^{\infty} H_n(\phi_h) \frac{(\phi -\phi_h)^n}{n!}.
\end{align}
By using Eqs. \eqref{eq:EOM1}, \eqref{eq:EOM2}, \eqref{eq:EOM3}, and \eqref{eq:EOM4} along with the condition $H(\phi_H) = 0$, we can solve for the expansion coefficients up to the desired orders. The lowest order results are shown below
\begin{align}
    &A_1(\phi_h) = - \frac{V(\phi_h)}{3 V'(\phi_h)}, \quad A_2(\phi_h) = -\frac{1}{6} + \frac{V(\phi_h) V''(\phi_h)}{6 V'(\phi_h)^2},\ldots,\\
    &B_1(\phi_h) = - \frac{V''(\phi_h)}{2 V'(\phi_h)}, \quad B_2(\phi_h) = \frac{1}{9} \left(-\frac{3 V^{(3)}(\phi_h)}{V'(\phi_h)}+\frac{V''(\phi_h) \left(3 V''(\phi_h)+2 V(\phi_h)\right)}{V'(\phi_h)^2}-2\right),\ldots\nn\\
    &H_1(\phi_h) = e^{2 B(\phi_h)} V'(\phi_h), \quad H_2(\phi_h) = \frac{1}{6} e^{2 B(\phi_h)} \left(8 V(\phi_h)-3 V''(\phi_h)\right)\nn, \ldots.
\end{align}
We can obtain the near-horizon expansion of $\tilde{A}, \tilde{B}$ in Eq. \eqref{eq:div-nearhor}. These functions have finite values near the boundary. For an operator with conformal weight $\Delta = 3$, the lowest-order coefficients are written as
\begin{align}
&\tilde{A}_1(\phi_h) = \frac{1}{\phi_h}- \frac{V(\phi_h)}{3 V'(\phi_h)}, \quad \tilde{A}_2(\phi_h) = -\frac{1}{6}- \frac{1}{\phi_h^2} + \frac{V(\phi_h) V''(\phi_h)}{6 V'(\phi_h)^2},\ldots,\\
    &\tilde{B}_1(\phi_h) = \frac{1}{\phi_h}- \frac{V''(\phi_h)}{2 V'(\phi_h)}, \nn\\
    &\tilde{B}_2(\phi_h) = \frac{1}{9} \left(-2 -\frac{9}{\phi_h^2}-\frac{3 V^{(3)}(\phi_h)}{V'(\phi_h)}+\frac{3 V''(\phi_h)^2}{V'(\phi_h)^2}+\frac{2 V(\phi_h) V''(\phi_h)}{V'(\phi_h)^2}\right),\ldots\nn\\
     &H_1(\phi_h) = \frac{e^{2 \tilde{B}(\phi_h)} V'(\phi_h)}{\phi_h^2}, \quad H_2(\phi_h) = \frac{1}{6 \phi_h^2} e^{2 \tilde{B}(\phi_h)} \left(8 V(\phi_h)-3 V''(\phi_h)\right)\nn, \ldots.
\end{align}
\section{Low temperature black holes}\label{sec:low-T BH}
In this appendix, we aim to find the black hole solutions in a low-temperature regime perturbatively. To achieve this, we solve the EKG equations perturbatively near the vacuum solution, which is known as the thermal gas. We find that the Gubser gauge is suitable for this purpose. We use the following ansatz
\begin{equation}\label{eq:low-T-ansatz}
    A(u)=A_{TG}(u)+\epsilon a(u),\quad B(u)=B_{TG}(u)+\epsilon b(u), \quad H(u)=H_{TG}(u)+\epsilon h(u),
\end{equation}
in the equations of motion, where $\epsilon$ is a small parameter that controls the perturbative expansion. We then expand the equations of motion to the first order in $\epsilon$. The thermal gas solution can be obtained by setting $\epsilon = 0$. It is given by
\bea
&&A_{TG}(u)=- \frac{1}{6} \int^u \, d\tilde{u} \frac{W(\tilde{u})}{W'(\tilde{u})} = \frac{1}{192} \left(\frac{1}{B_4}+96\right) \ln \left(1-4 B_4 u^2\right)-\frac{u^2}{48}-\ln (u),\nn\\
&&B_{TG}(u)= - \ln W'(u) =  -\ln \left(u(1-4 B_4 u^2)\right),\nn\\
&&H_{TG}(u)=1,
\eea
and the asymptotic boundary is at $u=0$. By imposing proper boundary conditions for $a, b, h$ in a large horizon radius regime we can find the following expression
\begin{equation}
    a(u)=\frac{f(u)}{8 f(u_H)},
    \quad b(u)=-\frac{f(u)}{2 f(u_H)},
    \quad h(u)=\frac{f(u)}{ f(u_H)},
\end{equation}
where  $f(u)$ is an auxiliary function. For $B_4=0$, it is given by 
\begin{equation}
    f(u)=1+e^{\frac{u^2}{6}} \left(1-\frac{u^2}{6}\right),
\end{equation}
and for $B_4\neq0$ it is 
\begin{align}
   f(u)&=\Gamma \left(-2-\frac{1}{48 B_4},\frac{1}{48 B_4}\right)-\Gamma \left(-2-\frac{1}{48 B_4},\frac{1-4 u^2 B_4}{48 B_4}\right)\nn \\
   &+48 B_4 \left(\Gamma \left(-1-\frac{1}{48 B_4},\frac{1-4 u^2 B_4}{48 B_4}\right)-\Gamma \left(-1-\frac{1}{48 B_4},\frac{1}{48 B_4}\right)\right).
\end{align}

\begin{figure}[htb]
    \centering
    \includegraphics[width=0.387\textwidth]{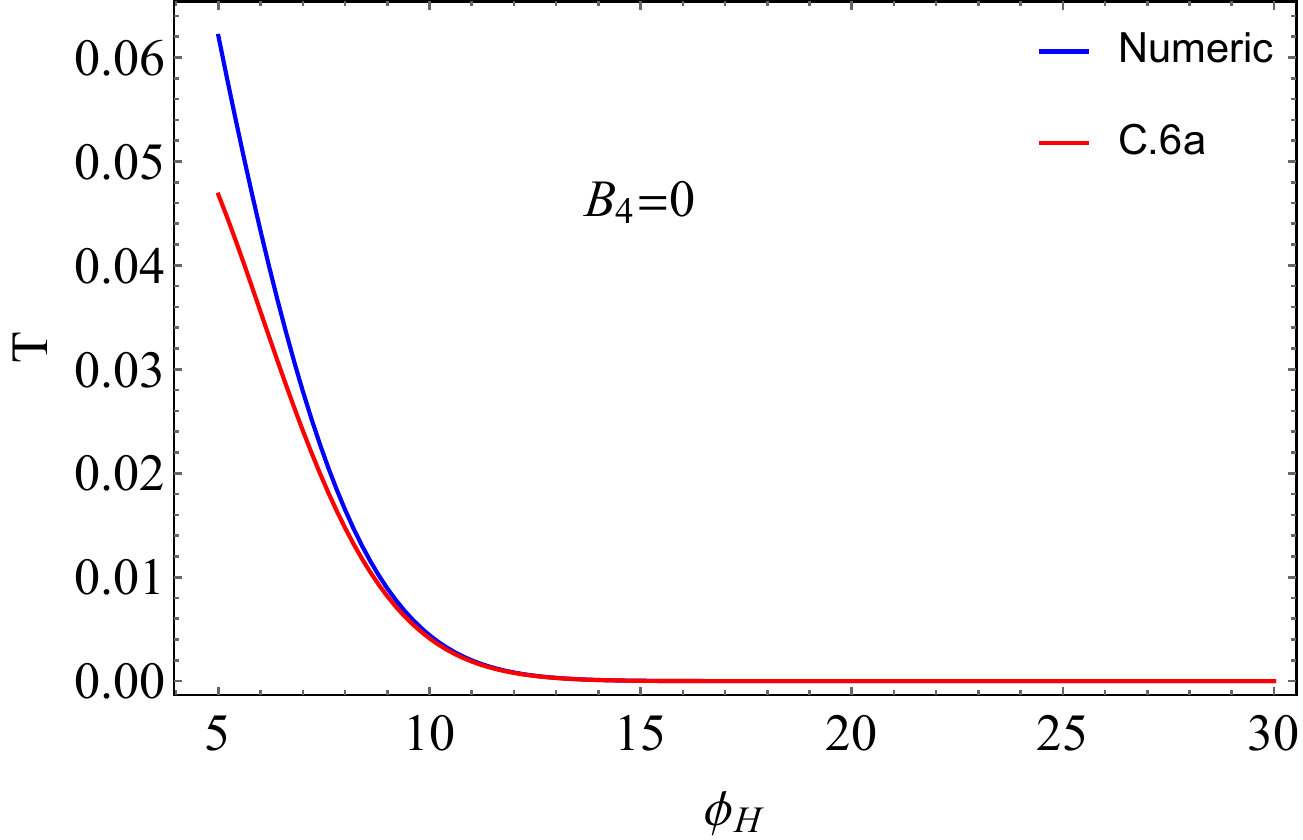}
    \includegraphics[width=0.405 \textwidth]{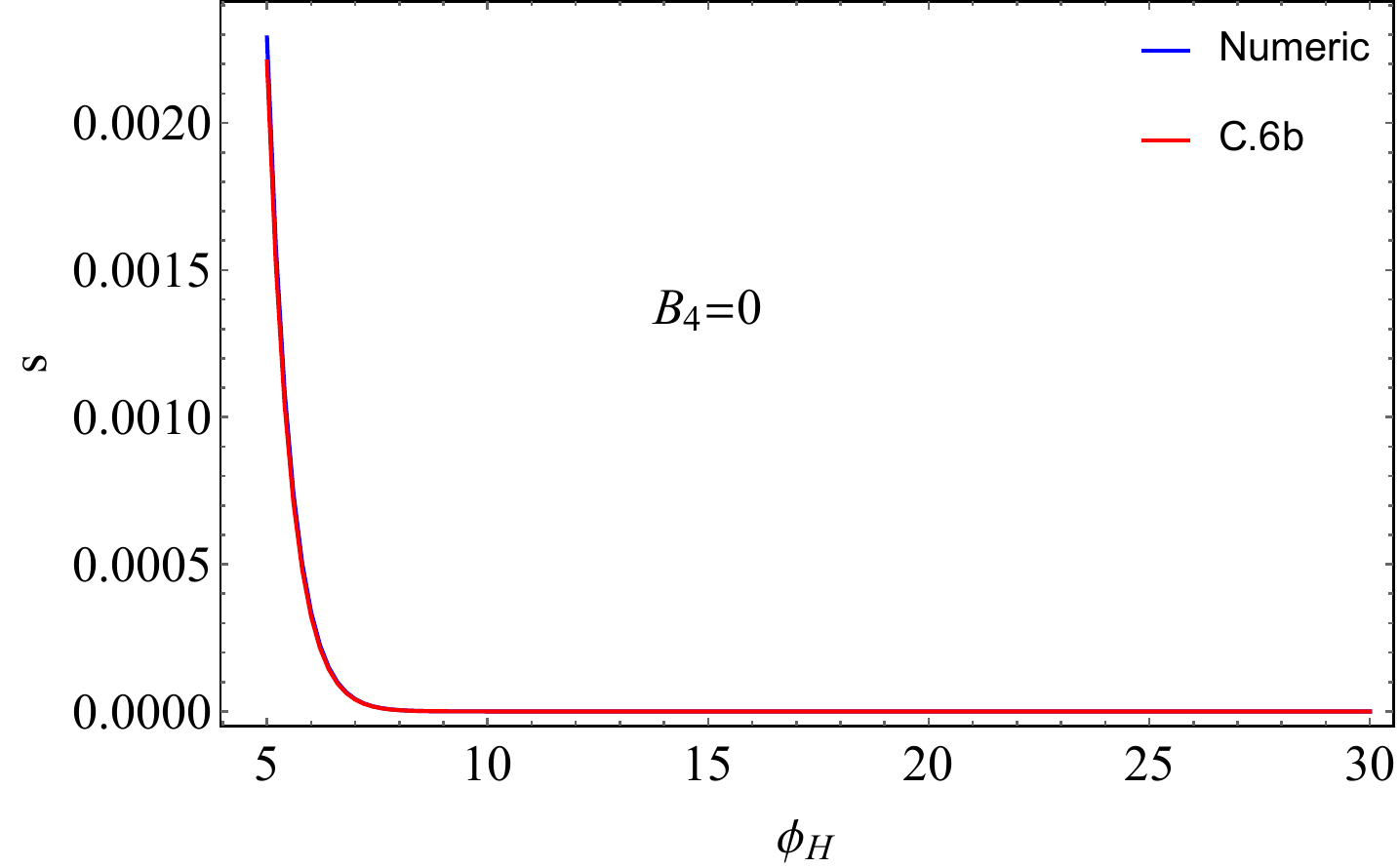}
    \includegraphics[width=0.4\textwidth]{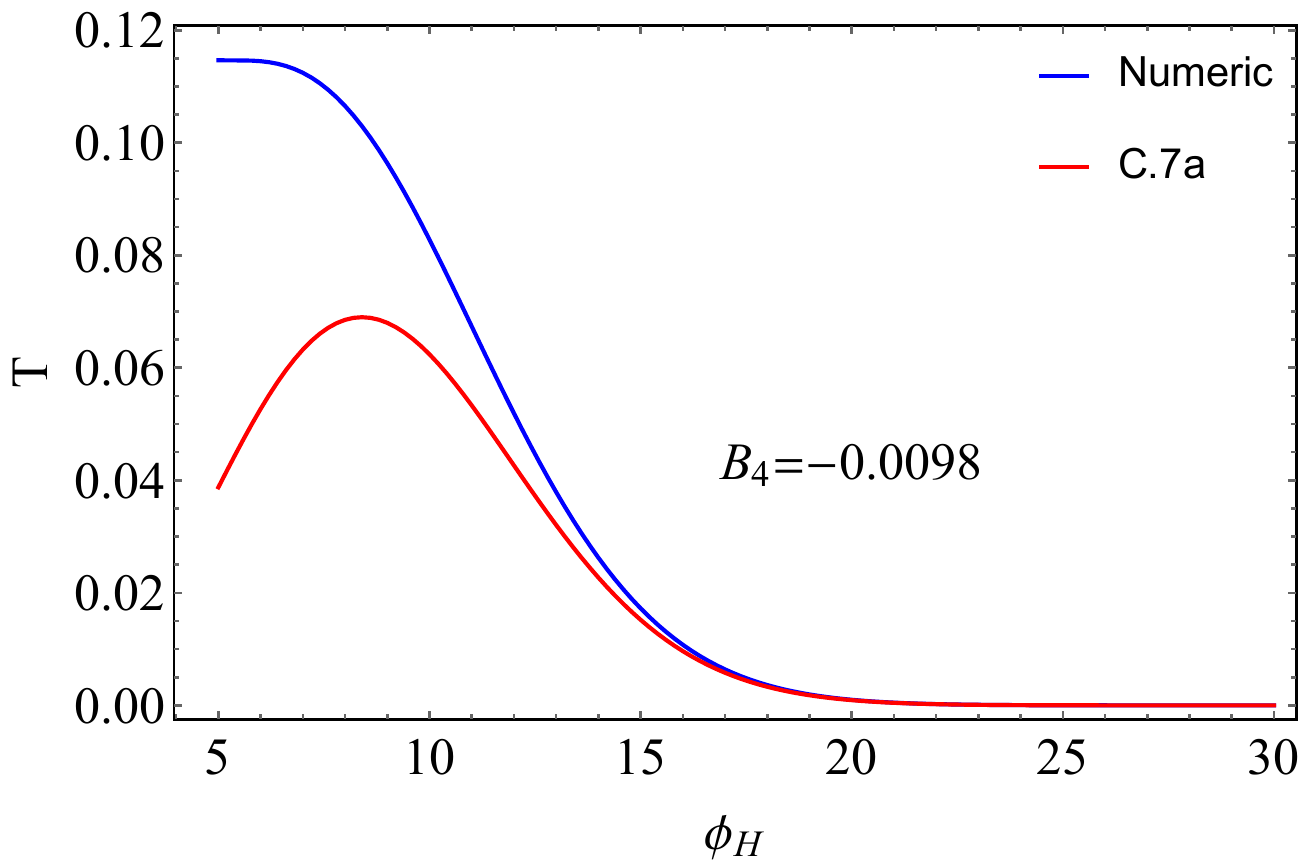}
    \includegraphics[width=0.4 \textwidth]{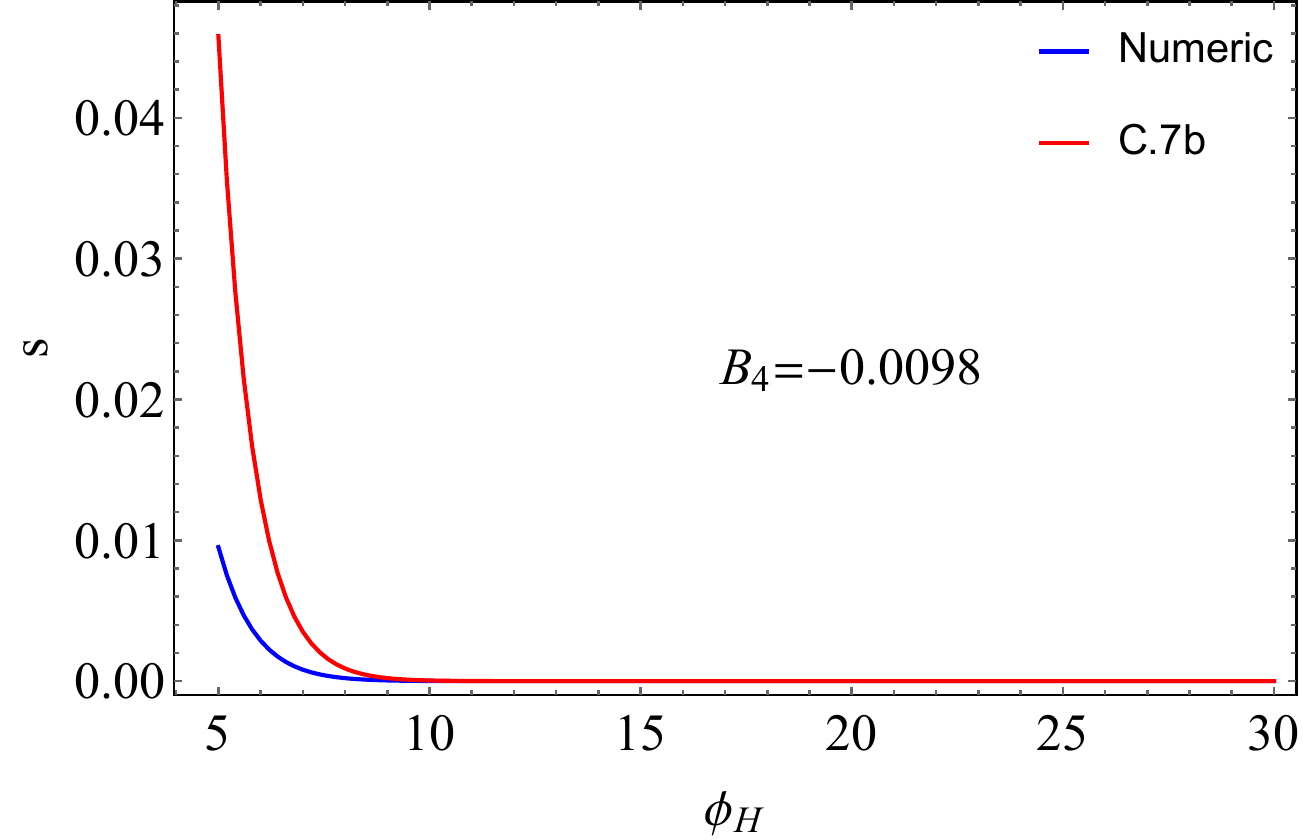}
    \includegraphics[width=0.389\textwidth]{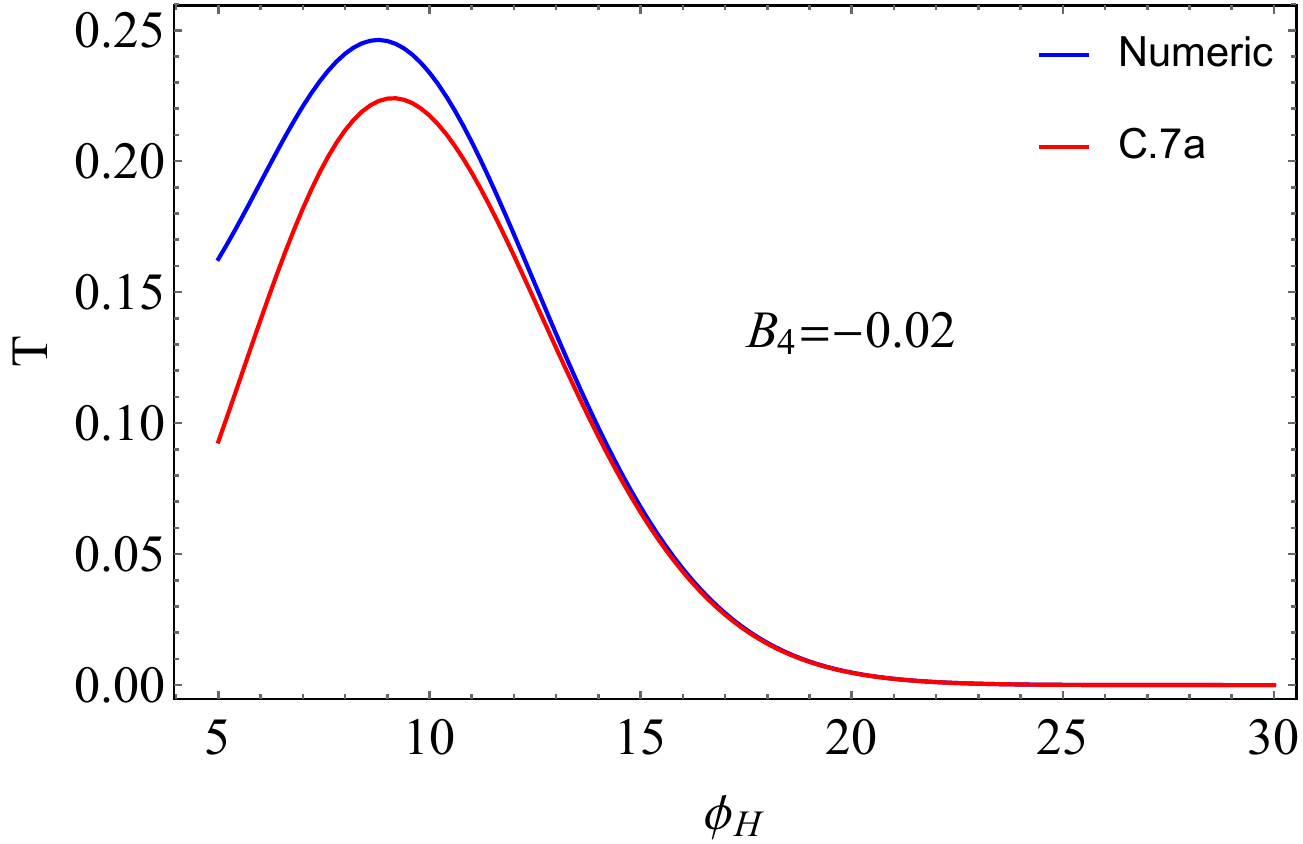}
    \includegraphics[width=0.4\textwidth]{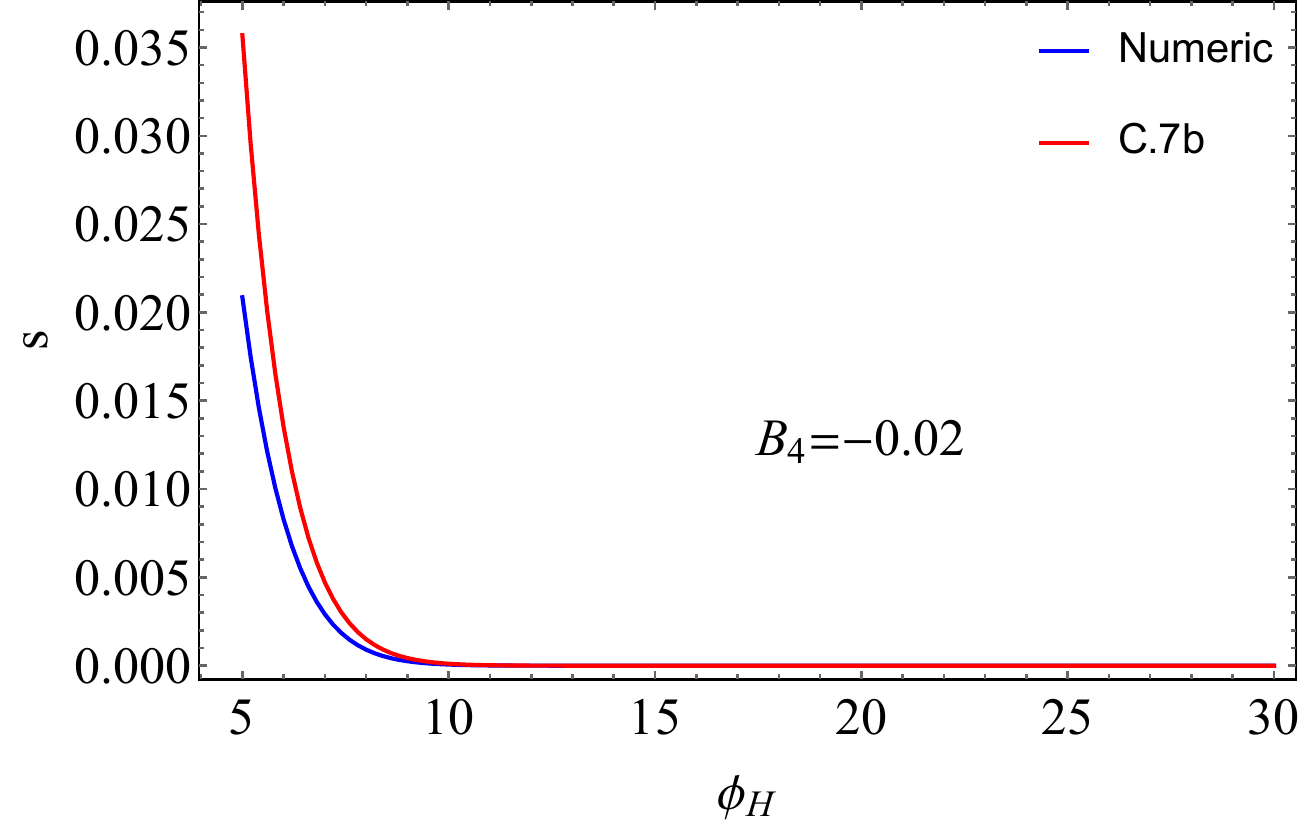}
    \caption{The left column of the plot shows the temperature, while the right column shows the entropy density. The first row corresponds to crossover ($B_4 = 0$), the second row is for second-order ($B_4 = - 0.00983491$), and the third row belongs to first-order ($B_4 = - 0.02$) phase transition.  The blue plots correspond to numeric results obtained from Eq. \eqref{eq: thermodynamics}, while the red plots result from Eqs. \eqref{eq:low-T B4is0} and \eqref{eq:low-T B4not0}.}
    \label{fig:low-T}
\end{figure}

As mentioned earlier, the low-temperature regime in all cases corresponds to large values of $\phi_H := u_H$. To understand the physics in this regime and also to cross-check our numerical results, in Fig.~\ref{fig:low-T} we show the temperature and the entropy density as functions of $\phi_H$. The red lines represent the analytical expressions in low-temperature limits, and the blue lines correspond to the numeric calculations. We summarize the analytical results in the following:\\
$B_4=0:$
\begin{subequations}\label{eq:low-T B4is0}
\begin{align}
& T=\frac{ \phi _H}{12 \pi }\exp\left({-\frac{\phi _H^2}{24}}\right),\\
& s= \frac{2 \pi}{\phi _H^3} \exp\left({-\frac{\phi _H^2}{8}}\right), \\
& c_s^2= \frac{1}{3}-\frac{8}{\phi _H^2}+\mathcal{O}\left({\phi_H}^{-3}\right),
\end{align}
\end{subequations}
$B_4\leq0\, (\text{and }  -\ln{\vert B_4\vert}\ll  \phi_H):$
\begin{subequations}\label{eq:low-T B4not0}
\begin{align}
& T= \frac{2^{\frac{1}{96 B_4}} \left(-B_4\right){}^{\frac{1}{192 B_4}+\frac{3}{2}} e^{-\frac{\phi _H^2}{48}} \phi _H^{\frac{1}{96 B_4}+4}}{3 \pi },\\
& s=2 \pi \frac{2^{\frac{1}{32 B_4}+3} e^{-\frac{\phi _H^2}{16}} \left(-B_4 \phi _H^2\right){}^{\frac{1}{64} \left(\frac{1}{B_4}+96\right)}}{\phi _H^3}, \\
& c_s^2= \frac{1}{3}-\frac{32}{\phi _H^2}-\frac{8}{B_4 \phi _H^4}+\mathcal{O}\left({\phi _H}^{-5}\right).
\end{align}
\end{subequations}

From Fig. \ref{fig:low-T}, we can infer that the analytical results converge to numerical results at large $\phi_H$. However, for cross-over transition, this convergence occurs for smaller $\phi_H$s, while for second- and first-order transitions, it happens for larger values of $\phi_H$.

\section{Radius of convergence in high temperatures}\label{sec: short}

In the spin-0 sector (sound channel), two coupled equations come from the metric perturbation and scalar field perturbation. At high-temperature regimes, the radius of convergence of the hydrodynamic series does not approach its CFT value that is computed in \cite{Grozdanov:2019uhi}. This is due to the coupling of the above two equations. The radius of convergence is computed by the collision of the hydro mode with the lowest scalar field non-hydro mode. One key observation is that the collision happens in complex momentum with the phase $\theta$ close to $\pi$ for $\phi_H= 0.1$ (independent of the value of $B_4$ in the potential) \footnote{With the SUGRA potential $V=-4-8 \cosh \left(\frac{\phi}{\sqrt{2}}\right)+\sinh ^2\left(\frac{\phi}{\sqrt{2}}\right)$ and with the simplest potential $V=-12-\frac{3}{2}\phi^2$ we found the same result for $\phi_H=0.1$. }
\be\label{eq: highT-sound-collision}
q^2_c=1.486 \exp(i \theta_c),\qquad \theta_c=0.988 \pi.
\ee

To understand the underlying mechanism, we can recall that the lowest collision between the non-hydro modes in the external scalar field channel and the spin-0 sector always occurs at $\theta=\pi$. This is also what happens for the lowest collision of the hydro-modes with each other in the spin-0 sector for AdS-Schwarzschild black hole \cite{Grozdanov:2019uhi}. Therefore, we can compare the QNMs in the spin-0 sector of an AdS-Schwarzschild with purely imaginary momenta $(\theta=\pi)$ and the QNMs of an external scalar field on the same geometry to understand the above observation. 

\begin{figure}[htb]
\centering
\includegraphics[width=0.45\textwidth, valign=t]{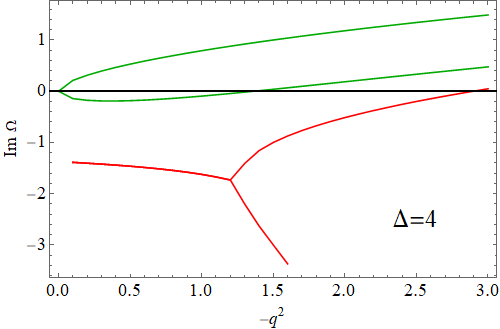} 
\hspace{0.25cm}
\includegraphics[width=0.45\textwidth, valign=t]{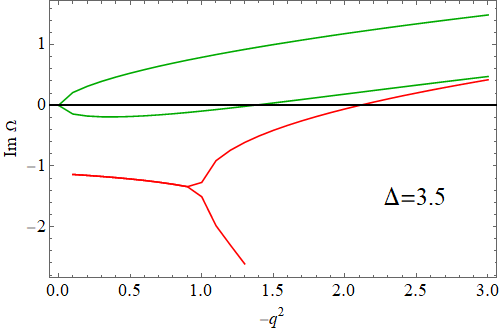}\\
\vspace{0.5cm}
\includegraphics[width=0.45\textwidth, valign=t]{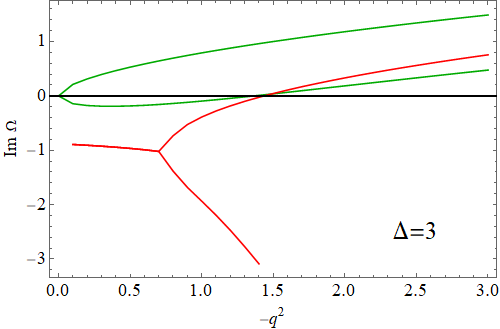} 
\hspace{0.25cm}
\includegraphics[width=0.45\textwidth, valign=t]{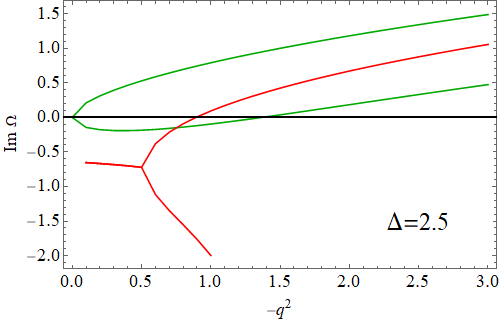}
\caption{ The dimensionless QNM frequencies $\Omega\equiv\frac{\omega}{2\pi T}$ as functions of purely imaginary momentum in units of $q\equiv \frac{k}{2\pi T}$. Green lines: sound mode. The green lines represent the sound mode, while the red lines depict the lowest non-hydro modes for an external scalar field with mass  $m^2=\Delta (\Delta-4)$.}
\label{fig: scalar}
\end{figure}

In Fig. \ref{fig: scalar}, we show the QNMs of an external scalar field (red lines) with various conformal weights (masses) and hydrodynamic sound modes (green lines) for purely imaginary momenta. Note that the radius of convergence of the sound hydro mode is $\vert q^2\vert=2$ \cite{Grozdanov:2019uhi}. Therefore, we look at the regime $\vert q^2\vert\leq2$. As we can see, there is no coincidence of the modes for $\Delta=(3.5, 4)$ in the selected range of $q^2$. However, for smaller $\Delta$'s, it happens. In particular, for $\Delta=3$, the modes coincide at $q^2=1.485$, which is very close to what we observe in our model \eqref{eq: highT-sound-collision}. The small deviation of \eqref{eq: highT-sound-collision}  is due to the coupling between the channels even in extremely high temperatures.

Another important point is that the linearized equation for the gravity modes in the sound channel is decoupled from the scalar field perturbation at high-temperature regimes, but not vice versa. We found similar behavior in the small momenta limit in our earlier studies \cite{Janik:2016btb}. The underlying message of this observation is as follows. If we want to study the perturbations around a state in the high-temperature regime of a deformed CFT, i.e.  $\mathcal{L}=\mathcal{L}_\textrm{CFT}+j^{4-\Delta} O_\Delta$, taking the high-temperature limit and investigating the linearized dynamics do not commute. In other words, one must find the linearized equations of the full theory and then take the high-temperature limit. What we found is that after these two steps, one still finds a coupling between the perturbations of the sound channel and the scalar channel.

\begin{figure}[htb]
\centering
\includegraphics[width=0.65\textwidth, valign=t]{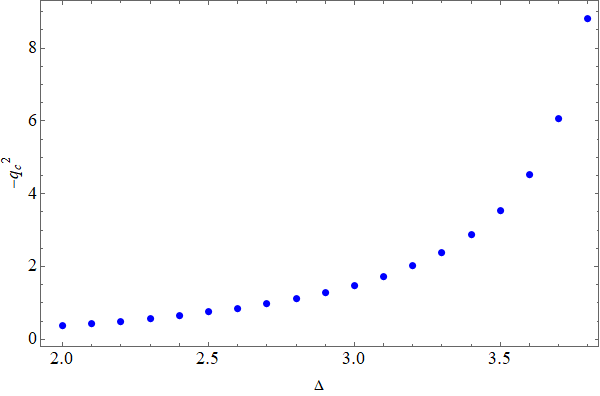}
\caption{
The critical value of the purely imaginary momentum as a function of conformal weight.}
\label{fig: AllDelta}
\end{figure}

In the left panel of Fig. \ref{fig: AllDelta}, we show the results for a full range of conformal weights, and in the right plot of this figure, we show the critical value of the complex momenta as a function of conformal weight. One interesting feature in Fig. \ref{fig: AllDelta} is that it seems that at $-q^2\rightarrow\infty$ the lowest non-hydro mode of an external scalar field with conformal weight $2\leq\Delta\leq4$ is bounded between the sound (gravity) hydro modes. Also, one may guess that the collision between the sound mode and the non-hydro scalar mode for $\Delta=4$ (massless scalar field) occurs at $\vert q_c\vert\gg2$. That can be seen in both the left and right panels. The left panel also shows that the non-hydro mode for $\Delta=2$ is asymptotic to the other sound hydro mode.

The radius of convergence fixes the applicability of the hydrodynamic expansions in the theory.
Since the radius of convergence even at high temperatures is fixed by the conformal dimension of the deformation term in the Lagrangian, one cannot simply take the high-temperature limit first and then apply the hydrodynamic description. Especially for $\Delta<2.71$, where $\vert q_c^2\vert<1$, the gradient expansion should be taken carefully. The smallest radius of convergence is associated with the operator saturating BF-bond with $\vert q_c^2\vert=0.365$.

\subsection{Large momenta}\label{sec:short-large-q}

\begin{figure}[htb]
\centering
\includegraphics[width=0.7\textwidth]{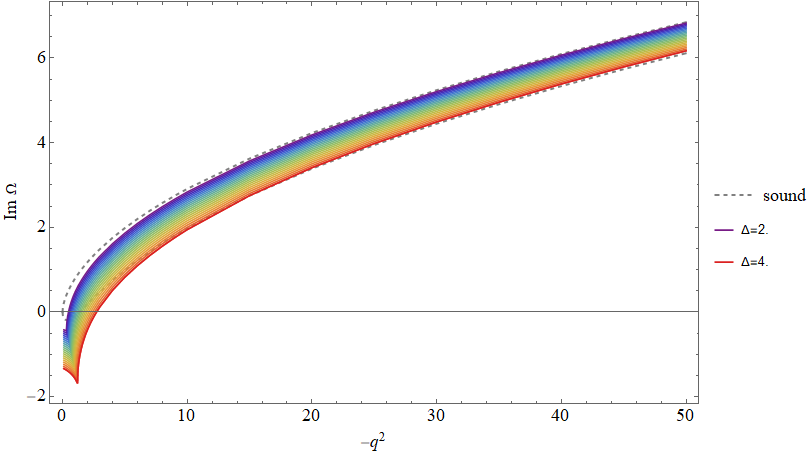}
\caption{ The plot of $\text{Im} \Omega$ in a large region of $-q^2$.}
\label{fig: AllDeltaHighq2}
\end{figure}

At large purely imaginary momenta, the sound hydro modes can provide some information about the non-hydro modes of an external scalar field.  Recall that $\Delta= 2$ is the BF-bound and $\Delta = 4$ corresponds to a marginal operator. However, this is not always the case. In extremely large momenta (e.g. $q^2=-1000$) all the lowest non-hydro modes of external scalar field approach to the upper dashed line in Fig. \ref{fig: AllDeltaHighq2}. 
\end{appendix}
\bibliographystyle{fullsort}
\bibliography{references}
\end{document}